\documentclass[12pt, draftclsnofoot, onecolumn]{IEEEtran}

\usepackage{graphicx} 
\usepackage{footnote}
\usepackage{amsmath}
\usepackage{mathtools}
\usepackage{amssymb} % gives real numbers R
\usepackage{enumerate}
\usepackage{bbm} % For indicator function notation
\usepackage{array}
\usepackage{gensymb}

\usepackage{subcaption}
\DeclareCaptionLabelSeparator{periodspace}{.\quad}
\usepackage{pgfplots}
\pgfplotsset{compat=1.11}

\usepackage{verbatim}

\newcommand{\matr}[1]{\mathbf{#1}}

\DeclarePairedDelimiterX\setc[2]{\{}{\}}{\,#1 \;\delimsize\vert\; #2\,} % Used for set notation (mathtools package)

\makeatletter
\newsavebox{\mybox}\newsavebox{\mysim}
\newcommand{\distras}[1]{%
  \savebox{\mybox}{\hbox{\kern3pt$\scriptstyle#1$\kern3pt}}%
  \savebox{\mysim}{\hbox{$\sim$}}%
  \mathbin{\overset{#1}{\kern\z@\resizebox{\wd\mybox}{\ht\mysim}{$\sim$}}}%
}% for i.i.d.

\usepackage{amsthm}

\newtheorem{theorem}{Theorem}
\newtheorem{corollary}{Corollary}
\newtheorem{lemma}{Lemma}

\theoremstyle{remark}
\newtheorem*{remark}{Remark}

\theoremstyle{definition}
\newtheorem{definition}{Definition}

\theoremstyle{definition}
\newtheorem{assumption}{Assumption}

\theoremstyle{definition}
\newtheorem{approximation}{Approximation}

\usepackage[cmintegrals]{newtxmath}
\usepackage{url}

\usepackage{caption}
\usepackage{setspace}

\begin{document}

% *******   TITLE   *******
\title{Characterizing the First-Arriving Multipath \\[-0.9ex] Component in 5G Millimeter Wave Networks:\\[-0.9ex] TOA, AOA, and Non-Line-of-Sight Bias \\[-1.0ex]
\thanks{This paper was presented in part at the 2019 IEEE Global Communications Conference, Waikoloa, HI \cite{Globecom}.\\[-4.1ex]}% <-this % stops a space
\thanks{The authors are with Wireless@VT in the Bradley Department of Electrical and Computer Engineering, Virginia Tech,\\[-1.0ex] Blacksburg, VA 24061 USA. Email: \{olone, hdhillon, buehrer\}@vt.edu\\[-4.3ex]}
\thanks{The work of C. E. O'Lone was supported by the Bradley Graduate Fellowship and by the Collins Aerospace Fellowship.  The \\[-1.0ex] work of H. S. Dhillon was supported by the U.S. NSF under Grant ECCS-1731711.\\[-4.3ex]}}
%\thanks{The work of H. S. Dhillon was supported by the U.S. NSF under Grant ECCS-1731711.}}
\author{Christopher~E.~O'Lone,~\IEEEmembership{Student~Member,~IEEE,}
		Harpreet~S.~Dhillon,~\IEEEmembership{Senior  \\[-0.8ex] Member,~IEEE,} 
		and~R.~Michael~Buehrer,~\IEEEmembership{Fellow,~IEEE} \vspace{-12pt}}% <-this % stops a space

\maketitle

% ******* ABSTRACT *******
\vspace{-47pt}
\begin{abstract}
\vspace{-10pt}

%	This paper presents an analytical propagation model for Fifth-Generation (5G) millimeter wave (mm-wave) cellular that allows for the derivation of important localization metrics.  
	This paper presents a stochastic geometry-based analysis of propagation statistics for 5G millimeter wave (mm-wave) cellular.  In particular, the time-of-arrival (TOA) and angle-of-arrival (AOA) distributions of the first-arriving multipath component (MPC) are derived.  These statistics find their utility in many applications such as cellular-based localization, channel modeling, and link establishment for mm-wave initial access (IA).
% Took out:  and channel modeling
Leveraging tools from stochastic geometry, a Boolean model is used to statistically characterize the random locations, orientations, and sizes of reflectors, \emph{e.g.}, buildings.  Assuming non-line-of-sight (NLOS) propagation is due to first-order (\emph{i.e.}, single-bounce) reflections, and that reflectors can either facilitate or block reflections, the distribution of the path length (\emph{i.e.}, absolute time delay) of the first-arriving MPC is derived.  This result is then used to obtain the first NLOS bias distribution in the localization literature that is based on the absolute delay of the first-arriving MPC for outdoor time-of-flight (TOF) range measurements.  This distribution is shown to match exceptionally well with commonly assumed gamma and exponential NLOS bias models in the literature, which were only attained previously through heuristic or indirect methods. Continuing under this analytical framework, the AOA distribution of the first-arriving MPC is derived, which gives novel insight into how environmental obstacles affect the AOA and also represents the first AOA distribution derived under the Boolean model.  
% Took out this line above at % sign: Due to the difficultly in measuring the NLOS bias phenomenon and consequent lack of empirical data, the derivation of such analytical bias models is of critical importance. 

\end{abstract}

% *******   KEYWORDS   *******
\vspace{-15pt}
\begin{IEEEkeywords}
\vspace{-11pt}
Localization, range measurement, non-line-of-sight (NLOS) bias, time-of-flight (TOF), time-of-arrival (TOA), angle-of-arrival (AOA), multipath component (MPC), stochastic geometry, Boolean model, Poisson point process (PPP), millimeter wave (mm-wave), first-order reflection, independent blocking.
% Kullback-Leibler (KL) divergence. 
% Cram\'{e}r-Rao lower bound
\end{IEEEkeywords}

% *******   INTRODUCTION   *******
\vspace{-15pt}
\section{Introduction} \label{Sec_Intro}
\vspace{-5pt}

	The past decade has seen tremendous advances in both the understanding and characterization of the mm-wave channel for 5G cellular.  These advances have been realized, in part, by the development of stochastic geometry-based analytical models.  One of the most tractable stochastic geometry tools employed to study mm-wave propagation, and consequently the most widely-utilized, is that of the \emph{Boolean model}, which is able to statistically capture the randomness in the locations, sizes, and orientations of environmental obstacles (\emph{e.g.}, buildings) \cite{Stoyan}.  The pioneering work first utilizing the Boolean model to study mm-wave propagation in cellular networks was conducted in \cite{Heath}.  While an excellent examination of the Boolean model's usefulness in studying blockage effects, the analysis in \cite{Heath} was focused on line-of-sight (LOS) links, and thus, the study of NLOS propagation under this model remained an open problem.

% In addition to empirical models generated via measurement campaigns, analytical models have also played a critical role in gaining insight into millimeter wave propagation.  Central to the development of many of these analytical models has been the use of stochastic geometry. 
	
	An important aspect of mm-waves is that diffraction effects are negligible while reflections dominate NLOS propagation \cite{AndrewsHeath}.  Thus, while \cite{Heath} did not study multipath effects under the Boolean model, other subsequent works have.  In \cite{Nor} for example, first-order reflections were incorporated into the Boolean model to derive the power delay profile (PDP). %, where the analysis was done under the assumption that all buildings have the same orientation per a channel realization.
% mention worse independent blocking assumption, not what we want, and under the condition 
The work in \cite{GDasConf} extends that of \cite{Nor} by considering buildings with random orientations and equipping the transmitter (Tx) and receiver (Rx) with directional antennas.  Channel characteristics were then derived under first-order reflections and independent blocking.  In \cite{Miaomiao}, the Boolean model under first-order reflections and independent blocking was used, along with a point process of transmitters, to ``quantify the total amount of network interference.''  Lastly, the work in \cite{Comms_Letter} also utilized a similar model to derive the probability an anchor-mobile pair can perform single-anchor localization.

	%While the above works all incorporate reflections into the Boolean model, they unfortunately do not derive the TOA and AOA statistics of the \emph{first-arriving} multipath component (MPC) experienced at the base station or mobile for a single link; which, as will be discussed shortly, is what we desire.  Additionally, many of the models above are particularly restrictive.  

	While many works incorporate NLOS propagation into the Boolean model, including the ones above, they unfortunately either have restrictive setups or they do not derive our metrics of interest, namely, the TOA and AOA statistics of the first-arriving MPC experienced at the mobile for a single link.  For example, the model in \cite{Nor} does not consider random orientations of buildings in a given Boolean model realization, and in \cite{GDasConf}, the assumption of directional antennas %at the base station and mobile 
pointed in \emph{fixed} directions makes it difficult to determine whether other NLOS paths are available from different directions.  Finally, while \cite{Aroon} does find the TOA of the first-arriving MPC, the model assumes one-dimensional reflectors and only the reflector closest to the mobile is responsible for facilitating the first-arriving reflection, an assumption that does not often hold.

	The obvious next question is: \emph{Why are the TOA and AOA statistics of the first-arriving MPC important?}  We begin with the TOA.  In addition to its use in establishing an absolute timing reference for channel modeling purposes, the TOA distribution of the first-arriving MPC is perhaps most useful for addressing the NLOS bias problem that arises in  localization.  The NLOS bias problem can be summarized as follows.  Consider a range measurement where the distance between a base station and a mobile is measured via the TOF of a signal transmitted by either node.  In a LOS scenario, multiplying the TOF by the speed of light yields the true base station-mobile separation distance.  However, the LOS path is often blocked and the resulting reflected, diffracted, or scattered signal will lead to an erroneously larger range estimate than the true separation distance.  This distance the signal travels \emph{in excess} of the true LOS distance is termed the \emph{NLOS bias}.  When these positively biased range measurements are used to perform localization, they ultimately lead to a biased/poor target position estimate.  
	
	For a noiseless NLOS, TOF range measurement, the range and bias are related simply by: 
\vspace{-14pt} 
\begin{align} \label{The_Range_Equation} 
R = d + B, \\[-7.5ex] \nonumber
\end{align} 
where $R$  is the distance (\emph{i.e.}, range) measured via the TOF, $d$ is the true separation distance, and $B$ is the NLOS bias ($P[B>0] = 1$) \cite{DrB}.  In many NLOS localization algorithms and analyses, having \emph{a priori} information regarding the statistics of the bias error (\emph{e.g.}, the distribution of $B$ in (\ref{The_Range_Equation})) can improve NLOS detection \cite{Cong}, improve estimator performance \cite{Reza}, and reduce the Cram\'{e}r-Rao Lower Bound on positioning error as well \cite{Qi_NLOS_Journal}.  Thus, obtaining an accurate distribution of the NLOS bias is of vital importance for geolocation in NLOS environments.

	% , can in practice reach up to a few hundred meters or even a few kilometers  \cite{killer_issue}, \cite{Qi_NLOS_Journal}

	 Although the NLOS bias problem has been around since the inception of range-based localization, there are, at present, no agreed upon statistical distributions characterizing the bias error in outdoor environments, such as urban canyons (see Sec. \ref{Sec_Discussion}).  While the localization literature does offer a variety of bias distributions, they are either: 1) chosen simply due to their tractability or desirable features such as having a positive support, \emph{e.g.}, half-Gaussian \cite{Mailaender}, Rayleigh \cite{GMix1}, positive uniform \cite{Unif}, and gamma \cite{Qi_2}; 2) chosen based on simple point scattering models from the channel modeling literature \cite[Sec. III-B]{Qi_NLOS_Journal}, \cite{Ertel}, \cite{Wu}; or 3) chosen indirectly based on the delay of the first-arriving MPC from empirical \emph{excess} delay LOS PDP models, such as the commonly used exponential distribution \cite{Chen}, \cite{Swaroop}, and not directly via the first-arriving MPC from \emph{absolute} delay NLOS PDPs.\footnote{\vspace{-7pt} \emph{Absolute} delays are measured w.r.t. the transmission time.  \emph{Excess} delays are measured w.r.t. the first-arriving detected signal.}  Since range measurements are often triggered on the first-arriving signal \cite{Qi_NLOS_Journal}, \cite{Qi_Multipath}, then if the LOS path is blocked, \emph{the first-arriving MPC (i.e., the first-arriving reflection assumed here) will be responsible for triggering the range measurement}.   Thus, in an NLOS scenario, obtaining the absolute delay (\emph{i.e.}, TOA or path length) distribution of the first-arriving MPC will yield the distribution of the range measurement, and subtracting from this the true base station-mobile separation distance will yield the distribution of the NLOS bias via (\ref{The_Range_Equation}). 
% (see \cite{Qi_NLOS_Journal}, \cite{Qi_Multipath} and the references therein) 
%	 From the above discussion, it should be clear that finding an accurate TOA distribution of the first-arriving MPC in an NLOS scenario will directly lead to an accurate distribution of the NLOS bias.  
Unfortunately, obtaining this TOA distribution empirically, through an outdoor measurement campaign, is a difficult task (see Sec. \ref{Sec_Discussion}).  Thus, an accurate \emph{analytical} solution is needed.  Towards this end, this paper derives the  TOA distribution of the first-arriving MPC under the Boolean model, which characterizes the TOA over \emph{all} random placements of environmental obstacles.\footnote{Although this distribution of the TOA of the first-arriving MPC is derived irrespective of blocking on the LOS path, and hence is the true TOA distribution of the first-arriving MPC, our system model allows for this distribution to also apply exclusively to scenarios where the LOS path is blocked (Sec. \ref{Sec_Assumptions}).  This is investigated further in Sec. \ref{Sec_Numerical_Results1}.}  
This result then yields the NLOS bias distribution; thus filling the void in the localization literature by offering the first bias distribution derived via the absolute delay of the first-arriving MPC and under a comprehensive stochastic geometry framework.

	We now address the AOA.  As mm-waves ``generally require high directionality to achieve a sufficient signal-to-noise-ratio (SNR),'' beam sweeping, a process whereby angular sections are tested and checked for a received signal, will be needed to establish links in the IA phase \cite{IAPaper}.  Since the first-arriving MPC is often likely to be the dominant MPC, understanding its AOA distribution will be important for developing techniques to improve the angular search space in the IA phase; reducing the time it takes to close a link.  Thus, this paper also derives the AOA distribution of the first-arriving MPC, which is the first AOA distribution derived under the Boolean model.   Additionally, this AOA distribution offers a distinct advantage over AOA distributions derived under older omni-directional scattering models, \emph{e.g.}, \cite{Ertel}, \cite{Swamy}, since point scatters can not capture blocking effects nor the dominant reflection effects of mm-waves.%Moreover, while many works before have analytically derived AOA distributions, these distributions were developed under primitive circular and elliptical omni-directional scattering models, \emph{e.g.}, \cite{Ertel}, \cite{Swamy}, that come from the older channel modeling literature.   These older models are at a distinct disadvantage when compared to the Boolean model since point scatters can not capture blocking effects nor the dominant reflection effects of mm-waves. %, and furthermore, the circular models can only capture local effects around the Tx and Rx.  
%Thus, this paper presents the first AOA distribution derived under the more comprehensive Boolean model which better captures mm-wave propagation effects.

% *** Additionally, this distribution could be used to aid sector-based localization and could be used in mm-wave channel simulators as well.

%	In light of the gaps in the localization literature regarding a proper/accurate NLOS bias characterization, as well as the lack of a characterization of the first-arriving MPC in the stochastic geometry literature, this paper offers the following contributions:
Given the gaps in the stochastic geometry and localization literature, our contributions are:
%This paper offers the following contributions:
\vspace{-2pt}
\begin{enumerate}
% 1) My statistical modeling setup - A framework to handle first-order reflections under the boolean model  which differs from those in the literature in that it (reflection hyperbola too) deals with easy to handle finite probability events, as opposed to intuition-depriving infinitesimal events.
% *** 2) I'm not mentioning the extension of the Lemma 5 results and Appendix A which I did...whatevs
\item The derivation of the TOA distribution of the first-arriving MPC under the Boolean model, both with and without blocking, which yields the NLOS bias distribution.
%which yields the first NLOS bias error model based on this desired metric for outdoor environments.
%analytically derived distribution of NLOS bias error for outdoor environments (and for mm-wave).
\item An analysis highlighting the close connection between this NLOS bias distribution and the exponential and gamma bias model assumptions in the localization literature.% , suggesting these decades-old assumptions should perhaps be the standard bias models moving forward.
%thus finally offering the first \emph{analytical} support for these choices. %This is also the first proposed analytical NLOS bias model for 5G mm-wave.
%\item A numerical analysis comparing the analytically derived bias distribution against those generated via simulation.  %This analysis highlights the accuracy of the model and discusses corner cases regarding model applicability. 
\item A discussion regarding the lack of outdoor measurement data characterizing NLOS bias and what information about the bias can be gleaned from measurements that do exist.  %This lack of measurement data emphasizes the importance of the analysis undertaken in this paper.
%\item Lastly, a discussion is presented regarding both the reasons surrounding a lack of outdoor measurement data on the NLOS bias and the information about the bias that can be gleaned from measurements that currently exist.  This lack of measurement data further emphasizes the importance of the analysis undertaken in this paper.
\item The derivation of the AOA distribution of the first-arriving MPC under the Boolean model, which represents the first AOA distribution derived under the Boolean model.
\item A numerical analysis of this distribution which reveals the connection between this AOA distribution and that derived under an elliptical, omni-directional scattering model.
%  and also reveals a fascinating connection between this AOA distribution and that derived under the elliptical omni-directional scattering models of the past.
\end{enumerate}

\vspace{-14pt}
\section{System Model} \label{Model_Assumptions}
\vspace{-4pt}

	This section first introduces the system model assumptions and then describes the Boolean model setup.  Next, a characterization of first-order reflections is given followed by a description of independent blocking.  Finally, an important result is presented regarding the number of reflectors facilitating visible (non-blocked) reflections. \emph{Common notation is given in Table \ref{Notation}.}

\vspace{-18pt}
\subsection{Assumptions} \label{Sec_Assumptions}
\vspace{-5pt}

\begin{assumption}[NLOS Propagation] \label{NLOS_Prop_Assump}
Only first-order specular reflections are considered. % for NLOS propagation. 
\end{assumption}

\vspace{-16pt}
\begin{remark}
First, specular reflections imply the angle-of-incidence (AOI) equals angle-of-reflection (AOR) at the point of incidence. Second, the effects of higher-order (\emph{i.e.} multiple-bounce) reflections are considered to be minimal due to increased pathloss and reflection losses, \emph{e.g.}, see \cite{Nor} and \cite{GDasConf}.  Third, diffraction effects are negligible at mm-wave frequencies \cite{AndrewsHeath} and hence are not considered.  Lastly, assuming only specular reflections implies that reflecting surfaces are sufficiently smooth such that scattering effects are negligible as well \cite{Nor}, \cite{SmoothSurfaces}.
%\footnote{While it is, of course, possible to detect higher-order reflections, detecting these reflections is rare due to the reasons stated and thus, these higher-order reflections are not considered in this analysis.}
\end{remark}

\vspace{-15pt}
\begin{assumption}[360$\degree$ Coverage] \label{Assumption_360}
%OLD: The base station and the mobile are both assumed to have 360$\degree$ coverage such that all possible reflection paths are illuminated.
The base station and mobile are equipped with either isotropic antennas or antenna arrays allowing 360$\degree$ beam sweeping, \emph{i.e.}, all reflection paths are illuminated.  %\cite{IAPaper} <--- No room for this but I do define beam sweeping and cite this in the last paragraph of the introduction
\end{assumption}

\vspace{-20pt}
\begin{assumption}[Independent Blocking] \label{Indep_Blocking_Assump}
Blocking on each received signal path is assumed to be independent.  Further, for each reflection, blocking on the incident path is assumed to be independent of blocking on the reflected path.
\end{assumption}

% 1) Reword assumption 3 since two reflectors can have same reflection path!!!  ---> Na these are technically separate reflection paths

\vspace{-15pt}
\begin{remark}
Treating blocking on each received signal path independently, \emph{i.e.}, \emph{independent blocking}, is a common assumption in the literature \cite{Nor}, \cite{GDasConf}, \cite{Miaomiao}.  Further, for each separate reflection path, treating blocking independently on the incident and reflected paths has been done previously in \cite{Comms_Letter}, and a similar treatment is also presented in \cite{IndepIndep}.  Sections \ref{Sec_Numerical_Results1} and \ref{Sec_Numerical_Results2} reveal that treating the incident and reflected paths independently matches true correlated blocking rather well.\footnote{By \emph{correlated blocking}, we mean the true blocking case that occurs in practice, where an obstruction can be responsible for blocking multiple paths at once (red oval, Fig. \ref{System_Model_Fig}).  This type of blocking is notoriously difficult to characterize analytically \cite{Aditya}.} 
\end{remark}

\vspace{-9pt}
\begin{remark}
Since blocking is independent on each received path, then blocking on the LOS path between the base station and mobile does not impact blocking on reflected paths.  Thus, whether the LOS path is blocked or not does not impact the analysis.  As such, we simply assume the LOS path is blocked for the NLOS bias analysis and can also simply ignore whether the LOS path is blocked or not in the AOA analysis.  This is explored further in Sections \ref{Sec_Numerical_Results1} and \ref{AOAsection}.
\end{remark}

\begin{table*}[t]% ---> puts at top %[h] 
\renewcommand{\arraystretch}{0.83}
\vspace{-3pt}
\caption{Summary of Notation} %title of the table
\label{Notation}
\vspace{-12pt}
\centering % centering table
%\fontsize{8pt}{9pt}   %The second argument to \fontsize is the baselineskip, set to 1.2 times the font size.
\footnotesize
\begin{tabular}{c l | c l } % creating eight columns
\hline\hline & & &\\[-1.3ex]%inserting double-line
\bfseries Symbol & \bfseries Description & \bfseries Symbol & \bfseries Description \\% [0.3ex]
\hline   & & &\\ [-1.2ex]% inserts single-line
$f_X(\cdot)$     & Probability distribution fn. (PDF) of RV $X$ & $F_X(\cdot)$   	& Cumulative distribution fn. (CDF) of RV $X$\\ 
$\textbf{Supp}(X)$ & Support of the RV $X$: $\{ x\in \mathbb{R} \,|\, f_X(x) > 0 \}$  &    $\mathbb{E}[X]$ 	& Expectation of the RV $X$ \\[0.5ex]
$P[A]$ &  Probability of event $A$  & $\lVert \cdot \rVert$  & The Euclidean norm  \\
$\delta(\cdot)$ & The Dirac delta function & $\mathbbm{1}[A]$ & Indicator Function, 1 if A true, 0 if A false  \\
$\mathbf{x}$ & Vector $\mathbf{x}$. All vectors are column vectors. & $[\mathbf{x}]_i$; $\mathbf{x} \cdot \mathbf{y}$  &  $i^\text{th}$ component of $\mathbf{x}$; Dot product of $\mathbf{x}$ with $\mathbf{y}$\\
$\mathbf{g}(x)$; $g(\mathbf{x})$  & Vector fn. of a scalar; Scalar fn. of a vector & $\matr{X}^T$; $\matr{X}^{-1}$  &	Transpose of matrix $\matr{X}$; Inverse of matrix $\matr{X}$ \\
$\mathbf{g}(\mathbf{x})$; $g(x)$ & Vector fn. of a vector; Scalar fn. of a scalar & $\mathbf{R}_\theta$  & The rotation matrix $\Bigl[\begin{smallmatrix}\cos \theta & \sin \theta \\ -\sin \theta & \cos \theta \end{smallmatrix}\Bigr]$ which \\
$\mathcal{L}_{[\mathbf{p}, \mathbf{q}]}$ & The set of points forming a \emph{line segment}& & rotates a vector clockwise by angle $\theta$  \\
&  between, and including, the points $\mathbf{p}$ and $\mathbf{q}$ & $\mathcal{Q}$; $\varnothing$ & Roman numeral set:$\{\text{I}, \text{II}, \text{III}, \text{IV}\}$; Empty set  \\
$\mu_n \big( \mathcal{A} \big)$ &  The $n$-dim. Lebesgue measure of set $\mathcal{A}$ & $\partial\mathcal{A}$ & Boundary of set $\mathcal{A}$ (closure minus interior) \\[0.5ex]
$\Phi$ & Set of points forming a Poisson Pt. Proc. & $\Phi(\mathcal{A})$ & The number of points of $\Phi$ in the set $\mathcal{A}$\\
$\mathcal{A} \oplus \mathcal{B}$ & Minkowski sum: For compact $\mathcal{A}, \mathcal{B} \subset \mathbb{R}^2$, & $\mathcal{A} \setminus \mathcal{B}$ & Set subtraction: $\{\mathbf{x} \in \mathcal{A} \,|\, \mathbf{x} \notin \mathcal{B} \}$  \\
& $\mathcal{A} \oplus \mathcal{B}\triangleq \big\{\mathbf{x}+\mathbf{y} \in \mathbb{R}^2 \,\big|\, \mathbf{x} \in \mathcal{A}, \mathbf{y} \in \mathcal{B} \big\}$ & $Q_\text{I}$   &  $1^\text{st}$ quadrant in $\mathbb{R}^2$:$\setc[\Big]{\! [x, y]^T \!\!\! \in \mathbb{R}^2 \!} {\! x \!\geq\! 0, y \!\geq\! 0 \!}$  \\
w.l.o.g.; s.t.  &  Without loss of generality; such that  & & Remaining quad.'s defined similarly.   \\
l.h.s.; r.h.s. & Left hand side; right hand side  & c.c.w.; w.r.t. & Counterclockwise; with respect to\\
\hline % inserts single-line
% --- Unused ---
% Angle of $\mathbf{x}$ w.r.t. $+x$ axis
\end{tabular}
\vspace{-20pt}
\end{table*}	

%$\delta(x)$    		& Dirac delta function & $u(x)$ & Unit step function, \emph{i.e.}, 0 for $x<0$ and 1 for $x \geq 0$ \\

\vspace{-19pt}
\subsection{Reflection Fundamentals Under the Boolean Model}	
\vspace{-3pt}

	This section first introduces the Boolean model setup and then uses this setup, along with the assumptions above, to derive results regarding first-order reflections.  We begin by formally defining our use of the term \emph{reflector}.

\begin{figure}[t]
\centering
\includegraphics [scale=0.41]{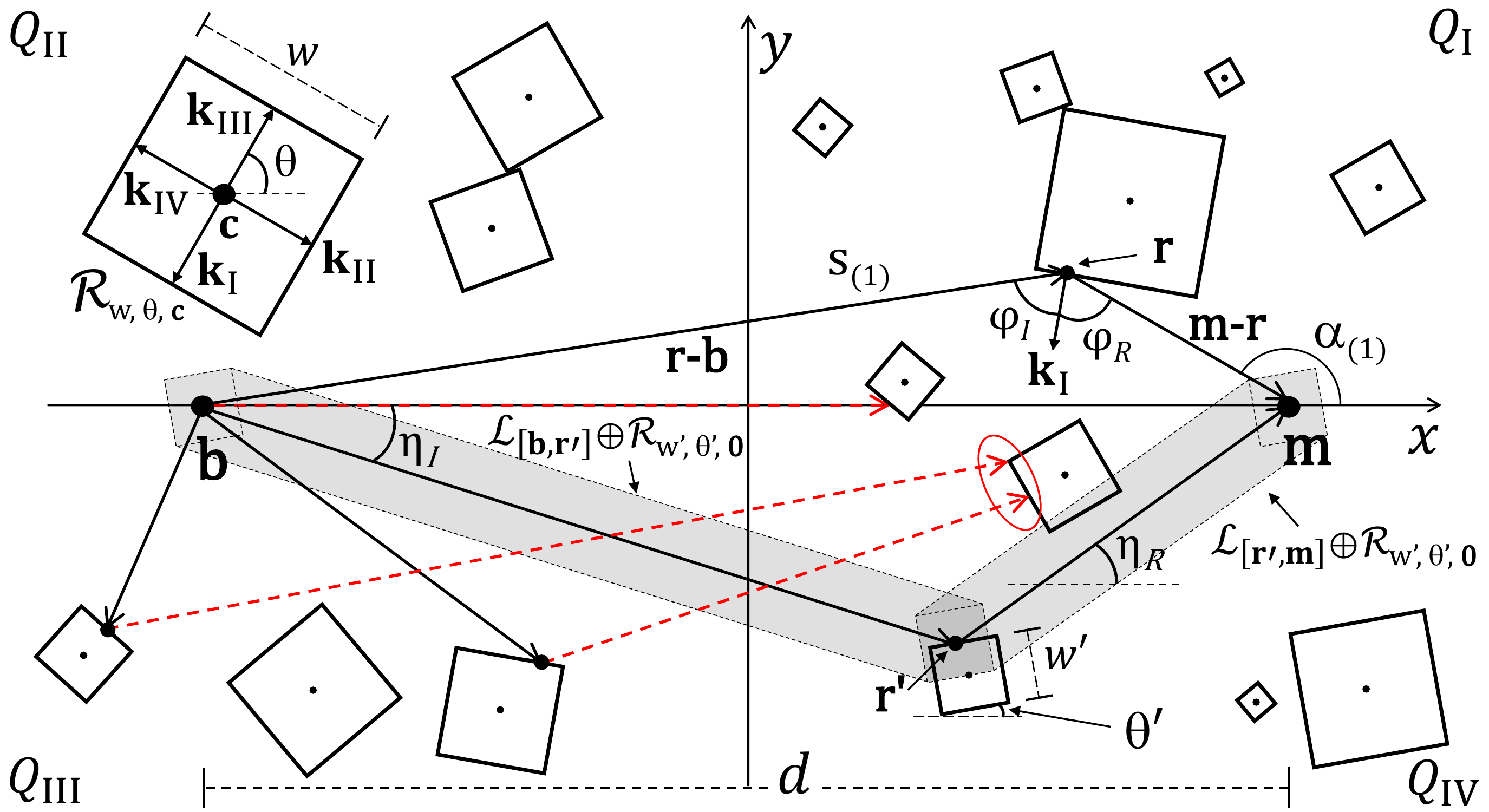}
\vspace{-5pt}
\caption{\textsc{System Model.} Depicted is a realization of the Boolean model over the test link setup, along with illustrations of various concepts and definitions from Sec. \ref{Model_Assumptions} such as: correlated blocking, independent blocking, reflection point characteristics, and a reflector's internal vectors.  The path length of the first-arriving reflection is $S_{(1)} = s_{(1)}$ and its AOA is $A_{(1)} = \alpha_{(1)}$. 
\\[-8ex]}
\label{System_Model_Fig}
\end{figure}

\vspace{-11pt}
\begin{definition}[Reflector $\mathcal{R}_{w, \theta, \mathbf{c}}$] \label{Reflector_Definition}
A \emph{reflector} $\mathcal{R}_{w, \theta, \mathbf{c}}$ is defined to be a square compact set with edge width $w \in (0, \infty)$, center point $\mathbf{c} \in \mathbb{R}^2$, and orientation $\theta \in (0, \pi/2)$, measured c.c.w. w.r.t. the $+x$-axis.  To aid in the analysis, we further define four internal vectors, $\mathbf{k}_\text{I}, \mathbf{k}_\text{II}, \mathbf{k}_\text{III}, \mathbf{k}_\text{IV}$, which are depicted in Fig. \ref{System_Model_Fig}, and of which $\mathbf{k}_\text{III}$ exhibits the reflector's orientation, $\theta$.
\end{definition}	

% *** TO DO ***
% Put this footnote in if reviewers ask...
% \footnote{As will soon be apparent, the orientation $\theta = \pi/2$ is  a pathological case under this setup and can be safely removed considering: 1) in reality, the probability a reflector is orientated \emph{exactly} parallel to the setup is zero, and 2) if one wants reflectors of this orientation, $\theta = \pi/2 \pm \epsilon$ for $\epsilon > 0$, will adequately suffice.}

\vspace{-12pt}
\begin{remark}
Although the term \emph{reflector} is used, a reflector $\mathcal{R}_{w, \theta, \mathbf{c}}$ can both facilitate a reflection and/or act as an impenetrable blockage.
\end{remark}

\vspace{-12pt}
\begin{definition}[Boolean Model of Reflectors, $\mathcal{B}$] \label{Def_BM}
Let $\Phi = \{\mathbf{c}_i\}_{i=1}^\infty$ be a \emph{homogeneous} PPP over $\mathbb{R}^2$ with intensity $\lambda>0$, let $\mathcal{U} = \{W_i\}_{i=1}^\infty$, $\mathcal{V} = \{\Theta_i\}_{i=1}^\infty$ be sequences of RVs representing the widths and orientations of reflectors, respectively, and let $\mathcal{U}, \mathcal{V}$, and $\Phi$ be mutually independent.  Next, let $\{W_i\}_{i=1}^\infty \! \distras{\text{\scriptsize i.i.d.}} \! f_W(w)$ and $\{\Theta_i\}_{i=1}^\infty  \! \distras{\text{\scriptsize i.i.d.}} \! f_\Theta(\theta)$, where $f_W = \text{unif}(w_{min}, w_{max}, n_w)$ and $f_\Theta = \text{unif}(\theta_{min}, \theta_{max}, n_\theta)$.\footnote{If $x,y \in \mathbb{R}$ with $x\leq y$ and $n>0$ is a finite integer, then we define ``$\text{unif}(x, y, n)$'' to be a discrete, uniform distribution with the support being the $n$ points equally spaced between (and including) $x$ and $y$.  The value of the PMF at each point in the support is consequently $1/n$.  Note, $x=y \implies n=1$ and $x<y \implies n>1$. Lastly, with the technical machinery developed in this paper, more general discrete distributions can be used as well.}  Then, the \emph{Boolean model of reflectors} is defined as: $\mathcal{B} \triangleq \bigcup_{i=1}^\infty \,  \mathcal{R}_{W_i, \Theta_i, \mathbf{c}_i}$.
\end{definition}

\vspace{-12pt}
\begin{remark}
Note, $\mathcal{B}$ is a \emph{random set} in $\mathbb{R}^2$.  A \emph{realization} of $\mathcal{B}$ is given when the widths, $\{W_i\}_{i=1}^\infty$, orientations, $\{\Theta_i\}_{i=1}^\infty$, and the PPP of center points, $\Phi$, are sampled according to the rules above.  We use `$\mathcal{B}$' for both the random set and its realization.  Its usage will be clear from context.
\end{remark}

\vspace{-12pt}
\begin{definition}[Test Link] \label{Def_TLS}
The \emph{test link} setup is defined to be that which places the base station at $\mathbf{b} \triangleq \big[- d/2, 0 \big]^T$ and the mobile at $\mathbf{m} \triangleq \big[d/2, 0 \big]^T$, for $d > 0$. (See Fig. \ref{System_Model_Fig}.)
\end{definition}

\vspace{-11pt}
\begin{remark}
The results derived in this paper apply w.l.o.g. to any translation and/or orientation of this test link setup due to the stationary and isotropic properties of the Boolean model \cite{Stoyan}. 
\end{remark}

% --- REFLECTOR AND SYSTEM MODEL ---
% --- NOTES ---
% 1) In MS Word: \mathcal{} = Lucia Handwriting OR Pristina
% 2) For symbols, use Symbols
% -------------------------------------------------------------------------------------------------		

% *** TO DO ***
% 1) Put in Figure reference below!!!

\vspace{-2pt}
	This Boolean model setup, along with Assumptions \ref{NLOS_Prop_Assump} and \ref{Assumption_360}, leads to some important geometric consequences regarding first-order reflections, which we now summarize.  We begin by considering a single reflector, $\mathcal{R}_{w, \theta, \mathbf{c}}$, with \emph{fixed} width and orientation, yet \emph{arbitrary} center point.  Next, let $\mathbf{r} \in \mathbb{R}^2$.  We say that $\mathbf{r}$ is a \emph{potential reflection point (PRP)} for $\mathcal{R}_{w, \theta, \mathbf{c}}$ if $\mathcal{R}_{w, \theta, \mathbf{c}}$ can be placed s.t. an edge of $\mathcal{R}_{w, \theta, \mathbf{c}}$ can intersect $\mathbf{r}$ to establish a first-order reflection at $\mathbf{r}$  (\emph{i.e.}, AOI$ \,=\, $AOR at $\mathbf{r}$). %, which are both measured w.r.t. a vector normal to the edge that would facilitate the reflection (see Fig. \ref{System_Model_Fig}).  
The following lemma lists all of the PRPs for $\mathcal{R}_{w, \theta, \mathbf{c}}$. %a given reflector.

\vspace{-8pt}
\begin{lemma}[Reflection Hyperbola] \label{Lemma_RH}
Let $\mathcal{H}_\theta$ be the set of all PRPs for reflector $\mathcal{R}_{w, \theta, \mathbf{c}}$. Then,
\vspace{-8pt}
\begin{align} \label{Lemma1_Equ}
\mathcal{H}_\theta = \Big\{ [x, y]^T \!\!\in \mathbb{R}^2  ~\Big|~ y^2 - x^2 + 2 \cot(2\theta) x y + d^2/4 = 0 \Big\}.
\end{align}
\end{lemma}

\vspace{-12pt}	
\begin{proof}
	First, note that $\mathbf{b}$ and $\mathbf{m}$ are always considered to be PRPs regardless of the reflector orientation, and thus trivially satisfy the lemma.\footnote{If an edge of $\mathcal{R}_{w, \theta, \mathbf{c}}$ were to intersect $\mathbf{b}$ or $\mathbf{m}$, to produce a reflection, this would, of course, be a pathological case, since there would be no incident or reflected path, respectively -- there would just be the LOS path between $\mathbf{b}$ and $\mathbf{m}$.  Thus, these two points are considered PRPs only to simplify the analysis.  This has no effect on results, as this event occurs with  zero probability.}  Next, we prove forward and reverse containment.
	
	\underline{\smash{($\subset$)}}:  Let $\mathbf{r} = [x, y]^T \! \in \mathcal{H}_\theta/\{\mathbf{b}, \mathbf{m}\}$.  Then $\theta \in (0, \pi/2) \!\implies\! \exists ! \, q \in \mathcal{Q}$ s.t. $\mathbf{r} \in Q_q$.  Next, let the vectors $\mathbf{r}-\mathbf{b}$ and $\mathbf{m}-\mathbf{r}$ represent the incident and reflected paths, respectively.  The AOI and AOR at $\mathbf{r}$ are given by  
\vspace{-9pt}
\begin{align*}
~~~~~~~~~~~~~~~~\varphi_I \triangleq \cos^{-1} \! \Bigg( \frac{\mathbf{k}_q \cdot (\mathbf{b}-\mathbf{r})} {\lVert \mathbf{k}_q \rVert \lVert \mathbf{b}-\mathbf{r} \rVert } \Bigg) ~~~~ \text{and} ~~~~ \varphi_R \triangleq \cos^{-1} \! \Bigg( \frac{\mathbf{k}_q \cdot (\mathbf{m}-\mathbf{r})} {\lVert \mathbf{k}_q \rVert \lVert \mathbf{m}-\mathbf{r} \rVert } \Bigg), ~~~~~~(\text{see Fig. \ref{System_Model_Fig}}) \\[-6.5ex]
\end{align*}
which are both measured w.r.t. the vector, $\mathbf{k}_q$, where $\mathbf{k}_q$ is normal to the edge that would facilitate the reflection in $Q_q$.  Since $\mathbf{r}$ is a PRP by implication, then $\varphi_I = \varphi_R$, and upon simplification, we have that for any $q \in \mathcal{Q}$, $y^2 - x^2 + 2 \cot(2\theta) x y + d^2/4 = 0$.  Thus, $\mathbf{r}$ is in the r.h.s. of (\ref{Lemma1_Equ}).

	\underline{\smash{($\supset$)}}:  Let $\mathbf{r} = [x, y]^T$ satisfy the r.h.s. of (\ref{Lemma1_Equ}), where $\mathbf{r} \neq \mathbf{b}, \mathbf{m}$ (these are already PRPs).  Since $\mathbf{b}$ and $\mathbf{m}$ are the only points on the $x$, $y$ axes that are in the r.h.s. of (\ref{Lemma1_Equ}), then $\exists ! q \in \mathcal{Q}$ s.t. $\mathbf{r} \in Q_q$, where $\mathbf{r}$ satisfies $y^2 - x^2 + 2 \cot(2\theta) x y + d^2/4 = 0$. For any $q$, we can work backwards from this equation to establish: $\frac{\mathbf{k}_q \cdot (\mathbf{b}-\mathbf{r})} {\lVert \mathbf{k}_q \rVert \lVert \mathbf{b}-\mathbf{r} \rVert }  =  \frac{\mathbf{k}_q \cdot (\mathbf{m}-\mathbf{r})} {\lVert \mathbf{k}_q \rVert \lVert \mathbf{m}-\mathbf{r} \rVert }$.  This implies $\varphi_I = \varphi_R$ at $\mathbf{r}$ and hence $\mathbf{r} \in \mathcal{H}_\theta$.
\end{proof}

\vspace{-8pt}
\begin{remark}
Note the following: 1) For $\mathbf{r} \in \mathbb{R}^2$, $\mathbf{r}$ is a PRP for $\mathcal{R}_{w, \theta, \mathbf{c}}$ if and only if $\mathbf{r} \in \mathcal{H}_\theta$; 2) the set condition in (\ref{Lemma1_Equ}) is a hyperbola, and thus we refer to the set of PRPs for $\mathcal{R}_{w, \theta, \mathbf{c}}$ as the \emph{reflection hyperbola for} $\mathcal{R}_{w, \theta, \mathbf{c}}$; 3) the reflector orientation, $\theta$, is only present in the ``$xy$'' term of the hyperbola equation, which implies that changing the orientation of the reflector results in a rotation of this hyperbola about the origin.  See Fig. \ref{Analytical_FW_Fig} for an example of $\mathcal{H}_{\theta = \pi/3}$.
%; and 4) $\mathbf{b}$ and $\mathbf{m}$ are \emph{not} the foci of this hyperbola, but rather lie on the hyperbola itself.
\end{remark}	

\vspace{-9pt}
\begin{remark}
Indeed, a reflector $\mathcal{R}_{w, \theta, \mathbf{c}}$ has uncountably many PRPs.  If $\mathcal{R}_{w, \theta, \mathbf{c}}$ were to actually intersect one of these PRPs with the appropriate edge, then a reflection would be established.\footnote{By \emph{appropriate edge}, we simply mean the edge oriented towards the base station and mobile that could facilitate reflections.  For PRPs in $Q_\text{I}$ only, for example, the appropriate edge facilitating reflections is the edge corresponding to the endpoint of $\mathbf{k}_\text{I}$.  For PRPs in $Q_\text{II}$ only, this would be the edge corresponding to the endpoint of $\mathbf{k}_\text{II}$.  Hence the internal vector labeling convention.}  We would then refer to this particular PRP as the \emph{reflection point (RP)}.  Thus, $\mathcal{R}_{w, \theta, \mathbf{c}}$ has many PRPs but can have only one RP.  \emph{Since this lemma states that all PRPs for $\mathcal{R}_{w, \theta, \mathbf{c}}$ lie on $\mathcal{H}_\theta$, then to check whether $\mathcal{R}_{w, \theta, \mathbf{c}}$ generates a reflection, one simply needs to check whether the appropriate edge of $\mathcal{R}_{w, \theta, \mathbf{c}}$ intersects the reflection hyperbola.}% for $\mathcal{R}_{w, \theta, \mathbf{c}}$.
\end{remark}

% Any point along the appropriate edge can be responsible for facilitating the reflection.

	We are oftentimes interested in reflection points corresponding to reflection paths less than or equal to a certain distance, $s$.  The terminology below aids in this characterization. 

\vspace{-10pt}
\begin{definition}[The $s$-Ellipse] \label{NLOSBE}
Under the test link setup, the \emph{s-ellipse} is defined as: $\mathcal{P}_s \triangleq \Big\{ [x,y]^T \in \mathbb{R}^2 ~\Big|~ x^2 \big/ u^2 + y^2 \big/ v^2 \leq 1 \Big\}$, where $u^2 = s^2 / 4$, $v^2 = (s^2 - d^2) / 4$, and $d < s < \infty$.   Further, for $s = d$ we set $\mathcal{P}_d \triangleq \lim_{s \to d} \mathcal{P}_s = \mathcal{L}_{[\mathbf{b}, \mathbf{m}]}$, and for $s = \infty$, we set $\mathcal{P}_\infty \triangleq \lim_{s \to \infty} \mathcal{P}_s = \mathbb{R}^2$. (See Fig. \ref{Analytical_FW_Fig}.) % $d \leq s \leq \infty$
\end{definition}	

%\begin{remark}
%This is an elliptical region since all reflections with a distance $s$ trace out an ellipse.  
%\end{remark}

\vspace{-7pt}
	Lastly, four PRPs exist which $\mathcal{R}_{w, \theta, \mathbf{c}}$ can intersect to generate reflections of \emph{exactly} $s$ meters. %These points are of particular importance and are highlighted below.
	
% *** TO DO ***
% 1) Put subscript $\theta_j$ on ALL boundary PRPs in WHOLE PAPER!!!???

\vspace{-11pt}
\begin{lemma}[Boundary PRPs] \label{Lemma_BPRPs}
Let $d \leq s < \infty$.  Consider the test link setup, a reflector $\mathcal{R}_{w, \theta, \mathbf{c}}$, and its corresponding reflection hyperbola, $\mathcal{H}_\theta$.  Then, $(\mathcal{H}_\theta \cap \partial \mathcal{P}_s) = \big\{ \mathbf{h}_\text{\emph{I}}, \mathbf{h}_\text{\emph{II}}, \mathbf{h}_\text{\emph{III}}, \mathbf{h}_\text{\emph{IV}} \big\}$, where
\vspace{-7pt} 
\begin{gather*}
\mathbf{h}_\text{\emph{I}} = \bigg[ \!\sqrt{z_\text{\emph{I, III}}}~\!, \frac{v}{u} \sqrt{u^2 - \!z_\text{\emph{I,III}}} \bigg]^T \! ,  ~
\mathbf{h}_\text{\emph{II}} =\bigg[ \!-\! \sqrt{z_\text{\emph{II, IV}}}~\!, \frac{v}{u} \sqrt{u^2 -\! z_\text{\emph{II,IV}}}\bigg]^T \! \!, ~~ \mathbf{h}_\text{\emph{III}} =-\mathbf{h}_\text{\emph{I}}, ~~ \mathbf{h}_\text{\emph{IV}} =-\mathbf{h}_\text{\emph{II}}, ~~~\text{and} \\
z_\text{\emph{I, III}} =\frac{s^4 \cot^2\!\theta} {4 \Big[ s^2 \csc^2\!\theta - d^2 \Big] }, ~~  z_\text{\emph{II, IV}} = \frac{ s^4 \tan^2\!\theta} {4 \Big[ s^2 \sec^2\!\theta - d^2 \Big] }.\\[-7ex]
\end{gather*}
The variables $u$ and $v$ are from Definition \ref{NLOSBE}.  (See Fig. \ref{Analytical_FW_Fig} for a depiction of these points.)
\end{lemma}

\vspace{-16pt}
\begin{proof}
Solve the system of two equations that define $\mathcal{H}_\theta$ and $\partial \mathcal{P}_s$. %(An alternate derivation of $\mathbf{h}_\text{II}$ is given in \cite{Nor}.)
\end{proof}

% Definitely keep \theta out of the subscript since that will be confusing with rotated coordinates later on
\vspace{-12pt}
\begin{remark}
The $\mathbf{h}_q$'s are functions of $s$, $d$, and $\theta$.  We write `$\mathbf{h}_q(s)$' to highlight the dependency on $s$, omit writing the dependency on $d$, and occasionally use a subscript $\theta$ to remind the reader when necessary.  The Roman numeral subscript on the $\mathbf{h}_q$'s denotes the quadrant in which it resides, for $s > d$.  For $s=d$, these points simplify to $\mathbf{b}$ and $\mathbf{m}$.  For $s=\infty$ not stated in the lemma, we set $\mathbf{h}_q(\infty) \triangleq \lim_{s \to \infty} \mathbf{h}_q(s) = [\pm \infty, \pm \infty]^T$, $\forall q \in \mathcal{Q}$, where `$\pm$' depends on the quadrant.
\end{remark}

\vspace{-15pt}
\subsection{Blocking} \label{Sec_Blocking}
\vspace{-4pt}

	This section briefly discusses what it means for a reflection point to be \emph{visible}, \emph{i.e.}, non-blocked under Assumption \ref{Indep_Blocking_Assump} (independent blocking).  As this treatment is analogous to that in \cite{Comms_Letter}, we only summarize the relevant results here, without proof.

\vspace{-11pt}	
\begin{definition}[Visible Reflection Point (VRP) for $\mathcal{R}_{w, \theta, \mathbf{c}}$] \label{Visible_Reflection_Point}
Let $\mathcal{B}$, $\mathcal{B}_1$, and $\mathcal{B}_2$ be realizations of i.i.d. Boolean models and let $\mathbf{r} \in \mathbb{R}^2$ be a RP for $\mathcal{R}_{w, \theta, \mathbf{c}} \subset \mathcal{B}$.  Then, the reflection path through $\mathbf{r}$ is \emph{visible} if $(\mathcal{B}_1 \cap \mathcal{L}_{[\mathbf{b}, \mathbf{r}]} ) \cup (\mathcal{B}_2 \cap \mathcal{L}_{[\mathbf{r}, \mathbf{m}]} ) = \varnothing$.  In this case, we say $\mathbf{r}$ is a \emph{visible RP for} $\mathcal{R}_{w, \theta, \mathbf{c}}$.
\end{definition}

\vspace{-15pt}
\begin{remark} 
As a simple example, if $\mathcal{B}$, $\mathcal{B}_1$, and $\mathcal{B}_2$ only contain reflectors of width $w^\prime$ and orientation $\theta^\prime$, then RP $\mathbf{r}^\prime$ in Fig. \ref{System_Model_Fig} is visible if no reflector center point from $\mathcal{B}_1$ falls in: $\mathcal{L}_{[\mathbf{b}, \mathbf{r}^\prime]} \oplus \mathcal{R}_{w^\prime, \theta^\prime, \mathbf{0}}$, and no reflector center point from $\mathcal{B}_2$ falls in: $\mathcal{L}_{[\mathbf{r}^\prime, \mathbf{m}]} \oplus \mathcal{R}_{w^\prime, \theta^\prime, \mathbf{0}}$.  When $\mathcal{B}$, $\mathcal{B}_1$, and $\mathcal{B}_2$ contain reflectors of many widths and orientations, the probability a RP is visible is given below. 
\end{remark}

\vspace{-17pt}
\begin{lemma}[Probability a Reflection Point is Visible \cite{Comms_Letter}] \label{Lemma_PRPV}
Consider the test link setup and let $\mathcal{B}$, $\mathcal{B}_1$, and $\mathcal{B}_2$ be i.i.d. Boolean models and $\mathbf{r}\in \mathbb{R}^2$ be a reflection point for $\mathcal{R}_{w, \theta, \mathbf{c}} \subset \mathcal{B}$.  Then, the probability that $\mathbf{r}$ is a VRP for $\mathcal{R}_{w, \theta, \mathbf{c}}$ is given by
\vspace{-10pt}
\begin{align*}
\rho(\mathbf{r}) &= P \big[(\mathcal{B}_1 \cap \mathcal{L}_{[\mathbf{b}, \mathbf{r}]} ) \cup (\mathcal{B}_2 \cap \mathcal{L}_{[\mathbf{r}, \mathbf{m}]} ) = \varnothing \big] = e^{-\lambda \mathbb{E}_{W, \Theta} \Big[ \mu_2 \big(\mathcal{L}_{[\mathbf{b}, \mathbf{r}]} \oplus \mathcal{R}_{W, \Theta, \mathbf{0}} \big) + \mu_2 \big(\mathcal{L}_{[\mathbf{r}, \mathbf{m}]} \oplus \mathcal{R}_{W, \Theta, \mathbf{0}} \big)    \Big] }, \\[-6.7ex]
\end{align*}
where $\forall \mathbf{p}, \mathbf{q} \in \mathbb{R}^2$, $\mathbf{p} \neq \mathbf{q}$
\vspace{-10pt}
\begin{align*}
\mu_2 \big(\mathcal{L}_{[\mathbf{p}, \mathbf{q}]} \oplus \mathcal{R}_{w, \theta, \mathbf{0}} \big) = 
\begin{cases}
\sqrt{2} \,w \lVert \mathbf{p} - \mathbf{q} \rVert \,\sin\big(\pi/4 + \theta - \eta \big) + w^2, & 0 \leq \theta - \eta \leq \pi/2 \\[-0.4ex]
\sqrt{2} \,w\lVert \mathbf{p} - \mathbf{q} \rVert\, \big\vert\sin\big( \! -\pi/4 + \theta - \eta \big)\big\vert + w^2, & \text{\emph{otherwise}} \\[-1.2ex]
\end{cases},
\vspace{-3pt}
\end{align*}
and $\eta = \tan^{-1}\big[([\mathbf{q}]_2 - [\mathbf{p}]_2)/([\mathbf{q}]_1 - [\mathbf{p}]_1)\big]$, and for $\mathbf{p} = \mathbf{q}$, $\mu_2 \big(\mathcal{L}_{[\mathbf{p}, \mathbf{q}]} \oplus \mathcal{R}_{w, \theta, \mathbf{0}} \big) = w^2$.
\end{lemma}

\vspace{-14pt}
\begin{proof}
Please refer to \cite{Comms_Letter}.  Note that the interplay between the slope of the line segment $\mathcal{L}_{[\mathbf{p}, \mathbf{q}]}$, given by $\eta$, and the orientation of the reflector, $\theta$, is what dictates how the Lebesgue measure of `$\mathcal{L}_{[\mathbf{p}, \mathbf{q}]} \oplus \mathcal{R}_{w, \theta}$' is evaluated -- hence the piecewise function.
\end{proof}

\vspace{-12pt}
\subsection{The Number of Reflectors Producing Visible Reflections}
\vspace{-3pt}

	Considering reflections with path lengths between $s_1$ and $s_2$ meters, this section presents the number of reflectors producing visible reflections, denoted by the RV: $V_{[s_1,\, s_2]}$, where $d \leq s_1 \leq s_2 \leq \infty$.  A subset of this metric, namely $V_{[d, s_2]}$ for $d = s_1 \leq s_2 < \infty$, was also studied in \cite{Comms_Letter} in the context of single-anchor localization.  However, the general metric is of particular interest to us here for two reasons.  The first is that it considers the infinite case, \emph{i.e.}, $V_{[d, \, \infty]}$, which is vitally important in the derivations that follow since this gives the \emph{total} number of reflectors producing visible reflections on \emph{all of} $\mathbb{R}^2$.  Obtaining this infinite case requires proving extra convergence results (Appendix \ref{LemmaProofExt}).  The second is that restricting our attention to reflections of distances $[s_1, \, s_2]$ aids in the derivation of the AOA.  Below we present the lemma for the general metric, $V_{[s_1, s_2]}$.  With the exception of the infinite case, the method behind the derivation is similar to that presented in \cite{Comms_Letter}. Thus, we only provide a proof sketch here to connect system model concepts to the analysis which follows and to give insight needed for subsequent derivations.

% --- AOA FIGURE ---
\begin{figure}[t]
\centering
\includegraphics [scale=0.46]{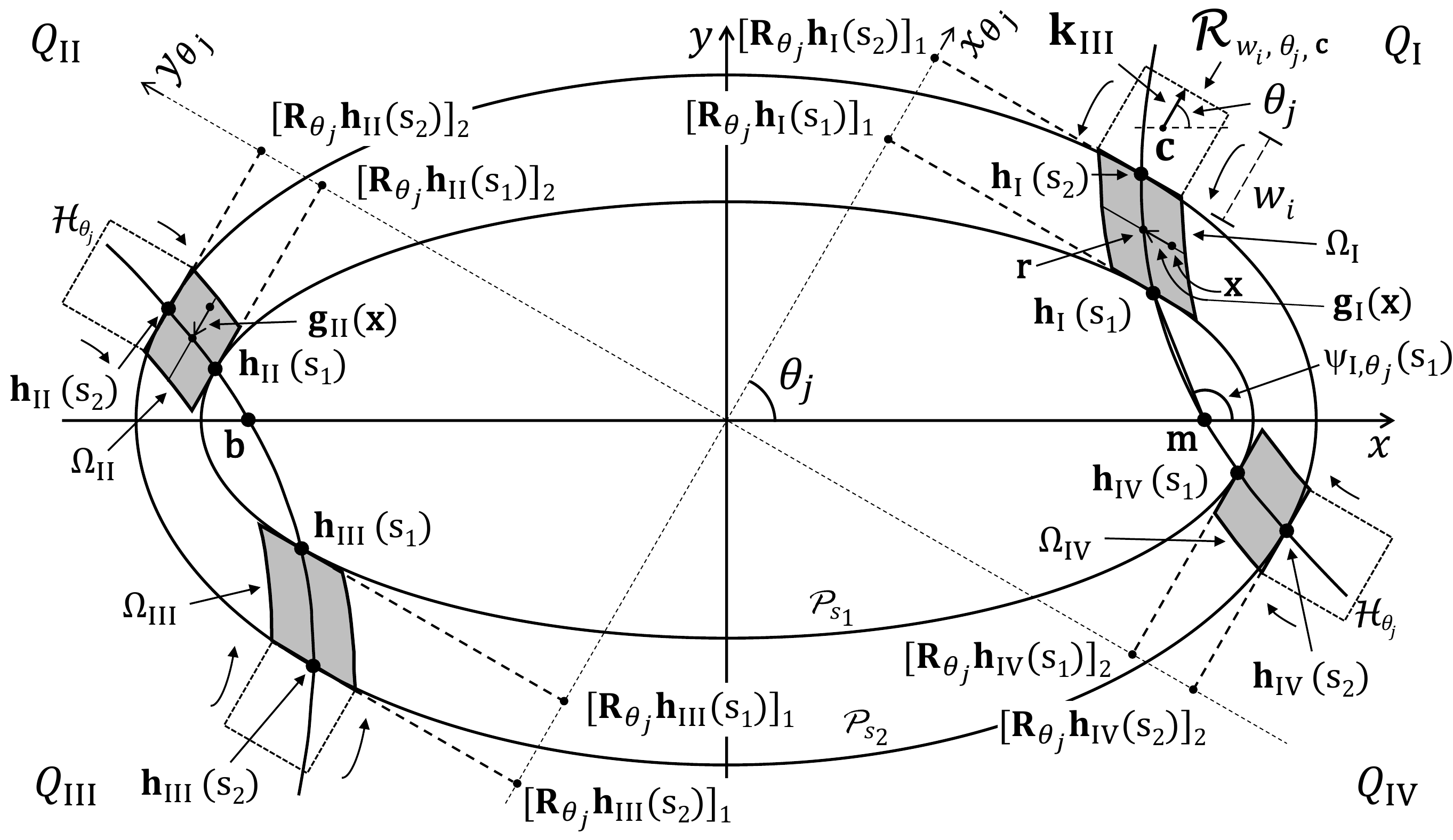}
\vspace{-10pt}
\caption{\textsc{Analytical Framework.}  The gray region in $Q_\text{I}$ is the region where the endpoint of $\mathbf{k}_\text{I}$ of reflector $\mathcal{R}_{w_i, \theta_j, \mathbf{c}}$ (see Fig. \ref{System_Model_Fig}) can lie in order to produce a first order reflection between $\mathbf{b}$ and $\mathbf{m}$ of distance in $[s_1, s_2]$.  If we let the edge corresponding to the endpoint of $\mathbf{k}_\text{I}$ be denoted as $\mathcal{E}_\text{I} \subset \partial \mathcal{R}_{w_i, \theta_j, \mathbf{c}}$, then this gray region is equivalently the region where the center point of edge $\mathcal{E}_\text{I}$ can lie for $\mathcal{R}_{w_i, \theta_j, \mathbf{c}}$ to produce a first order reflection.  This region can be thought of as being generated by sliding the $\mathcal{E}_\text{I}$-edge center point over $\mathcal{H}_{\theta_j}$ between $\mathcal{P}_{s_1}$ and $\mathcal{P}_{s_2}$, having the edge trace out the region.  Likewise for the other quadrants.  The AOA corresponding to a reflection at point $\mathbf{h}_\text{I}(s_1)$ is labeled $\psi_{\text{I}, \theta_j}(s_1)$ (see Definition \ref{sAOAs}).  In this figure, $\theta_j = \pi/3$.
\\[-8.5ex]}
\label{Analytical_FW_Fig}
\end{figure}
% --- NOTES ---
% 1) In MS Word: \mathcal{} = Lucia Handwriting OR Pristina
% 2) For symbols, use Symbols
% -------------------------------------------------------------------------------------------------	

% This Lemma implicitly assumes we have blocking!!!  (s_2 = infty is not valid for the no blocking case)
\vspace{-10pt}
\begin{lemma}[Number of Reflectors with VRPs] \label{Lemma_Vs}
Consider the test link setup and a deployment of reflectors under the Boolean model $\mathcal{B}$.  Let $V_{[s_1,\, s_2]}$ be the RV representing the number of reflectors with VRPs corresponding to distances between $s_1$ and $s_2$ meters, where $d \leq s_1 \leq s_2 \leq \infty$ (the case $s_1 = s_2 = \infty$ is disregarded).  Then, $V_{[s_1,\, s_2]} \!\distras{\,\,\,} \text{Poisson} \big(\hat{\lambda}(s_1,s_2) \big)$, where 
\vspace{-6pt}
\begin{align*}
\hat{\lambda}(s_1, s_2) = \frac{2 \lambda \mathbb{E}[W]}{n_\theta} \sum_{j=1}^{n_\theta} \left[ \int_{[\mathbf{R}_{\theta_j} \mathbf{h}_\text{\emph{I}}(s_1)]_1}^{[\mathbf{R}_{\theta_j} \mathbf{h}_\text{\emph{I}}(s_2)]_1}  \!\!\!\!\!\!\!  \rho\Big( \mathbf{R}^{-1}_{\theta_j} \mathbf{g}^*_\text{\emph{I}}(x_{\theta_j}) \Big) \text{\emph{d}}x_{\theta_j}    +   \int_{[\mathbf{R}_{\theta_j} \mathbf{h}_\text{\emph{II}}(s_1)]_2}^{[\mathbf{R}_{\theta_j} \mathbf{h}_\text{\emph{II}}(s_2)]_2} \!\!\!\!\!\!\!  \rho\Big( \mathbf{R}^{-1}_{\theta_j} \mathbf{g}^*_\text{\emph{II}}(y_{\theta_j}) \Big) \text{\emph{d}}y_{\theta_j} \right], \\[-7.1ex]
\end{align*}
and $\lambda$, $\mathbb{E}[W]$, $n_\theta$ are obtained from Definition \ref{Def_BM}, $\mathbf{h}_\text{\emph{I}}
(s)$ and $\mathbf{h}_\text{\emph{II}}(s)$ are from Lemma \ref{Lemma_BPRPs}, $\mathbf{b}$ and $\mathbf{m}$ are from Definition \ref{Def_TLS}, $\rho(\mathbf{r})$ is the probability that reflection point $\mathbf{r}$ is visible (Lemma \ref{Lemma_PRPV}), and%\footnote{The notation `$(s)$' for $\mathbf{h}_\text{I}(s)$ and $\mathbf{h}_\text{II}(s)$ simply serves as a reminder that these points are functions of $s$.}
\vspace{-6pt}
\begin{align} \label{Snapping_Functions}
\mathbf{g}^*_\text{\emph{I}}(x_{\theta_j}) = \Bigg[ x_{\theta_j}, \frac{-d^2 \sin(2\theta)}{8 x_{\theta_j}} \Bigg]^T \!,~~~ \mathbf{g}^*_\text{\emph{II}}(y_{\theta_j}) = \Bigg[ \frac{-d^2 \sin(2\theta)}{8 y_{\theta_j}},  y_{\theta_j} \Bigg]^T \!\!.
\end{align}
\end{lemma}
%\footnote{The notation `$(s)$' for $\mathbf{h}_\text{I}(s)$ and $\mathbf{h}_\text{II}(s)$ simply serves as a reminder that these points are functions of $s$.}

\vspace{-10pt}
\begin{proof}
	By independent thinning, we consider only reflectors in $\mathcal{B}$ with width, $w_i$, and orientation, $\theta_j$, where $i \!\in\! \{1, 2, \dots, n_w\}$ and $j \! \in \! \{1, 2, \dots, n_\theta\}$.  We denote this `thinned' Boolean model as $\mathcal{B}_{w_i, \theta_j}$, and its corresponding PPP of reflector center points as $\Phi_{w_i, \theta_j}$,  with intensity measure $\Lambda_{w_i, \theta_j} (B) \!=\! \frac{\lambda}{n_w n_\theta} \mu_2(B)$, for all Borel sets $B$.  Next, recall that: 1) a reflector $\mathcal{R}_{w_i, \theta_j, \mathbf{c}} \!\subset\! \mathcal{B}_{w_i, \theta_j}$ can produce a reflection iff its appropriate edge intersects $\mathcal{H}_{\theta_j}$; and 2) if a reflection is going to have a distance of $[s_1, s_2]$, then the RP must fall within $\mathcal{P}_{s_2} / (\mathcal{P}_{s_1}/ \partial \mathcal{P}_{s_1})$.  Thus, for reflectors in $\mathcal{B}_{w_i, \theta_j}$ to produce reflections of distances $[s_1, s_2]$, their center points, $\Phi_{w_i, \theta_j}$, must fall in the region
\vspace{-9pt}
\begin{align} \label{Reflection_Region}
\bigcup_{q \in \mathcal{Q}} \bigg[ \mathcal{E}_q \oplus \bigg( \mathcal{H}_{\theta_j} \cap Q_q \cap \Big( \mathcal{P}_{s_2} / (\mathcal{P}_{s_1}/ \partial \mathcal{P}_{s_1}) \Big) \bigg) \bigg] - \mathbf{k}_q,\\[-7.3ex] \nonumber
\end{align}
where $\mathcal{E}_q \subset \partial \mathcal{R}_{w_i, \theta_j, \mathbf{c}}$ is the set of points comprising the edge of the reflector corresponding to the endpoint of $\mathbf{k}_q$ (Fig. \ref{System_Model_Fig}).  (In this formulation, the center points of the $\mathcal{E}_q$'s are taken to be at the origin.)  For any $s_1, \, s_2$, the four quadrant portions of this region overlap on at most a null set, and thus, we may treat each separately and independently.

	Consider the $Q_\text{I}$ portion of this region: `$\mathcal{E}_\text{I} \oplus \big( \mathcal{H}_{\theta_j} \cap Q_\text{I} \cap \mathcal{P}_{s_2}/(\mathcal{P}_{s_1}/ \partial \mathcal{P}_{s_1}) \big) - \mathbf{k}_\text{I}$', which we write as $\Omega_\text{I} - \mathbf{k}_\text{I}$ to simplify notation.  We ultimately want the distribution of the number of center points of $\Phi_{w_i, \theta_j}$ in this region which correspond to reflectors producing \emph{visible} reflections.  To make this easier, we shift both $\Phi_{w_i, \theta_j}$ and $\Omega_\text{I} - \mathbf{k}_\text{I}$ by $\mathbf{k}_\text{I}$.  Thus, instead of referring to reflectors by their center point, $\mathbf{c}$, we now refer to them by the endpoint of their $\mathbf{k}_\text{I}$ vector, which is the center point of the $\mathcal{E}_\text{I}$-edge discussed above.  (The intensity measure of $\Phi_{w_i, \theta_j} + \mathbf{k}_\text{I}$ is also $\Lambda_{w_i, \theta_j}$.)
	
	Now, \emph{equivalently}, we seek the distribution of the number of $\mathcal{E}_\text{I}$-edge center points ($\mathcal{E}_\text{I}$-ECPs), $\Phi_{w_i, \theta_j} + \mathbf{k}_\text{I}$, in $\Omega_\text{I}$ (Fig. \ref{Analytical_FW_Fig}) which correspond to reflectors producing \emph{visible} reflections.  To obtain this distribution, we `thin' the PPP of $\mathcal{E}_\text{I}$-ECPs, $\Phi_{w_i, \theta_j} + \mathbf{k}_\text{I}$, over $\Omega_\text{I}$ to retain only those points which correspond to reflectors producing \emph{visible} reflections.  We denote this `thinned' point process as $\Phi_{\text{I}, v, w_i, \theta_j}$ and its intensity measure over $\Omega_\text{I}$ as $\Lambda_{\text{I}, v, w_i, \theta_j}(\Omega_\text{I}) = \int_{\Omega_\text{I}} \rho \big( \mathbf{g}_\text{I}(\mathbf{x}) \big) \text{d}\Lambda_{w_i, \theta_j}$.  Note that $\mathbf{r} = \mathbf{g}_\text{I}(\mathbf{x})$ is the function that maps the $\mathcal{E}_\text{I}$-ECP, $\mathbf{x}$, to the RP, $\mathbf{r}$, where the edge intersects $\mathcal{H}_{\theta_j}$ (Fig. \ref{Analytical_FW_Fig}), and that $\rho(\mathbf{r})$ is from Lemma \ref{Lemma_PRPV}.  Thus, the retention probability, $\rho \big( \mathbf{g}_\text{I}(\mathbf{x}) \big)$, is the probability that an $\mathcal{E}_\text{I}$-ECP corresponds to a reflector producing a visible reflection.  Since reflection paths are treated independently (Assumption \ref{Indep_Blocking_Assump}), this is an independent thinning. 
	% ... found by integrating the retention probability (\emph{i.e.}, the probability a point is retained) over $\Omega_\text{I}$ w.r.t. the intensity measure of the original point process $\Phi_{w_i, \theta_j} + \mathbf{k}_\text{I}$, \emph{i.e.},
	
	This integral is easily evaluated in a coordinate system rotated by $\theta_j$.  In this rotated system, let $\Omega_\text{I}^*$ be $\Omega_\text{I}$, $\mathbf{x}^* \!=\! [x_{\theta_j}, y_{\theta_j}]^T \!\in\! \Omega_\text{I}^*$ be an $\mathcal{E}_\text{I}$-ECP, and $\mathbf{r}^* \!= \mathbf{g}_\text{I}^*( \mathbf{x}^* )$ be the function that maps the $\mathcal{E}_\text{I}$-ECP, $\mathbf{x}^*$, to the RP, $\mathbf{r}^*$.  Note, $\mathbf{g}_\text{I}^*$, given in (\ref{Snapping_Functions}), is simply the reflection hyperbola expressed in rotated coordinates (now a rational function), and thus only depends on $x_{\theta_j}$ in $\mathbf{x}^* = [x_{\theta_j}, y_{\theta_j}]^T$.  Lastly, $\mathbf{r} = \mathbf{R}^{-1}_{\theta_j}\mathbf{r}^* = \mathbf{R}^{-1}_{\theta_j} \mathbf{g}_\text{I}^*( \mathbf{x}^* ) = \mathbf{R}^{-1}_{\theta_j} \mathbf{g}_\text{I}^*( x_{\theta_j} )$ and $\mathbf{r} = \mathbf{g}_\text{I}(\mathbf{x}) = \mathbf{g}_\text{I}(\mathbf{R}^{-1}_{\theta_j} \mathbf{x}^*)$ imply $\mathbf{g}_\text{I}(\mathbf{R}^{-1}_{\theta_j} \mathbf{x}^*) = \mathbf{R}^{-1}_{\theta_j} \mathbf{g}_\text{I}^*( x_{\theta_j} )$; and so applying the coordinate transformation $\mathbf{x} = \mathbf{R}^{-1}_{\theta_j} \mathbf{x}^*$, we have $\Lambda_{\text{I}, v, w_i, \theta_j}(\Omega_\text{I})$
\vspace{-7pt}
\begin{align} \label{IntOverOmega}
\stackrel{(a)}{=} \!\! \frac{\lambda}{n_w n_\theta}\!\! \int\limits_{\mathclap{ \Omega_\text{I}}} \!\!\! \rho\big(\mathbf{g}_\text{I}(\mathbf{x}) \big) \, \text{d}\mathbf{x}  \stackrel{(b)}{=}  \frac{\lambda}{n_w n_\theta}\!\! \int\limits_{\mathclap{ \Omega_\text{I}^*}} \!\!\! \rho \Big( \mathbf{g}_\text{I} \big(\mathbf{R}^{-1}_{\theta_j} \mathbf{x}^* \big) \!  \Big) \, \text{d} \mathbf{x}^*  \stackrel{(c)}{=}   \frac{\lambda w_i}{n_w n_\theta} \! \int_{[\mathbf{R}_{\theta_j} \mathbf{h}_\text{I}(s_1)]_1}^{[\mathbf{R}_{\theta_j} \mathbf{h}_\text{I}(s_2)]_1}  \!\!\!\!\!\!\!  \rho\Big( \mathbf{R}^{-1}_{\theta_j} \mathbf{g}^*_\text{I}(x_{\theta_j}\!) \Big) \text{d}x_{\theta_j}, \\[-7.5ex] \nonumber
\end{align}
where (a) follows from the definition of $\Lambda_{w_i, \theta_j}$ above, (b) by applying the coordinate transformation, and (c) by noting that the integral w.r.t. $y_{\theta_j}$ simplifies to $w_i$ and that the limits of the integral w.r.t. $x_{\theta_j}$ are obtained from Fig. \ref{Analytical_FW_Fig}.  The distribution of the number of reflectors from $\mathcal{B}_{w_i, \theta_j}$ which produce \emph{visible} reflections in $Q_\text{I}$ of distance $[s_1, s_2]$ is then  $\Phi_{\text{I}, v, w_i, \theta_j}(\Omega_\text{I}) \!\distras{\,\,\,} \text{Poisson} \big( \Lambda_{\text{I}, v, w_i, \theta_j}(\Omega_\text{I}) \big)$. % have to account for the fact that $\mathbf{m}$ is not in Q_I but this doesn't change the distribution at all.

%This thinning to retain only $\mathcal{E}_\text{I}$-ECPs corresponding to visible reflections is an independent thinning since we treat each reflection path independently.

	The same procedure above can be followed for the $Q_\text{II}$ portion of the region in (\ref{Reflection_Region}) by simply replacing `I' with `II' and by noting that $\mathbf{g}_\text{II}^*$, given in (\ref{Snapping_Functions}), depends on $y_{\theta_j}$.  Thus, $\Lambda_{\text{II}, v, w_i, \theta_j}(\Omega_\text{II}) = \frac{\lambda w_i}{n_w n_\theta} \! \int_{[\mathbf{R}_{\theta_j} \mathbf{h}_\text{II}(s_1)]_2}^{[\mathbf{R}_{\theta_j} \mathbf{h}_\text{II}(s_2)]_2}  \!  \rho\Big( \mathbf{R}^{-1}_{\theta_j} \mathbf{g}^*_\text{II}(y_{\theta_j}\!) \Big) \text{d}y_{\theta_j}$, and the number of reflectors from $\mathcal{B}_{w_i, \theta_j}$ with VRPs in $Q_\text{I}$ corresponding to reflections of distance $[s_1, s_2]$ is $\Phi_{\text{II}, v, w_i, \theta_j}(\Omega_\text{II}) \!\distras{\,\,\,} \text{Poisson} \big( \Lambda_{\text{II}, v, w_i, \theta_j}(\Omega_\text{II}) \big)$.  (Note, Appendix \ref{LemmaProofExt} verifies that the integrals in $\Lambda_{\text{I}, v, w_i, \theta_j}(\Omega_\text{I})$ and $\Lambda_{\text{II}, v, w_i, \theta_j}(\Omega_\text{II})$ converge for $s_2 = \infty$.)
	
%	Note that $\Lambda_{\text{I}, v, w_i, \theta_j}(\Omega_\text{I})$ and $\Lambda_{\text{II}, v, w_i, \theta_j}(\Omega_\text{II})$ hold for $d = s_1 \leq s_2 < \infty$.  To verify they hold for $s_2 = \infty$, we must show that their integral formulations converge.  This is done in Appendix \ref{LemmaProofExt}.
	
	Finally, the $Q_\text{III}$ portion of (\ref{Reflection_Region}) is symmetric with that of $Q_\text{I}$, and $Q_\text{IV}$ with that of $Q_\text{II}$ (Fig. \ref{Analytical_FW_Fig}), and so the number of reflectors producing visible reflections in these regions follow the same Poisson distributions.  The lemma follows by noting: 1) the four quadrant regions in (\ref{Reflection_Region}) are independent; 2) the original thinning of $\mathcal{B}$ to the $\mathcal{B}_{w_i, \theta_j}$'s is independent; and 3) the sum of independent Poisson RVs is Poisson with mean being the sum of the individual means.
\end{proof}	
	
% *** NOTE ***
% Mention that it doesn't matter whether we use brackets or parentheses (explain why)	
	
% *** NOTE ***
% 1) I needed to say this converges in distribution for some reason but I can't remember why at the moment (7/24/2020)	
	
\vspace{-8pt}	
\begin{remark}
It follows from the lemma that as $s_2 \rightarrow \infty$, $V_{[s_1, s_2]}$ converges \emph{in distribution} to $V_{[s_1, \infty]}$.
\end{remark}

\vspace{-14pt}
\section{The NLOS Bias Distribution} \label{Sec_DofNLOSbiasDist}
\vspace{-2pt}

	To characterize NLOS bias, this section derives the path length distribution of the first-arriving NLOS signal, with and without blocking.   Approximations and numerical results are then presented, followed by a discussion of the importance of having an analytically derived bias distribution. %and evidence from the literature supporting the bias results presented here. 
We start by introducing the following RVs:

\vspace{-9pt}
\begin{definition}[Distance Traversed by the $1^\text{st}$-Arriving NLOS Signal] \label{DefRVS1}
Let $S_{(1)}$ be the RV representing the path length, in meters, traveled by the first-arriving NLOS signal.  Note, $d \leq S_{(1)} < \infty$.%\footnote{Although $S_{(1)}$ is not itself an order statistic, we borrow the subscript notation `$(1)$' with parenthesis since the nature of this RV is reminiscent of order statistics.} 
\end{definition}

\vspace{-17pt}
\begin{definition}[NLOS Bias]
Let $B$ be the RV representing the distance, in meters, of the NLOS bias.  Note, $0 \leq B < \infty$. (This implies $\textbf{Supp}(B) = [0, \infty)$, and as a continuous RV, $P[B>0]=1$.)

% --- OLD FOOTNOTE ---
%\footnote{In the literature, bias is assumed to be a strictly positive value, yet is oftentimes modeled with a distribution whose support contains zero.  In this paper, we allow $B=0$ s.t. we properly capture it's support.  Since bias is a continuous r.v., $P[B=0]=0$ and so allowing for this condition is of no consequence and is simply done to clarify the exposition.}
\end{definition}

\vspace{-15pt}
\begin{remark}
With these RVs, we may state the simple relationship between the path length of the first-arriving NLOS signal and the NLOS bias by: $B = S_{(1)} - d$, which comes from (\ref{The_Range_Equation}) in Sec. \ref{Sec_Intro}.
\end{remark}

	With regards to $S_{(1)}$, the first and most important question one may ask is: \emph{Does this RV always exist? That is, under the system model, will there always be a reflection path, no matter how far out?}  We address this with regards to the two cases: with and without blocking.

	Without blocking, the answer is simple: \emph{Yes, $S_{(1)}$ always exists.} To see why, note that without blocking, the probability that any RP $\mathbf{r}$ is visible is one, \emph{i.e.} $\rho( \mathbf{r} ) = 1$, and hence, the integrands in $\hat{\lambda}(s_1, s_2)$ from Lemma \ref{Lemma_Vs} are one.  Therefore, \emph{without blocking}, Lemma \ref{Lemma_Vs} is valid for $d \leq s_1 \leq s_2 < \infty$ (for if $s_2=\infty$, then we would have $\hat{\lambda}(s_1, \infty) = \infty$).  Finally then, for this case without blocking, the probability that there will be at least one reflector producing a reflection on $\mathbb{R}^2$, \emph{i.e.}, the probability $S_{(1)}$ exists, is:
\vspace{-12pt} 
\begin{align*}
\lim_{s_2 \to \infty} P[V_{[d, s_2]} \geq 1] =  1 -  \lim_{s_2 \to \infty} P[V_{[d, s_2]} = 0] = 1 - \lim_{s_2 \to \infty} e^{-\hat{\lambda}(d, s_2)} = 1 - e^{-\infty} = 1. \\[-6.4ex]
\end{align*}

With blocking, the answer is unfortunately: \emph{No, $S_{(1)}$ does not always exist.}  Let us see why:

\vspace{-9pt}
\begin{corollary}[Lower Bound on the Probability of No Reflections, With Blocking] \label{LowerBound}
Consider the test link setup under the Boolean model.  Then, the probability that there are no reflectors producing visible reflections is lower bounded by $e^{-2}$, i.e., $P[V_{[d, \infty]} = 0] > e^{-2}$.
\end{corollary}

\vspace{-12pt}
\begin{proof}
Substituting the bound in (\ref{Int_Bnd_1}) for each of the integrals in the expression for $\hat{\lambda}(d, \infty)$ in Lemma \ref{Lemma_Vs} yields $\hat{\lambda}(d, \infty) < 2$, which implies $P[V_{[d,\infty]} = 0] = e^{-\hat{\lambda}(d, \infty)} > e^{-2}$, as desired. 
\end{proof}

\vspace{-7pt}
\begin{remark}
Note that this is a hard lower bound, \emph{i.e.}, under the independent blocking assumption, the probability that there are no reflections at all is always at least $e^{-2} \approx 0.135$, regardless of the density of reflectors, their size, and orientation. \emph{This corollary emphasizes the care one must take when adding blocking into stochastic propagation models.}
\end{remark}
% Thus, at least $13.5\%$ of the time, $S_{(1)}$ will not exist.
	
% *** TO DO ***
% 1) Since our independent blocking assumption tracks the true, correlated blocking assumption very well (see Sec. ___ ), then this bound is likely a good approximation for the true, correlated blocking case as well!

	As a consequence of this corollary, ensuring the existence of $S_{(1)}$, \emph{with blocking}, will require conditioning on the event $\{V_{[d, \infty]} \geq 1\}$, \emph{i.e.}, the event that there exists at least one reflector producing a visible reflection.  Thus, we can now establish the distribution of $S_{(1)}$.  \emph{\textbf{Before continuing, since the remainder of this section and the next is concerned with reflections of distance $\leq s$, for some $s$, we adopt the simplified notation:} $V_{[d, s]} \!\triangleq\! V_s$, \textbf{with mean} $\hat{\lambda}(d, s) \!\triangleq\! \hat{\lambda}(s)$}. %\emph{i.e.}, the first argument is always $d$, so we simply drop it.  
\vspace{-25pt}
\begin{theorem}[The Distribution of $S_{(1)}$, With Blocking] \label{S1wB}
Consider the test link setup under the Boolean model with independent blocking (Assumption \ref{Indep_Blocking_Assump}).  Then, the distribution of $S_{(1)}$, conditioned on there existing at least one reflector producing a visible reflection, is given by
\vspace{-8pt}
\begin{align*}
&CDF\!:  F_{S_{(1)}} (s_{(1)} \,|\, V_\infty \geq 1) = \frac{ 1 }{1 - e^{-\hat{\lambda}(\infty)} } \bigg( 1 - e^{-\hat{\lambda}\big(s_{(1)}\big)}  \bigg)\\[-5ex] \\
&PDF\!:  f_{S_{(1)}} (s_{(1)} \,|\, V_\infty \geq 1) = \frac{\lambda \mathbb{E}[W] s_{(1)}} {n_\theta \big( 1 \!- e^{-\hat{\lambda}(\infty)} \big)} e^{-\hat{\lambda}\big(s_{(1)}\big)} \!\! \sum_{j=1}^{n_\theta} \textstyle\Bigg[ \frac{\rho\big(\mathbf{h}_{\text{\emph{I}}, \theta_j}(s_{(1)}) \big)} {\sqrt{s_{(1)}^2 \!- d^2 \sin^2 \! \theta_j }} + \frac{\rho\big(\mathbf{h}_{\text{\emph{II}}, \theta_j}(s_{(1)}) \big)}{\sqrt{s_{(1)}^2 \!- d^2 \cos^2 \! \theta_j }} \Bigg],\\[-7ex]
\end{align*}
where all of the parameters are listed in Lemma \ref{Lemma_Vs} and $\textbf{\emph{Supp}}(S_{(1)}\,|\,V_\infty \geq 1) = [d, \infty)$.
\end{theorem}

\vspace{-14pt}	
\begin{proof}
Please refer to Appendix \ref{PfS1wB}.
\end{proof}

\vspace{-9pt}
\begin{remark}
%Conditioning on the event $\{V_\infty \geq 1 \}$ means that all visible reflections on $\mathbb{R}^2$ are considered when obtaining $S_{(1)}$.  
If one wants to account for Tx power, reflection losses, pathloss, and a signal detection threshold at the mobile, then this would be equivalent to strategically choosing a maximum distance, $s_{max}$, wherein only reflections that travel less than or equal to this distance are deemed detectable.  To obtain the distribution of $S_{(1)}$ in this case, we would restrict our attention to the region $\mathcal{P}_{s_{max}}$ and condition on the event $\{V_{s_{max}} \geq 1 \}$, rather than $\{V_\infty \geq 1 \}$. Consequently, all that would change in the above distribution is $\hat{\lambda}(\infty)$ being replaced with $\hat{\lambda}(s_{max})$, along with a new, restricted support: $[d, s_{max}]$.  Although our model can easily incorporate various channel parameters, we continue, however, with the most general case: assuming the first-arriving path can be detected regardless of its path length, \emph{i.e.}, conditioning on the event $\{V_\infty \geq 1 \}$. %which considers \emph{all} first-arriving reflections \emph{at any distance}. 
\end{remark}
% \emph{i.e.}, received signals less than a certain power are not detectable

	As a direct corollary, deriving the distribution of $S_{(1)}$, \emph{without blocking}, is straightforward since $S_{(1)}$ always exists, \emph{i.e.}, there is no need to condition on any event to guarantee existence. % Thus, we have:

\vspace{-8pt}
\begin{corollary}[The Distribution of $S_{(1)}$, Without Blocking] \label{Dist_S1_without_blocking}
Consider the test link setup under the Boolean model.  Then, the distribution of $S_{(1)}$ is given by
\vspace{-2pt}
\begin{align*}
%&CDF\!: F_{S_{(1)}}(s_{(1)}) = 1 \! - \exp\Bigg\{\!\! -\frac{\lambda \mathbb{E}[W]} {n_\theta} \! \sum_{j=1}^{n_\theta} \Bigg[ \! \sqrt{s_{(1)}^2 \!\!-d^2 \! \sin^2 \! \theta_j } \\
%&~~~~~~~~~~~~~~~~~~~~~ - d (\sin\theta_j + \cos\theta_j) + \sqrt{s_{(1)}^2 -d^2 \cos^2 \theta_j} \Bigg] \Bigg\} \\ \\
&CDF\!: F_{S_{(1)}}(s_{(1)}) = 1 \, - e^{\!\! -\frac{\lambda \mathbb{E}[W]} {n_\theta} \sum_{j=1}^{n_\theta} \left[ \! \sqrt{s_{(1)}^2 \!\!-d^2 \! \sin^2 \! \theta_j }  - d (\sin\theta_j + \cos\theta_j) + \sqrt{s_{(1)}^2 \!\!-d^2 \! \cos^2 \! \theta_j} \right] } \\ \\[-5.5ex]
%&PDF\!: f_{S_{(1)}}(s_{(1)}) = \frac{\lambda \mathbb{E}[W] s_{(1)}} {n_\theta}  \exp\Bigg\{\!\! -\frac{\lambda \mathbb{E}[W]} {n_\theta} \times\\
%& \sum_{j=1}^{n_\theta} \Bigg[ \! \sqrt{s_{(1)}^2 \!\!-d^2 \! \sin^2 \! \theta_j }  - d (\sin\theta_j \!+\! \cos\theta_j) + \! \sqrt{s_{(1)}^2 \!\!-d^2 \! \cos^2 \! \theta_j} \Bigg] \Bigg\} \times \\
%&\sum_{j=1}^{n_\theta} \Bigg[ \frac{1}{\sqrt{s_{(1)}^2 - d^2 \sin \theta_j } } + \frac{1}{\sqrt{s_{(1)}^2 - d^2 \cos \theta_j } } \Bigg]
&PDF\!: f_{S_{(1)}}(s_{(1)}) = \frac{\lambda \mathbb{E}[W] s_{(1)}} {n_\theta} \sum_{j=1}^{n_\theta} \textstyle\Bigg[ \frac{1}{\sqrt{s_{(1)}^2 - d^2 \sin^2 \theta_j } } + \frac{1}{\sqrt{s_{(1)}^2 - d^2 \cos^2 \theta_j } } \Bigg]~ \times \\[-1ex]
&~~~~~~~~~~~~~~~~~~~~~~~~~~~~~~~~~~~~~~~~~~~~~~~~~~~~~~e^{\!\! -\frac{\lambda \mathbb{E}[W]} {n_\theta} \sum_{j=1}^{n_\theta} \left[ \! \sqrt{s_{(1)}^2 \!\!-d^2 \! \sin^2 \! \theta_j }  - d (\sin\theta_j + \cos\theta_j) + \sqrt{s_{(1)}^2 \!\!-d^2 \! \cos^2 \! \theta_j} \right] },\\[-7ex]
\end{align*}
where the parameters are given in Definitions \ref{Def_BM} and \ref{Def_TLS} and $\textbf{\emph{Supp}}(S_{(1)}) = [d, \infty)$.
\end{corollary}

\vspace{-17pt}
\begin{proof}
Recalling the discussion above Corollary \ref{LowerBound}, we know that \emph{without blocking}, $\rho( \mathbf{r} ) = 1$ for any RP $\mathbf{r}$ and thus, Lemma \ref{Lemma_Vs} is valid for $d \leq s_1 \leq s_2 < \infty$.  Next, recalling the simplified notation above Theorem \ref{S1wB}, this implies that the number of reflectors producing reflections of distance $\leq s$, \emph{i.e.}, $V_s$, \emph{without blocking}, is valid for $d \leq s < \infty$.  Thus, knowing how $V_s$ changes for this case without blocking, we can now complete the derivation. Towards this end, we have: $F_{S_{(1)}}(s_{(1)}) = $
% So V_\infty would be Poisson Distributed but would have infinite mean, which can't happen (s=\infty gives a logical inconsistency, and so s cannot equal infinity)	
\vspace{-25pt}
\begin{align*}
~~~1 \! - P[S_{(1)} > s_{(1)}] \stackrel{(a)}{=} 1 \! - P[V_{s_{(1)}} = 0] = 1 \! - e^{-\hat{\lambda}\big(s_{(1)}\big)} \!\stackrel{(b)}{=} 1 \!- e^{-\frac{2 \lambda \mathbb{E}[W]}{n_\theta} \! \sum_{j=1}^{n_\theta} \Bigg[ \!\! \int\limits_{[\mathbf{R}_{\theta_j} \mathbf{m}]_1}^{[\mathbf{R}_{\theta_j} \mathbf{h}_\text{I}(s_{(1)})]_1}  \!\!\!\!\!\!\!\!\!\!  \text{d}x_{\theta_j}  ~+   \int\limits_{[\mathbf{R}_{\theta_j} \mathbf{b}]_2}^{[\mathbf{R}_{\theta_j} \mathbf{h}_\text{II}(s_{(1)})]_2} \!\!\! \!\!\!\!\!\!\!\!   \text{d}y_{\theta_j}\! \Bigg]}, \\[-6.7ex]
\end{align*}
where in (a), $V_{s_{(1)}}$ is the number of reflectors producing reflections (all are visible), and (b) follows from $\rho( \mathbf{r} ) = 1$ for any RP $\mathbf{r}$.  Lastly, $[\mathbf{R}_{\theta_j} \mathbf{m}]_1  = \frac{d}{2} \cos \theta_j$, $[\mathbf{R}_{\theta_j} \mathbf{h}_\text{I}\big(s_{(1)}\big)]_1 = \frac{1}{2} \sqrt{s_{(1)}^2 - d^2 \sin^2 \theta_j}$, $[\mathbf{R}_{\theta_j} \mathbf{b}]_2  = \frac{d}{2} \sin \theta_j$, $[\mathbf{R}_{\theta_j} \mathbf{h}_\text{II}\big(s_{(1)}\big)]_2  = \frac{1}{2} \sqrt{s_{(1)}^2 - d^2 \cos^2 \theta_j}$, which yields the CDF in the corollary.
	
	The PDF is obtained via differentiation w.r.t. $s_{(1)}$. The support follows from Definition \ref{DefRVS1}.  % Well technically follows from Lemma 5, and same for Theorem 1, but don't feel like explaining (why the case at s=infinity doesn't apply so we have d \leq s < infinity), so this is fine.
\end{proof}

\vspace{-11pt}
\begin{remark}
Since this corollary does not consider blocking, and since independent blocking was our main approximation, this corollary represents a true, approximation-free derivation of the $S_{(1)}$ (\emph{i.e.}, the NLOS bias) distribution under the Boolean model with first-order reflections.
% with first-order reflections.  
Further, this distribution, without blocking, offers a simple, closed-form approximation of the distribution of $S_{(1)}\,|\,\{V_\infty \geq 1\}$, with blocking, in cases where the reflector/blockage density is low.

\end{remark}

\vspace{-7pt}
\begin{remark}
In a similar vein to the remark following Theorem \ref{S1wB}, we could incorporate Tx/Rx parameters and channel effects for this distribution as well via a restriction to the region $\mathcal{P}_{s_{max}}$ and conditioning on the event $\{V_{s_{max}} \geq 1\}$.  The support would also change accordingly.
\end{remark}

\vspace{-19pt}
\subsection{Exponential Family Approximations for NLOS Bias}
\vspace{-3pt}

	The distributions of $S_{(1)}$, both with and without blocking, appear to manifest a form that resembles an exponential distribution, or perhaps that of a distribution from an exponential family.  Consequently, this section attempts to bridge the gap between this analysis and the prevailing localization literature by demonstrating that the distributions derived here, via analysis, match commonly used NLOS bias distributions assumed in the literature.  %We show that the bias distribution derived here can be well-approximated for non-blocking case by an exponential distribution, and for the blocking case by a gamma distribution. 
	
\subsubsection{The Non-Blocking Case}
	
	Here, we attempt to approximate the distribution of the bias, $B$, where $B = S_{(1)} - d$ and the distribution of $S_{(1)}$ is from Corollary \ref{Dist_S1_without_blocking}.  Although not obvious, the argument of the exponential of the CDF in Corollary \ref{Dist_S1_without_blocking} is nearly linear in $s_{(1)}$.  This suggests a close connection with a true, exponential distribution, as we see below:
	
\vspace{-10pt}
\begin{approximation}[Distribution of Bias, Without Blocking] \label{Approx_without_blocking}
Consider the test link setup under the Boolean model and the distribution of $S_{(1)}$ from Corollary \ref{Dist_S1_without_blocking}, with the NLOS bias given by $B = S_{(1)} - d$.  Then, the exponential distribution approximation, $\tilde{F}_B$, of the NLOS bias distribution, $F_B$, is: $B \distras{\,\,} F_B(b) \approx \tilde{F}_B(b) = Exp \big(2 \lambda \, \mathbb{E}[W] \big) = 1 - e^{-2 \lambda \, \mathbb{E}[W]\, b}$, where $\textbf{Supp}(B) = [0, \infty)$, which applies to the exponential distribution approximation, $\tilde{F}_B$, as well.
\end{approximation}

\vspace{-11pt}
\begin{proof}
	We begin by noting some facts about the distribution of $F_B$.  First, from Corollary \ref{Dist_S1_without_blocking}, we know that $F_{S_{(1)}}(s_{(1)}) = 1 - e^{-g(s_{(1)})}$, where $-g(s_{(1)})$ is the large argument of the exponential in Corollary \ref{Dist_S1_without_blocking}, and that $\textbf{Supp}(S_{(1)}) = [d, \infty)$.  Next, since $B = S_{(1)} - d$, then $F_B(b) = F_{S_{(1)}}(b + d)$, which implies $F_B(b) = 1 - e^{-g(b+d)}$, where $\textbf{Supp}(B) = [0, \infty)$.
Thus, the goal here is to find an exponential distribution approximation for $F_B$, \emph{i.e.}, $1 - e^{-g(b+d)} = F_B(b) \approx \tilde{F}_B(b) = 1 - e^{-\alpha b}$, where we need to find a suitable $\alpha$.  In other words, we would like an $\alpha$ s.t. `$\alpha b$' approximates `$g(b+d)$' as well as possible.  This is done via a heuristic argument based on asymptotics.
	
	To determine $\alpha$, first note that for $b=0$, $\alpha b = 0$ and $g(b+d) = 0$.  Next, although not obvious, $g(b+d)$ becomes linear in $b$ when $b \gg 0$.  Thus, for our approximation, $\alpha b \approx g(b+d)$, we set $\alpha$ equal to the slope of $g(b+d)$ when $b \gg 0$, that is: $\alpha = \lim_{b \to \infty} \frac{\partial}{\partial b} \big[ g(b+d) \big]$
\vspace{-8pt}
\begin{align*}
&= \lim_{b \to \infty} \frac{\partial}{\partial b} \Bigg[ \frac{\lambda \mathbb{E}[W]}{n_\theta} \sum_{j=1}^{n_\theta}   \sqrt{b^2 + 2bd + d^2 \cos^2 \theta_j } - d(\sin \theta_j + \cos \theta_j)  + \sqrt{b^2 + 2bd + d^2 \sin^2 \theta_j }  \Bigg] \\
&= \lim_{b \to \infty} \!\frac{\lambda \mathbb{E}[W]}{n_\theta} \! \sum_{j=1}^{n_\theta} \textstyle \left[ \frac{b+d}{\sqrt{b^2 + 2bd + d^2 \cos^2 \theta_j }} + \frac{b+d}{\sqrt{b^2 + 2bd + d^2 \sin^2 \theta_j }} \right] =2 \lambda \mathbb{E}[W].\\[-7ex]
\end{align*}
This completes the approximation.
\end{proof}

\begin{figure}[t]
\centering
% --- APPROXIMATION OF BIAS DISTRIBUTION WITHOUT BLOCKING ---
\begin{minipage}[t]{0.48\textwidth}
\centering
\includegraphics [scale=0.54]{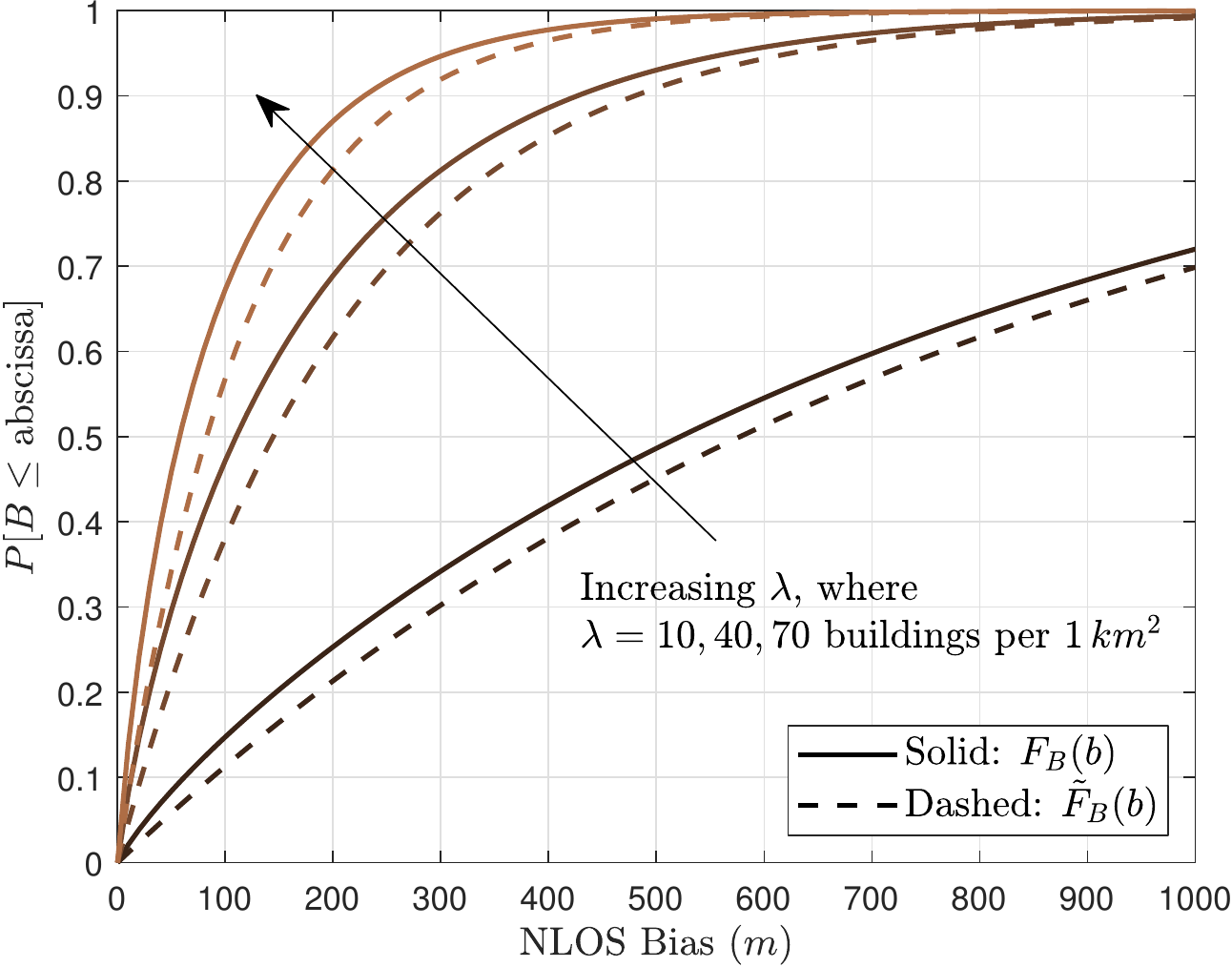}
\vspace{-9pt}
\caption{\textsc{Approximation of Bias Distribution Without Blocking.}  This figure plots the NLOS bias distribution, derived via Corollary \ref{Dist_S1_without_blocking} (Solid), against the NLOS bias distribution approximation in Approx. \ref{Approx_without_blocking} (Dashed).  This comparison was made for a test link setup with $d = 200m$, and for a Boolean model with reflector widths and orientations distributed as: $f_W = \text{unif}(w_{min} =20m, w_{max} = 100m, n_w = 5)$ and $f_\Theta = \text{unif}(\theta_{min} = 10\degree, \theta_{max} = 80\degree, n_\theta = 8)$.  
\\[-8ex]}
\label{Approx1_Comparison}
\end{minipage}\hfill
\begin{minipage}[t]{0.48\textwidth}
\centering
% --- MOMENT MATCHED DISTRIBUTIONS FIGURE ---
\includegraphics [scale=0.54]{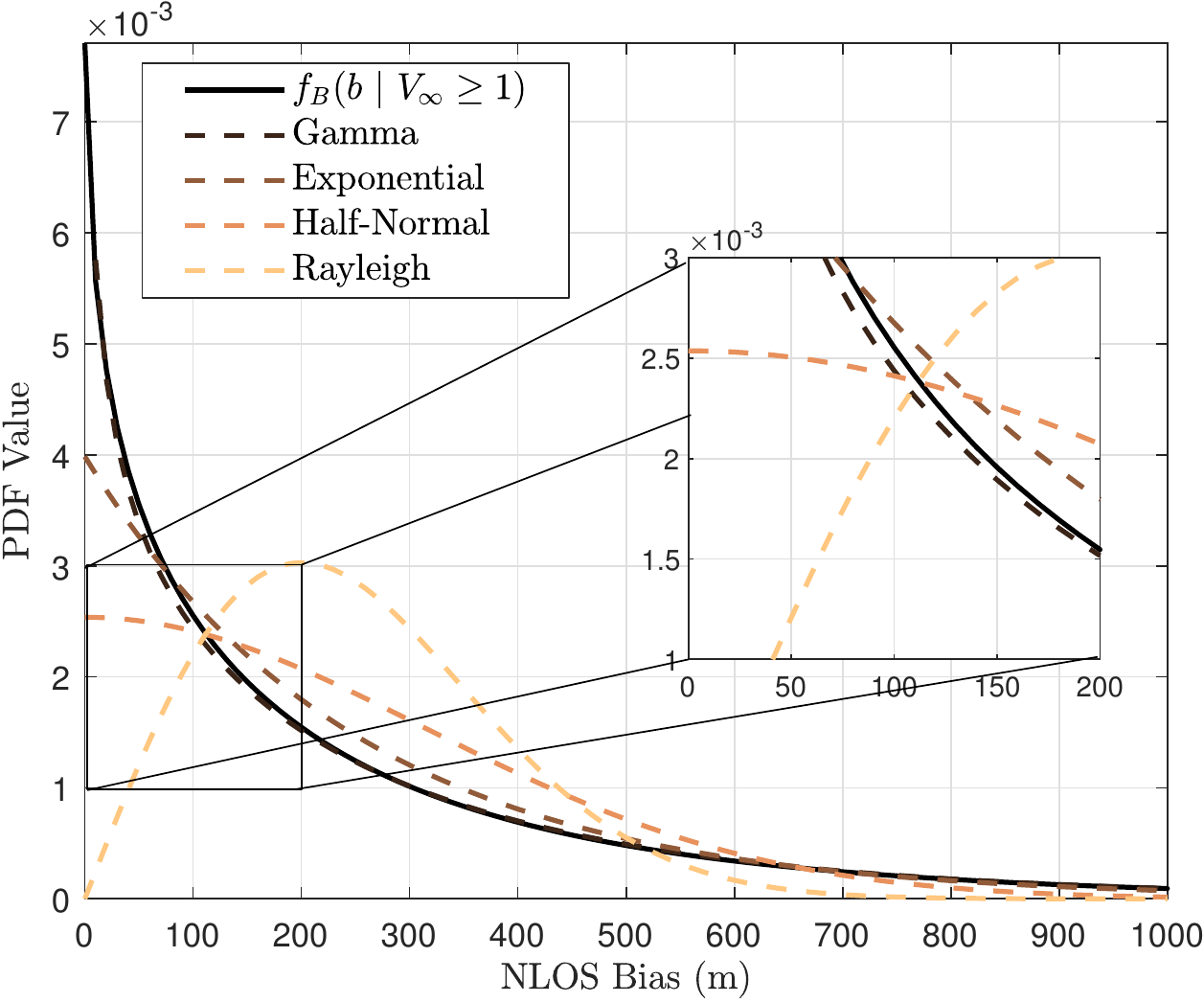}
\vspace{-9pt}
\caption{\textsc{Approximations of Bias Distribution With Blocking.} This figure plots the NLOS bias distribution with blocking, derived via Theorem \ref{S1wB}, against that of four common exponential family distributions used to model NLOS bias in the localization literature. 
\\[-8ex]}
\label{Moment_Matched_Dists}
\end{minipage}	
\end{figure}

\vspace{-9pt}
\begin{remark}
This exponential distribution approximation of the NLOS bias distribution, for the non-blocking case, is depicted in Fig. \ref{Approx1_Comparison}.  In addition to being a good approximation for the true NLOS bias, this exponential approximation is also desirable due to its simplicity.
\end{remark}

% --- APPROXIMATIONS ---

\subsubsection{The Blocking Case}
	We now turn to the blocking case, which implies we need to condition on the event $\{V_\infty \geq 1\}$.  Thus, NLOS Bias is given by $B\,|\,\{V_\infty \geq 1\} = S_{(1)}\,|\,\{V_\infty \geq 1\} - d$, and so it follows from Theorem \ref{S1wB} that $f_B(b \,|\, V_\infty \geq 1) = f_{S_{(1)}}(b + d \,|\, V_\infty \geq 1)$ and $\textbf{Supp}(B\,|\,V_\infty\geq1) = [0, \infty)$.  Since we cannot make an obvious connection here between $F_B(b \,|\, V_\infty \geq 1)$ and an exponential distribution, as was done without blocking, we take a different approach and compare $f_B(b \,|\, V_\infty \geq 1)$ to various, common distributions of NLOS bias used in the literature.  
	
% Note: I didn't need to put the conditioning on X, i.e., X | V_\infty \geq 1, but I did it anyway because it seems to make the discussion clearer

	This comparison is done via the use of the Kullback-Leibler (KL) divergence \cite{CoverAndThomas}.  Specifically, we examine the KL divergence from $B\,|\, \{V_\infty \geq 1\}$ to $X\,|\, \{V_\infty \geq 1\}$, \emph{i.e.}, 
\vspace{-6pt}
\begin{align} \label{KLDivDef}
D(X \,\lvert \rvert\, B) = \int_0^\infty \!\!\! f_X(x\,|\, V_\infty \geq 1) \,\ln \Bigg( \frac{f_X(x\,|\, V_\infty \geq 1)}{f_B(x \,|\, V_\infty \geq 1)} \Bigg) \, \text{d}x ,\\[-6.7ex] \nonumber
\end{align}
where we choose $X\,|\, \{V_\infty \geq 1\}$ to be distributed by one of the four common NLOS bias distributions: gamma, exponential, half-normal, and Rayleigh.\footnote{These distributions share the same support as $B\,|\, \{V_\infty \geq 1\}$.  The conditioning of the RVs $B$ and $X$ on $\{V_\infty \geq 1\}$ is omitted in (\ref{KLDivDef}) and Table \ref{KLD} for notational simplicity. Literature references for these common bias distributions are given in Section \ref{Sec_Intro}.} First, the parameters of these four comparison distributions were found via `moment-matching,' \emph{i.e.}, the moments of $B\,|\, \{V_\infty \geq 1\}$, computed numerically via $f_B(b \,|\, V_\infty \geq 1)$ using the vales for $d$, $f_W$, and $f_\Theta$ from Fig. \ref{Approx1_Comparison}, and for $\lambda = 10, 40, 70$ buildings/$km^2$ (Table \ref{KLD}), were matched to the necessary moments of $X \,|\, \{V_\infty \geq 1\}$ to obtain the distributions' parameter values.\footnote{We choose $\alpha$ and $\beta$ in $\Gamma(\alpha, \beta)$ to be the shape and rate parameters, respectively, and $\beta^\prime$ in $Exp(\beta^\prime)$ to be the rate parameter.} Once the parameters of the comparison distributions were found, then the KL divergences from $B\,|\, \{V_\infty \geq 1\}$ to $X\,|\, \{V_\infty \geq 1\}$ were computed numerically, with the results given in Table \ref{KLD}.

	Table \ref{KLD} reveals that, overall, the gamma distribution provides the best match with our distribution of bias, and, of the single-parameter families, the exponential distribution offers the best approximation.  For the $\lambda = 40$ case from Table \ref{KLD}, we plot the distribution of $B \,|\, \{V_\infty \geq 1\}$ and of $X\,|\,\{V_\infty \geq 1\}$, for the various moment-matched, comparison distributions, in Fig. \ref{Moment_Matched_Dists}.  From the figure, it is clear that the gamma and exponential distributions offer great approximations, with the gamma distribution approximation being virtually indistinguishable from our analytically derived bias distribution.

\begin{table}[t]
\caption{Kullback-Leibler Divergence from $B$ to $X$}
\renewcommand{\arraystretch}{0.89}
\centering
\footnotesize 
\vspace{-12pt}
\begin{tabular}{c | cccc} 
\hline\hline \\[-1.5ex]
\# of Buildings&\multicolumn{4}{c}{$D(X \,\lvert \rvert\, B)$ in nats, where $X$ is distributed by} \\
per $km^2$, $\lambda$ & $\Gamma(\alpha, \beta)$ &  $Exp(\beta^\prime)$  & $\frac{1}{2}\mathcal{N}(\sigma)$ & $Rayleigh(\sigma^\prime)$\\
\hline \\[-1.3ex] 
10 & 0.0101 & 0.0238 & 0.1221 & 0.4178 \\ 
40 & 0.0045 & 0.0181 & 0.1104 & 0.4041 \\
70 & 0.0022 & 0.0117 & 0.0954 & 0.3793 \\
\hline
\end{tabular}\vspace{-23pt}
\label{KLD}
\end{table}

	Motivated by these results, we present a simple, yet accurate, gamma distribution approximation of the NLOS bias for the general blocking case, via moment matching.
\vspace{-10pt}
\begin{approximation}[Distribution of Bias, With Blocking]
Consider $S_{(1)}\,|\, \{V_\infty \geq 1\}$ from Theorem \ref{S1wB} for a given test link setup under a Boolean model with set parameters, where its first and second moments, $\mathbb{E}[S_{(1)} \,|\, V_\infty \geq 1]$, $\mathbb{E}[S_{(1)}^2 \,|\, V_\infty \geq 1]$, are computed numerically for this particular setup using $f_{S_{(1)}}(s_{(1)} \,|\, V_\infty \geq 1)$.  Next, the NLOS bias is $B\,|\,\{V_\infty \!\geq\! 1\} = S_{(1)}\,|\,\{V_\infty \!\geq \!1\} - d$, and hence
\vspace{-11pt}
\begin{align*}
\mathbb{E}[B \,|\, V_\infty \!\geq 1] = \mathbb{E}[S_{(1)} |\, V_\infty \! \geq 1] - d \,, ~~~
\mathbb{E}[B^2 |\, V_\infty \!\geq 1] = \mathbb{E}[S_{(1)}^2 |\, V_\infty \!\geq 1] - 2 d \mathbb{E}[S_{(1)} |\, V_\infty\! \geq 1] + d^2\!. \\[-7ex]
\end{align*}
%\begin{align*}
%\mathbb{E}[B \,|\, V_\infty \!\geq 1] &= \mathbb{E}[S_{(1)} |\, V_\infty \! \geq 1] - d, ~~~\text{and} \\
%\mathbb{E}[B^2 \,|\, V_\infty \!\geq 1] &= \mathbb{E}[S_{(1)}^2 |\, V_\infty \!\geq 1] - 2 d \mathbb{E}[S_{(1)} |\, V_\infty\! \geq 1] + d^2. 
%\end{align*}
Then, the gamma distribution approximation of the NLOS bias distribution, $f_B$, is given by
\vspace{-14pt}
\begin{align*}
B\,|\, \{ V_\infty \geq 1 \} \distras{\,\,} f_B(b \,|\, V_\infty \geq 1) \approx \tilde{f}_B(b \,|\, V_\infty \geq 1 ) = \Gamma( \alpha, \beta),\\[-7ex]
\end{align*}
where $\textbf{Supp}(B\,|\, V_\infty \geq 1) = [0, \infty)$, which applies to $\tilde{f}_B$ as well, and 
\vspace{-7pt}
\begin{align} \label{parameters}
\alpha = \frac{\Big(\mathbb{E}[B \,|\, V_\infty \geq 1] \Big)^2}{\mathbb{E}[B^2 \,|\, V_\infty \geq 1] - \Big(\mathbb{E}[B \,|\, V_\infty \geq 1] \Big)^2} ~~~\text{and}~~~
\beta = \frac{\mathbb{E}[B \,|\, V_\infty \geq 1]}{\mathbb{E}[B^2 \,|\, V_\infty \geq 1] - \Big(\mathbb{E}[B \,|\, V_\infty \geq 1] \Big)^2}, \\[-6.8ex] \nonumber
\end{align}
are the shape and rate parameters, respectively.
\end{approximation}

\vspace{-12pt}
\begin{proof}
	Since $B\,|\,\{V_\infty \geq 1\} = S_{(1)}\,|\,\{V_\infty \geq 1\} - d$ and from Theorem \ref{S1wB}, $\textbf{Supp}(S_{(1)}\,|\, V_\infty \geq 1) = [d, \infty)$, then clearly $\textbf{Supp}(B\,|\,V_\infty \geq 1) = [0, \infty)$, which we apply to the gamma distribution approximation as well.
	
	Next, we perform moment matching to obtain our gamma distribution approximation of the NLOS bias distribution.  Towards this end, we note that in general, if $X \distras{\,\,} \Gamma(\alpha^\prime, \beta^\prime)$, then we may write the parameters, $\alpha^\prime$ and $\beta^\prime$, in terms of the moments of $X$ as follows:
	\vspace{-7pt}
\begin{align} \label{aprime_bprime}
\alpha^\prime = \frac{(\mathbb{E}[X])^2}{\mathbb{E}[X^2] - (\mathbb{E}[X])^2}, ~~\text{and} ~~~  \beta^\prime = \frac{\mathbb{E}[X]}{\mathbb{E}[X^2] - (\mathbb{E}[X])^2}. \\[-6.5ex] \nonumber
\end{align}
Since we aim for a gamma distribution approximation, \emph{i.e.}, $\tilde{f}_B(b \,|\, V_\infty \geq 1) = \Gamma ( \alpha , \beta )$, of the NLOS bias distribution, $f_B$, we can obtain its parameters by matching the moments of $f_B$ with the moments of $\tilde{f}_B$.  Thus, we write the parameters, $\alpha$ and $\beta$, from $\tilde{f}_B(b \,|\, V_\infty \geq 1) = \Gamma ( \alpha , \beta )$, in terms of the moments of $\tilde{f}_B$, as in (\ref{aprime_bprime}) above.  Then, matching moments, we substitute the moments of $f_B$ in for those of $\tilde{f}_B$ to obtain $\alpha$ and $\beta$ from (\ref{parameters}).  This completes the approximation.
\end{proof}  

%\begin{remark}
%As an example, the gamma distribution in Fig. \ref{Moment_Matched_Dists} was computed using this approximation.
%\end{remark}

\vspace{-2pt}
\subsubsection{Summary}

% *** NOTE: I could get rid of this section if need be. ***

	This section demonstrated that the analytically-derived distribution of NLOS bias, both with and without blocking, are well-approximated by a gamma and exponential distribution, respectively.  %The fact that an exponential distribution offers a good approximation in the special, non-blocking case follows from the void probability of a PPP, which itself has an exponential form and was used, implicitly, in the derivation of Corollary \ref{Dist_S1_without_blocking}.  This is more clearly seen, however, in \cite[Theorem 5]{Globecom}.  
While a gamma distribution might offer an even better approximation of the bias (than the exponential) in the non-blocking case, as it has two parameters to modify, we note that the exponential approximation is not only sufficient, but it also maintains an elegant simplicity, as evidenced in Approximation \ref{Approx_without_blocking}.  

% also does not stray far from intuition, since an exponential distribution is just a special case of the gamma distribution, \emph{i.e.}, $\Gamma(\alpha = 1, \beta) = Exp(\beta)$.
% i.e., don't need two parameters in the non-blocking case for the approximation

	\emph{Since the exponential and gamma NLOS bias models presented here are derived via the absolute delay of the first-arriving MPC (the most accurate method for determining NLOS bias to date), then it is fascinating to find that we have arrived at two NLOS bias models that have been assumed in the localization literature, via indirect or heuristic methods, for decades. % (see Sec. \ref{Sec_Intro}).  
Thus, this analysis suggests that these two bias models were indeed good assumptions and should perhaps be the standard bias models moving forward, especially for 5G mm-wave.}

\vspace{-15pt}
\subsection{Numerical Results} \label{Sec_Numerical_Results1}
\vspace{-4pt}
% NOTE: These bias distributions are CLOSE, since bias has a HUGE range of possible values (see Qi Journal)

\begin{figure}[t]
\centering
% --- VARYING LAMBDA FIGURE ---
\begin{minipage}[t]{0.48\textwidth}
\centering
\includegraphics [scale=0.54]{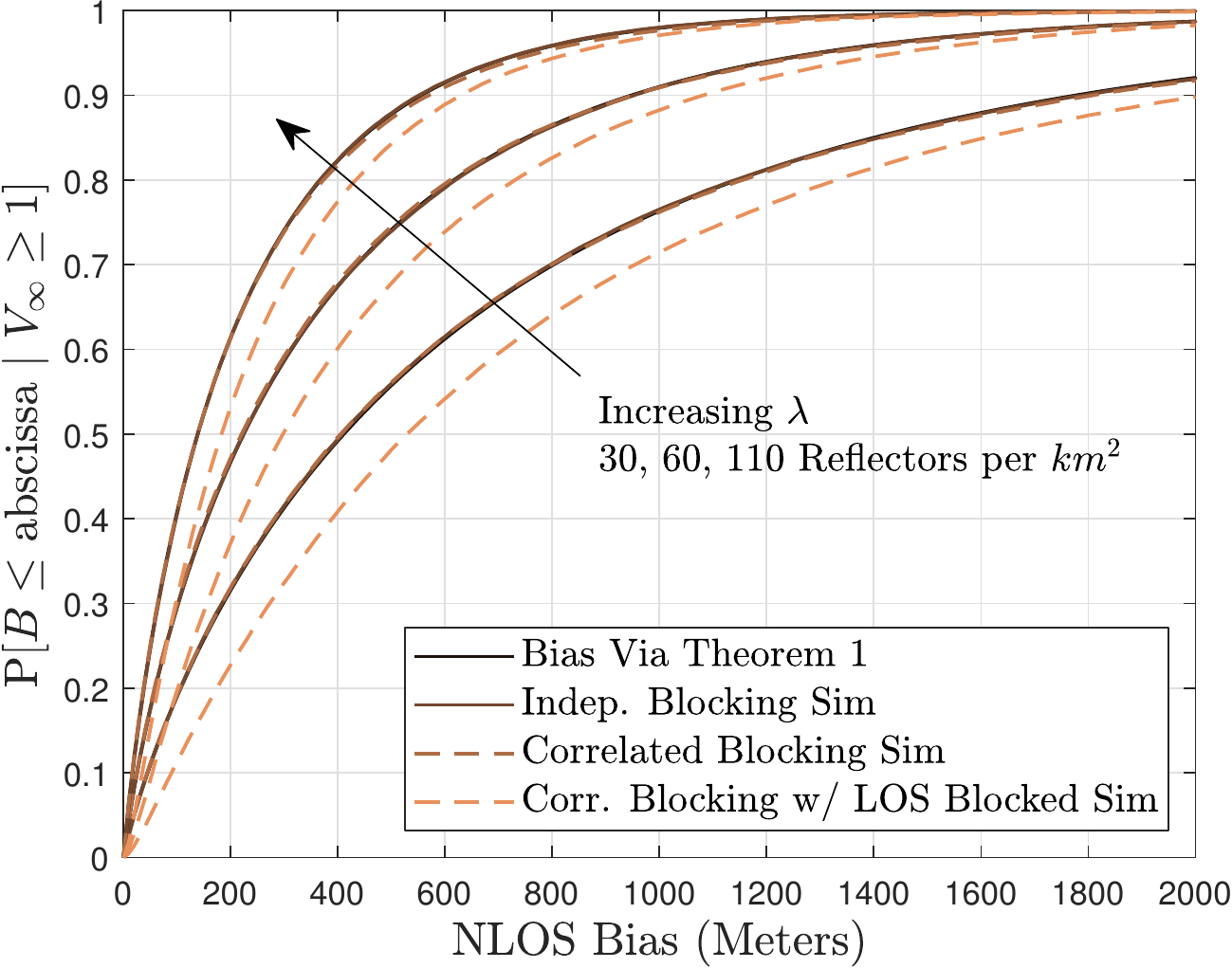}
\vspace{-9pt}
\caption{\textsc{NLOS Bias Distributions: Varying Reflector Density.}  These results were generated for a test link setup with $d = 350m$.  The reflector widths were sampled from $f_W = \text{unif}(w_{min} =10m, w_{max} = 40m, n_w = 4)$, and orientations from $f_\Theta = \text{unif}(\theta_{min} = 10\degree, \theta_{max} = 80\degree, n_\theta = 8)$.  Note, for each $\lambda$, the first three CDFs listed in the legend overlap each other.
\\[-8.9ex]}
\label{Varying_Lambda}
\end{minipage}\hfill
% --- VARYING d FIGURE ---
\begin{minipage}[t]{0.48\textwidth}
\centering
\includegraphics [scale=0.54]{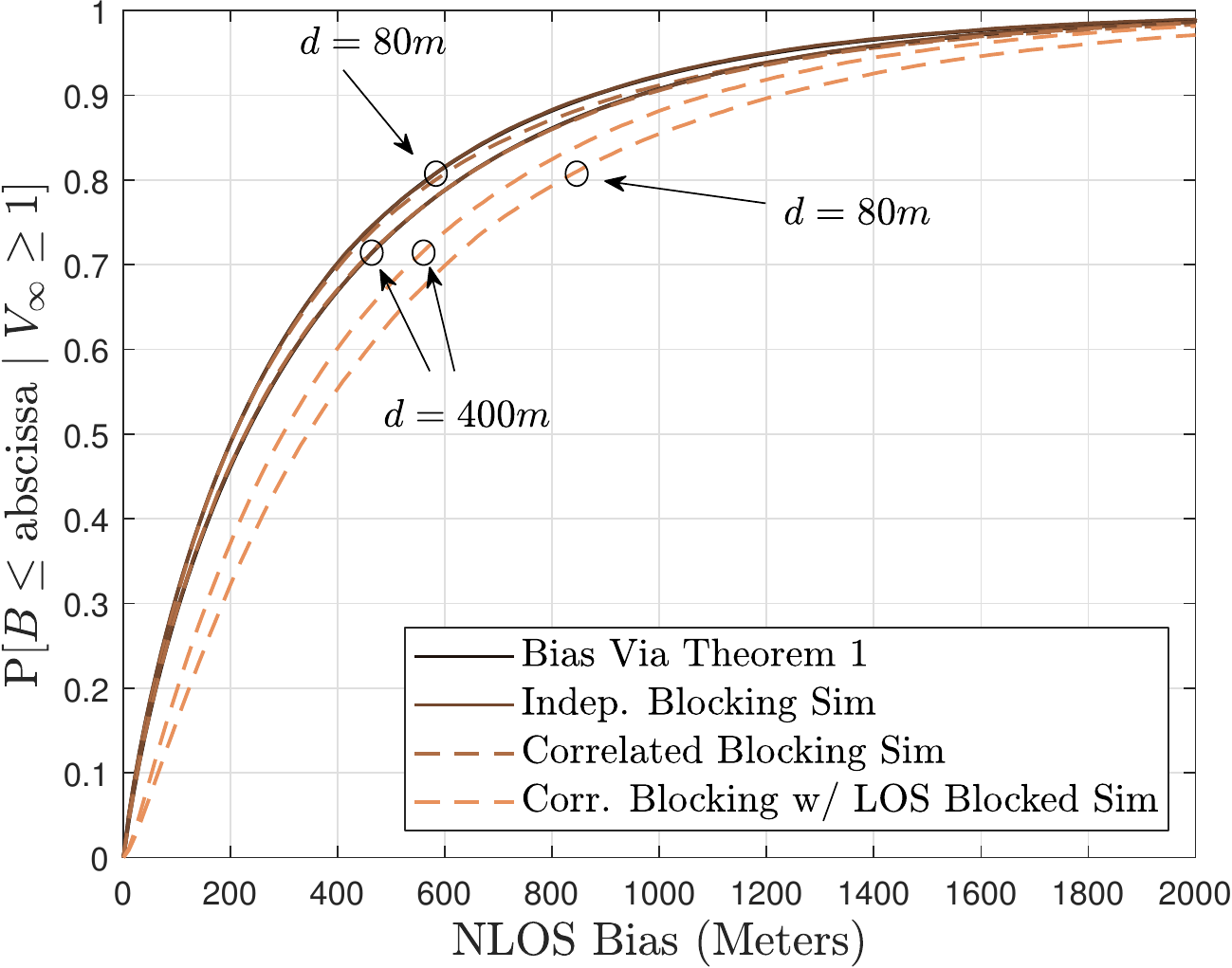}
\vspace{-9pt}
\caption{\textsc{NLOS Bias Distributions: Varying Separation Distance.}  These results were generated for a reflector density of $\lambda = 60$ reflectors per $km^2$.  The reflector widths and orientations were sampled from the same distributions listed in Fig. \ref{Varying_Lambda}.  Note that for each separation distance, $d$, the first three CDFs listed in the legend overlap, save for the slight deviation in the`\emph{Correlated Blocking Sim}' CDF for $d = 80m$.
\\[-8.9ex]}
\label{Varying_ds}
\end{minipage}
\end{figure}

	Here, we compare our analytically derived NLOS bias distribution against three separate NLOS bias distributions generated via simulation.  For our analytically derived bias, $B\,|\,\{V_\infty \geq 1\} = S_{(1)}\,|\,\{V_\infty \geq 1\} - d$, its distribution is given by $F_B(b \,|\, V_\infty \geq 1) = F_{S_{(1)}}(b + d \,|\, V_\infty \geq 1)$, where the CDF of $S_{(1)}\,|\,\{V_\infty \geq 1\}$ is given in Theorem \ref{S1wB}.  This is labeled as `\emph{Bias via Theorem \ref{S1wB}}' in Figs. \ref{Varying_Lambda} and \ref{Varying_ds}.  Next, the three comparison bias distributions were generated over $10^5$ Boolean model realizations where the path length of the first-arriving NLOS path was recorded in each realization.\footnote{\vspace{-7pt}Note that only Boolean model realizations with at least one non-blocked reflection were used.}   We now briefly detail how these three distributions were generated.

% --- NUMERICAL RESULTS ---
% --------------------------------	

	For the distribution labeled `\emph{Indep. Blocking Sim}' in Figs. \ref{Varying_Lambda} and \ref{Varying_ds}, a Boolean model of reflectors is placed over the test link setup and all reflection paths, without regards to blocking, are noted.  Then, each reflection path is checked for blockages by checking whether the incident and reflected paths are blocked using separate Boolean models (Definition \ref{Visible_Reflection_Point}).  Thus, the steps taken to simulate this distribution exactly matches our analytical approach.  This simply serves as an extra check on the analysis.  From Figs. \ref{Varying_Lambda} and \ref{Varying_ds}, we see that, indeed, this does match our analytical distribution (it is not visible, as it exactly overlaps with the analytical bias CDF).   
	
	For the distribution labeled `\emph{Correlated Blocking Sim}' in Figs. \ref{Varying_Lambda} and \ref{Varying_ds}, a Boolean model of reflectors is placed as above.  However, now each reflection path is checked for blockages using this \emph{same} Boolean model.  This represents true, correlated blocking on reflection paths.  \emph{In almost all cases plotted, this virtually overlaps the previous two distributions, indicating that the independent blocking assumption accurately captures correlated blocking on reflection paths.}
	
	Finally, we conduct a simulation similar to `\emph{Correlated Blocking Sim}' above, but with the extra restriction that only Boolean model realizations where the LOS path is blocked by at least one reflector are considered.  This departs from the Boolean model assumption, due to the forced conditioning, and also represents an extreme case of correlated blocking.  The bias distribution generated in this case is labeled `\emph{Corr. Blocking w/ LOS Blocked Sim}' in Figs. \ref{Varying_Lambda} and \ref{Varying_ds}.

	For the first result in Fig. \ref{Varying_Lambda}, these four distributions above were plotted for different reflector densities.  As the reflector density increases, the bias distributions shift to the left -- a trend that matches intuition.  Additionally, the close match with the `\emph{Corr. Blocking w/ LOS Blocked Sim}' case indicates that the analytical bias model, which assumes independent blocking, can reasonably capture the effect of forced blockages.  We explore this case further below, along with it's match with the analytical bias and possible corner-cases.

% --- NEW PARAGRAPH ---
	In Fig. \ref{Varying_ds}, the density of reflectors, as well as their size and orientation distributions, remained constant and the only parameter that changed was the base station-mobile separation distance, $\!d$.  We can see that as the base station and mobile begin to close in on each other, the `\emph{Corr. Blocking w/ LOS Blocked Sim}' CDF begins to deviate slightly from the other three CDFs.  This occurs due to the conditioning in the `\emph{Corr. Blocking w/ LOS Blocked Sim}' case, where at least one reflector is forced between the base station and mobile; blocking the LOS path.  As $d$ decreases, the buildings appear larger in relation to this base station-mobile separation distance, and thus, forcing at least one large building in between the two introduces significant correlated blocking.  For example, when $d = 80m$, if the largest reflector, $w = 40m$, is placed appropriately, it can take up to $\sim\!\!70\%$ of the separation distance, thus having the potential to block many reflection paths at once. For the reader accustomed to examining positioning error distributions, bias error, relatively speaking, is significantly larger, often on the order of hundreds of meters to a few kilometers, \cite{killer_issue}, \cite{Qi_NLOS_Journal}.  Thus, despite the amount of correlated blocking introduced in the $d=80m$ case, the deviation of the `\emph{Corr. Blocking w/ LOS Blocked Sim}' CDF from the other three CDFs is surprisingly small.  Consequently, the independent blocking assumption holds reasonably well in these cases of significant correlated blocking.  That being said, placing the base station and mobile at the extreme ends of a large building, so that the building covers $\sim\!\!100\%$ of the separation distance, will almost certainly cause the independent blocking assumption to break down and the accuracy of the analytical bias distribution to degrade.  As with any analytical model, it is always important to be aware of such model limitations and `corner cases'.

\vspace{-17pt}
\subsection{Discussion} \label{Sec_Discussion}
\vspace{-4pt}

	To the authors' knowledge, no outdoor measurement campaigns characterizing NLOS bias currently exist.  This observation was noted in 1996 \cite{GSM},
%\begin{quote}
%Unfortunately the NLOS error has not been subject to research before.  Many impulse response measurements have been made, but the focus has been on the shape of the impulse response, not on the absolute timing.
%\end{quote}
and was echoed again in 2007 \cite{Mailaender}:
\begin{quote}
At the present time, very little is known about the statistics of the NLOS variables [bias] in realistic propagation environments, and there are no established models.
\end{quote}
and we believe this lack of measurement data characterizing NLOS bias still exists to this day.  

	Appropriately characterizing the NLOS bias outdoors requires measuring the absolute delay of the first-arriving MPC under NLOS conditions.  This is difficult for a number of reasons, the first of which is the need for highly-accurate, nanosecond-level (or less) synchronization between the Tx and Rx.  Not many non-tethered channel sounders exist that offer enough bandwidth, along with the accurate synchronization necessary, to perform the absolute TOF measurements needed \cite{George_Channel_Sounder}.  Additionally, the measurements themselves are tedious due to the need for Rubidium clocks at the Tx and Rx which require a ``synchronization training period'' of an hour or more and which can fall out of synchronization just as quickly \cite{George_Channel_Sounder}.  Furthermore, the Tx and Rx would require accurate GPS positioning in order to extract the bias, which can be hard to obtain depending on the measurement environment, \emph{e.g.}, urban canyons.  Finally, these measurements would need to occur, for a given Tx-Rx separation distance, over many realizations of the surrounding environment in order to generate statistics of the bias.  Thus, deriving accurate \emph{analytical} bias models, such as those presented here, is necessary due to these difficulties in \emph{empirically} characterizing the NLOS bias.

	Given this lack of data with which to compare our bias distributions against, it is reasonable to ask: ``What \emph{can} be gleaned about NLOS bias from other (semi-related) measurement-based models that \emph{already} exist in the literature?'' 
% This information is more important for localization community than it is for channel modeling community.
To answer this, we attempt to glean insight into the nature of NLOS bias by examining a mm-wave channel model of excess multipath delays derived from outdoor LOS measurements.  From the model in \cite{Samimi}, excess delays of MPCs from LOS PDPs are sampled from an exponential distribution, which was derived via a fit to measurement data.  Supposing these multipath delays are independent, then the excess delay of the \emph{first-arriving} multipath component is also exponentially distributed.\footnote{The first order statistic of i.i.d. exponentially distributed RVs is also exponentially distributed.}  Since the measurements are LOS, then to obtain the absolute delay, one can simply add on the LOS TOF, which is a simple shift of the exponential distribution.  Finally, if we assume that there is a blockage that 1) removes the LOS component, and 2) is not significantly correlated with the first-arriving reflection path, then there is reason to believe that this shifted exponential distribution can reasonably represent the absolute delay of the first-arriving MPC in a NLOS scenario.  (Similar reasoning is also given in \cite{Swaroop}.)  This evidence suggests that the distribution for the path length (\emph{i.e.}, absolute delay) of the first-arriving MPC presented here is at least ``in-line'' with what one would expect in reality.

	We conclude this section by noting that Theorem \ref{S1wB}, \emph{i.e.}, the distribution of the first-arriving MPC, can also be used in channel simulators \cite{NYUSIM2} to provide a distribution from which to sample an absolute timing reference for excess delay PDPs in NLOS scenarios.

% MY REASONING HERE WAS RIGHT :)
% Quote from ``UWB Geolocation Techniques for IEEE 802.15.4a Personal Area Networks''
%Considering the IEEE channel model [11], we see that multipath arrival times follow a Poisson distribution. In other words, the time difference between any two paths is an exponentially distributed random variable. The IEEE channel measurements provide the mean of this exponential random variable in different scenarios. Assuming that we are able to detect the first arriving path in the NLOS situation by the algorithm in Section II, the absent LOS path can be considered as the preceding path of that first arriving NLOS path. Hence, the NLOS error can be modeled as an exponentially distributed random variable as specified by the channel measurements.
% This was the same argument used in Swaroop Venkatesh's dissertation as well!

% ******* AOA of First-Arriving Reflection *******
\vspace{-7pt}
\section{The Angle-of-Arrival of the First-Arriving Reflection} \label{AOAsection}
\vspace{-1pt}

% --- INTRO PARAGRAPH ---

	This section derives the AOA distribution of the first-arriving reflected path, \emph{with blocking}.  We begin with some important AOAs which correspond to the boundary PRPs from Lemma \ref{Lemma_BPRPs}.
	
\vspace{-8pt}	
\begin{definition}[The $s$-Meter AOAs for $\mathcal{R}_{w, \theta, \mathbf{c}}$] \label{sAOAs}
Recall from Lemma \ref{Lemma_BPRPs} that there are precisely four PRPs that reflector $\mathcal{R}_{w, \theta, \mathbf{c}}$ can intersect to produce a reflection of exactly $s$ meters, where $d < s < \infty$.  These PRPs are labeled, $\mathbf{h}_{\text{I}, \theta}(s)$, $\mathbf{h}_{\text{II}, \theta}(s)$, $\mathbf{h}_{\text{III}, \theta}(s)$, $\mathbf{h}_{\text{IV}, \theta}(s)$.  If $\partial \mathcal{R}_{w, \theta, \mathbf{c}}$ were to intersect $\mathbf{h}_{q, \theta}(s)$ ($q \in \mathcal{Q}$) to produce a reflection, then we label the AOA at $\mathbf{m}$ of the reflected path, $\mathcal{L}_{[\mathbf{h}_{q, \theta}(s), \mathbf{m}]}$, by $\psi_{q, \theta}(s)$, which is measured in radians c.c.w. w.r.t. the $+x$-axis.  We call these four AOAs the \emph{$s$-meter AOAs for $\mathcal{R}_{w, \theta, \mathbf{c}}$}.  The $Q_\text{I}$ $s_1$-meter AOA for $\mathcal{R}_{w, \theta_j, \mathbf{c}}$ is depicted in Fig. \ref{Analytical_FW_Fig}. 
\end{definition}

\vspace{-8pt}
	Observe, for example, the $Q_\text{I}$ $s_1$-meter AOA for $\mathcal{R}_{w, \theta_j, \mathbf{c}}$, \emph{i.e.}, $\psi_{\text{I}, \theta_j}(s_1)$, in Fig. \ref{Analytical_FW_Fig}.  Note that as $s_1$ increases, $\mathbf{h}_{\text{I}, \theta_j}(s_1)$ tracks along the reflection hyperbola, $\mathcal{H}_{\theta_j}$, as the $s_1$-ellipse boundary, $\partial \mathcal{P}_{s_1}$, expands out.  Consequently, the corresponding AOA, $\psi_{\text{I}, \theta_j}(s_1)$, changes as well.  Hence, the $s$-meter AOAs for $\mathcal{R}_{w, \theta, \mathbf{c}}$ are \emph{functions} of $s$ (parameterized by $\theta$).  Their behavior as functions of $s$ is of particular importance in subsequent derivations and so we present the following lemma.
	
\vspace{-7pt}	
\begin{lemma}[The $s$-meter AOA Functions for $\mathcal{R}_{w, \theta, \mathbf{c}}$]  \label{Psi_Fns}
The $s$-meter AOA functions for $\mathcal{R}_{w, \theta, \mathbf{c}}$, i.e., $\alpha = \psi_{q, \theta}(s)$, for $q \in \mathcal{Q}$, are given in Table \ref{sAOA_functions}, along with their derivatives and inverses.
\end{lemma}

\begin{table*}[t]% ---> puts at top %[h]
\renewcommand{\arraystretch}{.75}
\caption{The $s$-Meter AOA Functions for $\mathcal{R}_{w, \theta, \mathbf{c}}$, \emph{i.e.}, $\alpha = \psi_{q,\theta}(s)$, Their Derivatives, and Respective Inverses}
\label{sAOA_functions}
\vspace{-8pt}
\centering % centering table
%\fontsize{9pt}{10pt}
\footnotesize
\scalebox{1.04}{%
\begin{tabular}{ll | ll} % creating columns
\hline\hline  
& & & \\[-1.2ex]
$\psi_{\text{I},\theta}(s) =\cos^{-1}\! \!\Big(\! -\!\frac{d \sin\theta}{s} \Big) \!+\! \theta \!-\! \frac{\pi}{2}$, & $d \leq s < \infty$ & $\psi_{\text{I},\theta}^\prime(s) = - \frac{d \sin\theta}{s \sqrt{s^2 - d^2\sin^2\theta}}$, & $d \leq s < \infty$ \\ & & & \\[-1ex]
$\psi_{\text{I},\theta}^{-1} (\alpha) = \frac{d \sin\theta}{\sin\big(\alpha - \theta\big)}$, & $\theta < \alpha \leq 2\theta$ & $\big(\psi_{\text{I},\theta}^{-1}\big)^\prime (\alpha) =-\psi_{\text{I},\theta}^{-1}(\alpha) \cot\big(\alpha \!- \theta \big)$, & $\theta < \alpha \leq 2\theta$ \\[-1.5ex]
& & & \\
\hline 
& & & \\[-1.2ex]
$\psi_{\text{II},\theta}(s) =\cos^{-1}\!\! \Big(\! -\!\frac{d \cos\theta}{s} \Big) \!+\! \theta$, & $d \leq s < \infty$ & $\psi_{\text{II},\theta}^\prime(s) = - \frac{d \cos\theta}{s \sqrt{s^2 - d^2\cos^2\theta}}$, & $d \leq s < \infty$ \\ & & & \\[-1ex]
$\psi_{\text{II},\theta}^{-1} (\alpha) = \frac{-d \cos\theta}{\cos\big(\alpha - \theta\big)}$, & $\frac{\pi}{2} + \theta < \alpha \leq \pi$ & $\big(\psi_{\text{II},\theta}^{-1}\big)^\prime (\alpha) =\psi_{\text{II},\theta}^{-1}(\alpha) \tan\big(\alpha \!- \theta \big)$, & $\frac{\pi}{2} + \theta < \alpha \leq \pi$ \\[-1.5ex]
& & & \\
\hline 
& & & \\[-1.2ex]
$\psi_{\text{III},\theta}(s) =\cos^{-1}\!\! \Big(\frac{d \sin\theta}{s} \Big) \!+\! \theta \!+\! \frac{\pi}{2}$, & $d \leq s < \infty$ & $\psi_{\text{III},\theta}^\prime(s) = \frac{d \sin\theta}{s \sqrt{s^2 - d^2\sin^2\theta}}$, & $d \leq s < \infty$ \\ & & & \\[-1ex]
$\psi_{\text{III},\theta}^{-1} (\alpha) = \frac{d \sin\theta}{\sin\big(\alpha - \theta\big)}$, & $\pi \leq \alpha < \pi + \theta$ & $\big(\psi_{\text{III},\theta}^{-1}\big)^\prime (\alpha) =-\psi_{\text{III},\theta}^{-1}(\alpha) \cot\big(\alpha \!- \theta \big)$, & $\pi \leq \alpha < \pi + \theta$
 \\[-1.5ex]
& & & \\
\hline 
& & & \\[-1.2ex]
$\psi_{\text{IV},\theta}(s) =\cos^{-1}\!\! \Big( \frac{d \cos\theta}{s} \Big) \!+\! \theta \!+\! \pi$, & $d \leq s < \infty$ & $\psi_{\text{IV},\theta}^\prime(s) =  \frac{d \cos\theta}{s \sqrt{s^2 - d^2\cos^2\theta}}$, & $d \leq s < \infty$ \\ & & & \\[-1ex]
$\psi_{\text{IV},\theta}^{-1} (\alpha) = \frac{-d \cos\theta}{\cos\big(\alpha - \theta\big)}$, & $\pi \!+\!  2\theta\! \leq \alpha \!< \!\frac{3\pi}{2}\!+\! \theta$ & $\big(\psi_{\text{IV},\theta}^{-1}\big)^\prime (\alpha) =\psi_{\text{IV},\theta}^{-1}(\alpha) \tan\big(\alpha \!- \theta \big)$, & $\pi \!+\!  2\theta\! \leq \alpha \!<\! \frac{3\pi}{2}\!+\! \theta$ \\
\hline\hline\\[-3ex]
% --- Unused ---
% Angle of $\mathbf{x}$ w.r.t. $+x$ axis
\end{tabular}}
\vspace{-20pt}
\end{table*}

\vspace{-12pt}
\begin{proof}
We present only the derivation of the $Q_\text{I}$ $s$-meter AOA function for $\mathcal{R}_{w, \theta, \mathbf{c}}$, \emph{i.e.} $\psi_{\text{I}, \theta}(s)$, as the other quadrant $s$-meter AOA functions follow similarly. %, and the derivatives and inverses follow trivially.  
We begin by recalling from Definition \ref{sAOAs} that $\psi_{\text{I}, \theta}(s)$ is the angle of the slope of the reflection path $\mathcal{L}_{[\mathbf{h}_{\text{I}, \theta}(s), \mathbf{m}]}$.  Thus, letting $\alpha = \psi_{\text{I}, \theta}(s)$, we have the relationship: $\tan(\alpha) = \frac{-[\mathbf{h}_{\text{I}, \theta}(s)]_2 } { \frac{d}{2} - [\mathbf{h}_{\text{I}, \theta}(s)]_1}$, which leads to the following: 
%$\tan(\alpha) = -[\mathbf{h}_{\text{I}, \theta}(s)]_2 \, / \, \Big( \frac{d}{2} - [\mathbf{h}_{\text{I}, \theta}(s)]_1 \Big)$
\vspace{-10pt}
\begin{align} \label{tan_expression}
\alpha = \psi_{\text{I}, \theta}(s) =
\begin{cases}
\tan^{-1} \!\left[ \frac{d^2 - s^2} {d \sqrt{s^2 \csc^2 \theta - d^2} - s^2 \cot\theta} \right], &\! \frac{d^2 - s^2} {d \sqrt{s^2 \csc^2  \theta - d^2} - s^2 \cot\theta} \geq 0 \\
\tan^{-1} \!\left[ \frac{d^2 - s^2} {d \sqrt{s^2 \csc^2  \theta - d^2} - s^2 \cot\theta} \right] + \pi, &\! \frac{d^2 - s^2} {d \sqrt{s^2 \csc^2 \theta - d^2} - s^2 \cot\theta} < 0 \\[-0.2ex]
\end{cases},~~ \text{for}~~ d < s < \infty.
%\!\stackrel{(b)}{=}\!
%\begin{cases}
%\!\tan^{-1}\!\! \left[ \!\frac{d^2 - s^2} {d \sqrt{s^2 \csc^2 \!\theta - d^2} - s^2 \!\cot\theta} \!\right], & \!\!\! s \!>\! d \tan \theta \\
%\!\tan^{-1}\!\! \left[ \!\frac{d^2 - s^2} {d \sqrt{s^2 \csc^2 \!\theta - d^2} - s^2 \!\cot\theta} \!\right]\! \!+\! \pi, &\! \!\! s \!\leq \!d \tan \theta \\[-0.2ex]
%\end{cases},
\vspace{-16pt}
\end{align}
This follows from simplifying the r.h.s. of the above relationship using Lemma \ref{Lemma_BPRPs}, taking $\tan^{-1}\!(\cdot)$ of both sides, and noting that for $-\infty \leq x < 0$, $\textbf{Range}\big(\!\tan^{-1}(x) \big) = [\pi/2, 0)$, and so $\pi$ must be added when the slope of the reflection path is negative.  Finally, taking $\cos^{-1} \!\big(\! \cos ( \cdot ) \big)$ of the r.h.s. in (\ref{tan_expression}) and simplifying yields the expression for $\psi_{\text{I}, \theta}(s)$ in Table \ref{sAOA_functions} for $d < s < \infty$.\footnote{Note, in Table \ref{sAOA_functions} we allow $s=d$.  This corresponds to a RP at $\mathbf{h}_{\text{I}, \theta}(d) = \mathbf{m}$, and since there is no reflection path, the d-meter AOA, $\psi_{\text{I}, \theta}(d)$, is not defined.  In this case, we simply take the AOA to be the limiting case, \emph{i.e.}, $\psi_{\text{I}, \theta}(d) = \lim_{s \to d} \psi_{\text{I}, \theta}(s) = 2 \theta$, which is obtained from (\ref{tan_expression}) via L'H\^{o}pital's rule and by noting the conditions' dependency on $\theta$.  Note that $\psi_{\text{I}, \theta}(d)$ in the Table \ref{sAOA_functions} expression conveniently yields the same value.  A similar argument is used to handle the $s=d$ case in $Q_\text{IV}$ as well.} 
\end{proof}

\vspace{-17pt}
\begin{definition}[AOA of the $1^\text{st}$-Arriving Reflection]
Let $A_{(1)}$ be the RV representing the AOA, in radians, of the first-arriving reflection, measured c.c.w. w.r.t. the $+x$-axis. Note, $0 < A_{(1)} < 2\pi$. 
\end{definition}

\vspace{-16pt}
\begin{remark}
Although the distribution of $S_{(1)}$ was derived assuming the LOS path was blocked, Assumption \ref{Indep_Blocking_Assump} asserts that $f_{S_{(1)}}(s_{(1)} \,|\, V_\infty \geq 1)$ does not change when blocking on the LOS path is ignored.  Consequently, moving forward, we simply assume $f_{S_{(1)}}(s_{(1)} \,|\, V_\infty \geq 1)$ was derived irrespective of what happens on the LOS path.  Since we also seek to derive the distribution of $A_{(1)}$ irrespective of what happens on the LOS path, then the distributions of $S_{(1)}$ and $A_{(1)}$ will both characterize the first-arriving reflection \emph{under the same conditions}.  This implies that $S_{(1)}$ and $A_{(1)}$ each describe different properties of the \emph{same} first-arriving reflection path; hence, $A_{(1)}$ is subject to the same existence issues as $S_{(1)}$, and so must be conditioned on $\{V_\infty \geq 1\}$ as well.  %Hence, this implies they both characterize the same reflection path.
\end{remark}

\vspace{-18pt}
\begin{lemma}[$A_{(1)}$ Conditional Distribution Given $S_{(1)}$] \label{CondAOA}
Consider the test link setup and Boolean model.  Then, the conditional PDF of $A_{(1)}$ given a first-arriving reflection of distance $S_{(1)}$ is: $f_{A_{(1)}}(\alpha_{(1)} \,|\, S_{(1)}, V_\infty \geq 1) = \frac{1}{\sum_{j^\prime=1}^{n_\theta} \sum_{q^\prime \in \mathcal{Q}} \omega_{q^\prime, \theta_{j^\prime}}(s_{(1)})}  \sum_{j=1}^{n_\theta} \sum_{q \in \mathcal{Q}} \omega_{q, \theta_j}(s_{(1)}) \, \delta \Big( \alpha_{(1)} - \psi_{q, \theta_j}(s_{(1)}) \Big)$, where $\omega_{\text{\emph{I}}, \theta_j}(s_{(1)})$, $\omega_{\text{\emph{II}}, \theta_j}(s_{(1)})$, $\omega_{\text{\emph{III}}, \theta_j}(s_{(1)})$, and $\omega_{\text{\emph{IV}}, \theta_j}(s_{(1)})$, are given, respectively, by:
% --- NOTE ---
% 1) The omega_q actually simplify to 1/sqrt but whatever, fixing this will require taking up too much space in the document!
\vspace{-4pt}
\begin{align*}
&\rho\big(\mathbf{h}_{\text{\emph{I}}, \theta_j}(s_{(1)})\big) \textstyle\frac{s_{(1)} \big(\psi^{-1}_{\text{\emph{I}}, \theta_j}\big)^\prime \big(\psi_{\text{\emph{I}}, \theta_j}(s_{(1)})\big)}{2\sqrt{s_{(1)}^2 - d^2 \sin^2\!\theta_j}} \psi_{\text{\emph{I}}, \theta_j}^\prime (s_{(1)}), ~~~~~~~~ \rho\big(\mathbf{h}_{\text{\emph{II}}, \theta_j}(s_{(1)})\big) \textstyle\frac{s_{(1)} \big(\psi^{-1}_{\text{\emph{II}}, \theta_j}\big)^\prime \big(\psi_{\text{\emph{II}}, \theta_j}(s_{(1)})\big)}{2\sqrt{s_{(1)}^2 - d^2 \cos^2\!\theta_j}} \psi_{\text{\emph{II}}, \theta_j}^\prime (s_{(1)}) \\
&\rho\big(\mathbf{h}_{\text{\emph{III}}, \theta_j}(s_{(1)})\big) \textstyle\frac{s_{(1)} \big(\psi^{-1}_{\text{\emph{III}}, \theta_j}\big)^\prime \big(\psi_{\text{\emph{III}}, \theta_j}(s_{(1)})\big)}{2\sqrt{s_{(1)}^2 - d^2 \sin^2\!\theta_j}} \psi_{\text{\emph{III}}, \theta_j}^\prime (s_{(1)}), ~~~ \rho\big(\mathbf{h}_{\text{\emph{IV}}, \theta_j}(s_{(1)})\big) \textstyle\frac{s_{(1)} \big(\psi^{-1}_{\text{\emph{IV}}, \theta_j}\big)^\prime \big(\psi_{\text{\emph{IV}}, \theta_j}(s_{(1)})\big)}{2\sqrt{s_{(1)}^2 - d^2 \cos^2\!\theta_j}} \psi_{\text{\emph{IV}}, \theta_j}^\prime (s_{(1)}),
\end{align*}
$\mathbf{h}_{q, \theta_j}$, $\rho(\cdot)$, and $\psi_{q, \theta_j}$ from Lemmas \ref{Lemma_BPRPs}, \ref{Lemma_PRPV}, and \ref{Psi_Fns}, respectively, and $\textbf{\emph{Supp}}(A_{(1)} \,|\, S_{(1)}, V_\infty \!\geq\! 1) \!=\! \big\{\psi_{q, \theta_j}(s_{(1)})\big\}$ for $q \!\in\! \mathcal{Q}$, $j \!\in\! \{1, \dots, n_\theta\}$.
\end{lemma}

\vspace{-10pt}
\begin{proof}
Let $S_{(1)} = s_{(1)}$ and consider all of the boundary PRPs, $\mathbf{h}_{q, \theta_j}(s_{(1)})$ ($q \in \mathcal{Q}$,  $j \in \{1, \dots, n_\theta\}$) that reflectors of $\mathcal{B}$ can intersect to produce the first-arriving reflection path of distance $s_{(1)}$.  (Letting $s_1 = s_{(1)}$, Fig. \ref{Analytical_FW_Fig} gives an example depiction of four of these PRPs, \emph{i.e.}, those associated with a reflector of orientation $\theta_j$.  The PRPs for the reflectors of other orientations would be placed along $\partial \mathcal{P}_{s_{(1)}}$ as well, if depicted.)  Recall from Definition \ref{sAOAs}, that associated with each of these PRPs is an AOA for the reflected path, $\psi_{q, \theta_j}(s_{(1)})$. Since there are `$4n_\theta$' potential AOAs, we must determine the probability $A_{(1)}$ equals any one of them.
% it is tempting to say that the probability $A_{(1)}$ equals any one of them is $1/(4n_\theta)$, but this would be naive.   Looking at the four $\mathbf{h}_{q, \theta_j}(s_{(1)})$'s in Fig. \ref{Analytical_FW_Fig} for example, clearly, each one of these is visible with a different probability, $\rho \big( \mathbf{h}_{q, \theta_j}(s_{(1)}) \big)$, and so the probability $A_{(1)}$ equals any of their corresponding AOAs, $\psi_{q, \theta_j}(s_{(1)})$, should at least take into account this visibility probability.  
Thus, we seek a conditional distribution of the form stated in the lemma, where each possible AOA, $\psi_{q, \theta}(s_{(1)})$, has associated with it a weighting factor, $\omega_{q, \theta_j}(s_{(1)})$, where $P[A_{(1)} = \psi_{q, \theta}(s_{(1)}) \,|\, S_{(1)}, V_\infty] = \frac{\omega_{q, \theta_j}(s_{(1)})}{\sum_{j^\prime=1}^{n_\theta} \sum_{q^\prime \in \mathcal{Q}} \omega_{q^\prime, \theta_{j^\prime}}(s_{(1)})}$.  %Our aim now is to determine these weighting factors.  

	Correctly determining these weighting factors requires conditioning on $S_{(1)}$ being within an infinitesimal sliver:  $f_{A_{(1)}}(\alpha_{(1)} \,|\, S_{(1)}, V_\infty \geq 1) = \lim_{\Delta s_{(1)} \to 0} f_{A_{(1)}}(\alpha_{(1)} \,|\, s_{(1)} \leq S_{(1)} \leq s_{(1)} + \Delta s_{(1)}, V_\infty \geq 1)$.  In so doing, by ignoring zero probability events in this conditioning, we can choose $\Delta s_{(1)}$ s.t. \emph{one and only one} reflector produces a reflection with distance in $[s_{(1)}, s_{(1)} + \Delta s_{(1)}]$, which is that producing the first-arriving reflection.  This implies we have conditioned on one and only one reflector having a VRP in `$\mathcal{P}_{s_{(1)}+\Delta s_{(1)}} / (\mathcal{P}_{s_{(1)}} / \partial \mathcal{P}_{s_{(1)}})$.'  Now, there are `$4n_w n_\theta$' mutually exclusive ways, or ``sub-events,'' in which this reflector can produce a VRP in `$\mathcal{P}_{s_{(1)}+\Delta s_{(1)}} / (\mathcal{P}_{s_{(1)}} / \partial \mathcal{P}_{s_{(1)}})$,' with the typical sub-event being the center point of edge $\mathcal{E}_q$ of $\mathcal{R}_{w_i, \theta_j, \mathbf{c}}$ (for $q \in \mathcal{Q}, i \in \{1, \dots, n_w\}, j \in \{1, \dots, n_\theta\}$) falling in $\Omega_q = \mathcal{E}_q \oplus \big( \mathcal{H}_{\theta_j} \cap Q_q \cap \mathcal{P}_{s_{(1)}+\Delta s_{(1)}}/(\mathcal{P}_{s_{(1)}}/ \partial \mathcal{P}_{s_{(1)}}) \big)$ (see proof of Lemma \ref{Lemma_Vs} for $\mathcal{E}_q$ definition).  Letting $s_1 = s_{(1)}$ and $s_2 = s_{(1)}+\Delta s_{(1)}$, Fig. \ref{Analytical_FW_Fig} depicts four of these sub-events for $\mathcal{R}_{w_i, \theta_j, \mathbf{c}}$.  Examining Fig. \ref{Analytical_FW_Fig}, we note that as $\Delta s_{(1)} \to 0$, the $\mathcal{E}_\text{I}$ edge center point of $\mathcal{R}_{w_i, \theta_j, \mathbf{c}}$ falling in $\Omega_\text{I}$ implies that the RP for the first-arriving reflection is at $\mathbf{h}_{\text{I}, \theta_j}(s_{(1)}) \implies A_{(1)} = \psi_{\text{I}, \theta_j}(s_{(1)})$.  Thus, we now have a way for determining whether $A_{(1)}$ equals any of the `$4n_\theta$' potential AOAs, since this occurs if the reflector producing the first-arriving reflection falls in one of the corresponding  sub-events, \emph{i.e.}, $\Omega_\text{I}, \Omega_\text{II}, \Omega_\text{III}, \Omega_\text{IV}$ for $\mathcal{R}_{w_i, \theta_j, \mathbf{c}}$.  This yields $f_{A_{(1)}}(\alpha_{(1)} \,|\, s_{(1)} \leq S_{(1)} \leq s_{(1)} + \Delta s_{(1)}, V_\infty \geq 1) =$
\vspace{-7pt}
\begin{align} \label{Expanded_Conditioning}
\frac{\frac{1}{n_w} \frac{1}{n_\theta} \!\! \sum\limits_{j=1}^{n_\theta} \sum\limits_{q \in \mathcal{Q}} \sum\limits_{i=1}^{n_w} \! w_i \Gamma_{q, \theta_j} (s_{(1)})  \, \delta \Big( \alpha_{(1)} - \psi_{q, \theta_j}(s_{(1)}) \! \Big)}       {\frac{1}{n_w} \frac{1}{n_\theta} \! \sum_{j^\prime=1}^{n_\theta} \sum_{q^\prime \in \mathcal{Q}} \sum_{i^\prime=1}^{n_w} \! w_{i^\prime} \Gamma_{q^\prime, \theta_{j^\prime}}(s_{(1)})}  =  \frac{\sum\limits_{j=1}^{n_\theta} \sum\limits_{q \in \mathcal{Q}} \!\!\Gamma_{q, \theta_j} (s_{(1)})  \, \delta \Big( \alpha_{(1)} - \psi_{q, \theta_j}(s_{(1)}) \!\Big)}    {\sum_{j^\prime=1}^{n_\theta} \sum_{q^\prime \in \mathcal{Q}} \Gamma_{q^\prime, \theta_{j^\prime}}(s_{(1)})}, \\[-6.3ex] \nonumber
\end{align}
where $\Gamma_{\text{I}, \theta_j}(s_{(1)})$, $\Gamma_{\text{II}, \theta_j}(s_{(1)})$, $\Gamma_{\text{III}, \theta_j}(s_{(1)})$, and $\Gamma_{\text{IV}, \theta_j}(s_{(1)})$ are given, respectively, by
\vspace{-4pt}
\begin{align*}
\int\limits_{[\mathbf{R}_{\theta_j} \mathbf{h}_{\text{I}, \theta_j} (s_{(1)})]_1}^{[\mathbf{R}_{\theta_j} \mathbf{h}_{\text{I}, \theta_j}(s_{(1)} + \Delta s_{(1)})]_1}  \!\!\!\!\!\!\!\!\!\!\!\!\!\!\!\!\!\!\!\!\!\!   \rho\Big( \mathbf{R}^{-1}_{\theta_j} \mathbf{g}^*_\text{I}(x_{\theta_j}\!) \!\Big) \text{d}x_{\theta_j}, ~  \!\!\!\!\!\!\!\!\!\!\!\!\!\!\!\!\!\!\!\! \int\limits_{[\mathbf{R}_{\theta_j} \mathbf{h}_{\text{II}, \theta_j}(s_{(1)})]_2}^{[\mathbf{R}_{\theta_j} \mathbf{h}_{\text{II}, \theta_j}(s_{(1)} + \Delta s_{(1)})]_2}  \!\!\!\!\!\!\!\!\!\!\!\!\!\!\!\!\!\!\!\!\!\!   \rho\Big( \mathbf{R}^{-1}_{\theta_j} \mathbf{g}^*_\text{II}(y_{\theta_j}\!) \!\Big)\text{d}y_{\theta_j},  ~  \!\!\!\!\!\!\!\!\!\!\!\!\!\!\!\!\!\!\!\!\! \int\limits_{[\mathbf{R}_{\theta_j} \mathbf{h}_{\text{III}, \theta_j}(s_{(1)} + \Delta s_{(1)})]_1}^{[\mathbf{R}_{\theta_j} \mathbf{h}_{\text{III}, \theta_j}(s_{(1)})]_1}  \!\!\!\!\!\!\!\!\!\!\!\!\!\!\!\!\!\!\!\!\!\!\!   \rho\Big( \mathbf{R}^{-1}_{\theta_j} \mathbf{g}^*_\text{III}(x_{\theta_j}\!)\! \Big) \text{d}x_{\theta_j},  ~  \!\!\!\!\!\!\!\!\!\!\!\!\!\!\!\!\!\!\!\!\!\! \int\limits_{[\mathbf{R}_{\theta_j} \mathbf{h}_{\text{IV}, \theta_j}(s_{(1)} + \Delta s_{(1)})]_2}^{[\mathbf{R}_{\theta_j} \mathbf{h}_{\text{IV}, \theta_j}(s_{(1)})]_2}  \!\!\!\!\!\!\!\!\!\!\!\!\!\!\!\!\!\!\!\!\!\!\!   \rho\Big( \mathbf{R}^{-1}_{\theta_j} \mathbf{g}^*_\text{IV}(y_{\theta_j}\!) \!\Big) \text{d}y_{\theta_j}. \\[-6.2ex]
\end{align*}
Here, each sub-event from above is weighted by `$\frac{1}{n_w} \frac{1}{n_\theta} w_i \Gamma_{\!q, \theta_j} \!(\!s_{(1)}\!)$,' where `$\frac{1}{n_w} \frac{1}{n_\theta}$' is the probability $\mathcal{R}_{w_i, \theta_j, \mathbf{c}}$ is selected and `$w_i \Gamma_{\!q, \theta_j}\! (\!s_{(1)}\!)$' is the probability its $\mathcal{E}_q$ edge center point falls in $\Omega_q$ to create a \emph{visible} reflection, as opposed to the other $\Omega_q$ regions (see Fig. \ref{Analytical_FW_Fig}).  Note that `$w_i \Gamma_{\!\text{I}, \theta_j}\! (\!s_{(1)}\!)$' was derived in the same manner as in (\ref{IntOverOmega}), and simply computes the area of $\Omega_\text{I}$, where each $\mathbf{x} \in \Omega_\text{I}$ is weighted by the probability its corresponding edge produces a visible reflection.  
% *** TO DO ***
% 1) Is this explanation above good??? ---> YES I BELIEVE SO! (10/16/2020)
We refer the reader to the proof of Lemma \ref{Lemma_Vs} for further details.  The other `$w_i \Gamma_{\!q, \theta_j} (\!s_{(1)}\!)$' are derived similarly.

	Next, it's easiest to work with these integrals when they are evaluated w.r.t. a common AOA variable.  Thus, consider the transformations: for $\Gamma_{\text{I}, \theta_j} (s_{(1)})$, let $x_{\theta_j} = \big[ \mathbf{R}_{\theta_j} \mathbf{h}_{\text{I}, \theta_j} \big( \psi_{\text{I}, \theta_j}^{-1} (\alpha) \big) \big]_1$, for $\Gamma_{\text{II}, \theta_j} (s_{(1)})$, let $y_{\theta_j} = \big[ \mathbf{R}_{\theta_j} \mathbf{h}_{\text{II}, \theta_j} \big( \psi_{\text{II}, \theta_j}^{-1} (\alpha) \big) \big]_2$, for $\Gamma_{\text{III}, \theta_j} (s_{(1)})$, let $x_{\theta_j} = \big[ \mathbf{R}_{\theta_j} \mathbf{h}_{\text{III}, \theta_j} \big( \psi_{\text{III}, \theta_j}^{-1} (\alpha) \big) \big]_1$, and for $\Gamma_{\text{IV}, \theta_j} (s_{(1)})$, let $y_{\theta_j} = \big[ \mathbf{R}_{\theta_j} \mathbf{h}_{\text{IV}, \theta_j} \big( \psi_{\text{IV}, \theta_j}^{-1} (\alpha) \big) \big]_2$. For $\Gamma_{\text{I}, \theta_j} (s_{(1)})$, this transformation implies  $\text{d} x_{\theta_j} = \big[ \mathbf{R}_{\theta_j} \frac{\text{d}}{\text{d} \alpha} \mathbf{h}_{\text{I}, \theta_j} \big( \psi_{\text{I}, \theta_j}^{-1} (\alpha) \big) \big]_1 \text{d} \alpha$, which yields
\vspace{-6pt}
\begin{align} \label{ApproximationsIntheLimit}
&\Gamma_{\text{I}, \theta_j} (s_{(1)}) \stackrel{(a)}{=} \int_{\psi_{\text{I}, \theta_j}(s_{(1)})}^{\psi_{\text{I}, \theta_j}(s_{(1)} + \Delta s_{(1)})}  \!\!\!\!\!\!\!\!\!\!\!\!\!\!\!  \rho\Big( \mathbf{h}_{\text{I}, \theta_j} \big( \psi_{\text{I}, \theta_j}^{-1} (\alpha) \big) \!\Big) \Big[ \mathbf{R}_{\theta_j} \textstyle\frac{\text{d}}{\text{d} \alpha} \mathbf{h}_{\text{I}, \theta_j} \big( \psi_{\text{I}, \theta_j}^{-1} (\alpha) \big) \Big]_{\! 1} \text{d} \alpha \\
&\!\!\stackrel{(b)}{=} \! \! \rho\Big( \mathbf{h}_{\text{I}, \theta_j} \!\big( s_{(1)} \big) \!\Big) \! \Big[ \mathbf{R}_{\theta_j} \textstyle\frac{\text{d}}{\text{d} \alpha} \!\big[ \mathbf{h}_{\text{I}, \theta_j} \!  \big( \psi_{\text{I}, \theta_j}^{-1} \! (\alpha) \big) \! \big]_{\alpha = \psi_{\text{I}, \theta_j}(s_{(1)})} \!  \Big]_{\! 1} \!  \Big( \psi_{\text{I}, \theta_j}\!(s_{(1)} \! + \! \Delta s_{(1)}) - \psi_{\text{I}, \theta_j}\!(s_{(1)}) \! \Big) \! \stackrel{(c)}{=} \! \omega_{\text{I}, \theta_j} \!( s_{(1)}) \Delta s_{(1)}, \nonumber \\[-6.5ex] \nonumber
\end{align}
where (a) applies the transformation, (b) is the evaluation of the integral as ``base-times-height'' when driving $\Delta s_{(1)} \to 0$, and (c) follows by simplifying $[ \cdot ]_1$ and by the definition of the derivative of $\psi_{\text{I}, \theta_j}(s_{(1)})$ when driving $\Delta s_{(1)} \to 0$ ($\omega_{\text{I}, \theta_j}( s_{(1)})$ is given in the lemma statement).  Following the same procedure, we obtain: $\Gamma_{\text{II}, \theta_j} (s_{(1)}) \!=\! \omega_{\text{II}, \theta_j}( s_{(1)}) \Delta s_{(1)}$, $\Gamma_{\text{III}, \theta_j} (s_{(1)}) \!=\! \omega_{\text{III}, \theta_j}( s_{(1)}) \Delta s_{(1)}$, and $\Gamma_{\text{IV}, \theta_j} (s_{(1)}) = \omega_{\text{IV}, \theta_j}( s_{(1)}) \Delta s_{(1)}$.  Finally, substituting these into (\ref{Expanded_Conditioning}), the $\Delta s_{(1)}$'s cancel, and taking $\lim_{\Delta s_{(1)} \to 0}$ ensures our approximations (b) and (c) in (\ref{ApproximationsIntheLimit}) are, in fact, equalities in the limit.
\end{proof}

\vspace{-15pt}
\begin{theorem}[$A_{(1)}$ Marginal Distribution] \label{DistA_1}
Consider the test link setup and Boolean model.  Then, the marginal PDF of $A_{(1)}$ is given by:
\vspace{-25pt}
\begin{align*}
f_{\! A_{(1)}} (\alpha_{(1)} \,|\, V_\infty \! \geq \! 1) = \! \sum\limits_{j = 1}^{n_\theta} \sum\limits_{q \in \mathcal{Q}} \left[ \! \frac{\Big(\psi_{q, \theta_j}^{-1}\Big)^\prime \!\! (\alpha_{(1)}) \, \omega_{q, \theta_j} \Big( \psi_{q, \theta_j}^{-1} (\alpha_{(1)}) \! \Big) f_{S_{(1)}}\Big( \psi_{q, \theta_j}^{-1} (\alpha_{(1)}) \, \Big|\, V_\infty \! \geq \! 1 \Big)}{\sum_{j^\prime = 1}^{n_\theta} \sum_{q^\prime \in \mathcal{Q}} \omega_{q^\prime, \theta_{j^\prime}} \Big( \psi_{q, \theta_j}^{-1} (\alpha_{(1)}) \Big)}  \beta_{q, \theta_j} (\alpha_{(1)}) \right], 
\end{align*}
where $\beta_{\text{\emph{I}}, \theta_j} (\alpha_{(1)}) = - \mathbbm{1}[\theta_j < \alpha_{(1)} \leq 2\theta_j]$, $\beta_{\text{\emph{II}}, \theta_j} (\alpha_{(1)}) = - \mathbbm{1}[\frac{\pi}{2} + \theta_j < \alpha_{(1)} \leq \pi]$, $\beta_{\text{\emph{III}}, \theta_j} (\alpha_{(1)}) = \mathbbm{1}[\pi \leq \alpha_{(1)} < \pi + \theta_j]$, $\beta_{\text{\emph{IV}}, \theta_j} (\alpha_{(1)}) = \mathbbm{1}[\pi + 2\theta_j \leq \alpha_{(1)} < \frac{3\pi}{2} + \theta_j]$, $\psi_{q, \theta_j}^{-1}$ and $\Big( \psi_{q, \theta_j}^{-1} \Big)^\prime$are given in Lemma \ref{Psi_Fns}, $\omega_{q, \theta_j}$ from Lemma \ref{CondAOA}, the distribution of $S_{(1)}\,|\, \{V_\infty \geq 1 \}$ from Theorem \ref{S1wB}, and $\mathbf{Supp}(A_{(1)} \, | \, \{ V_\infty \geq 1 \}) = \bigcup_{j = 1}^{n_\theta} \Big[ (\theta_j, 2\theta_j] \cup (\frac{\pi}{2} + \theta_j, \pi + \theta_j) \cup [\pi + 2\theta_j, \frac{3\pi}{2} + \theta_j) \Big]$.
\end{theorem}

\vspace{-12pt}
\begin{proof}
Please refer to Appendix \ref{PfDistA_1}.
\end{proof}

% *** TO DO ***
% 1) Put plots in degrees!

% --- NUMERICAL RESULTS ---
\begin{figure}
\centering
\begin{subfigure}{0.83\textwidth}
\centering
\includegraphics[scale=1, width=\linewidth]{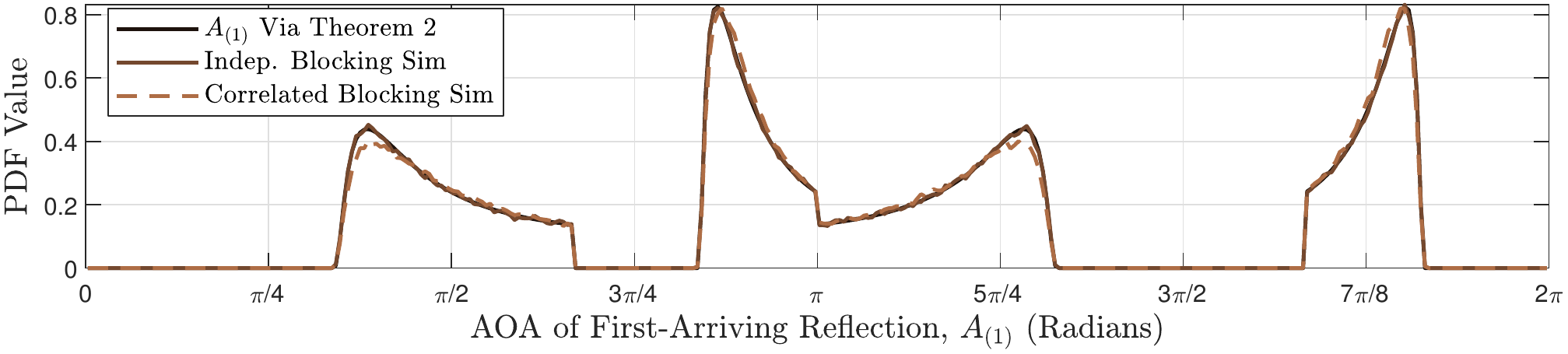}
\vspace{-20pt}
\caption{Reflector orientations: $\textbf{Supp}(\Theta) = \{60^\circ\}$ \\[1.5ex]} \label{AOA:a}
\end{subfigure}
\\
\begin{subfigure}{0.83\textwidth}
\centering
\includegraphics[width=\linewidth]{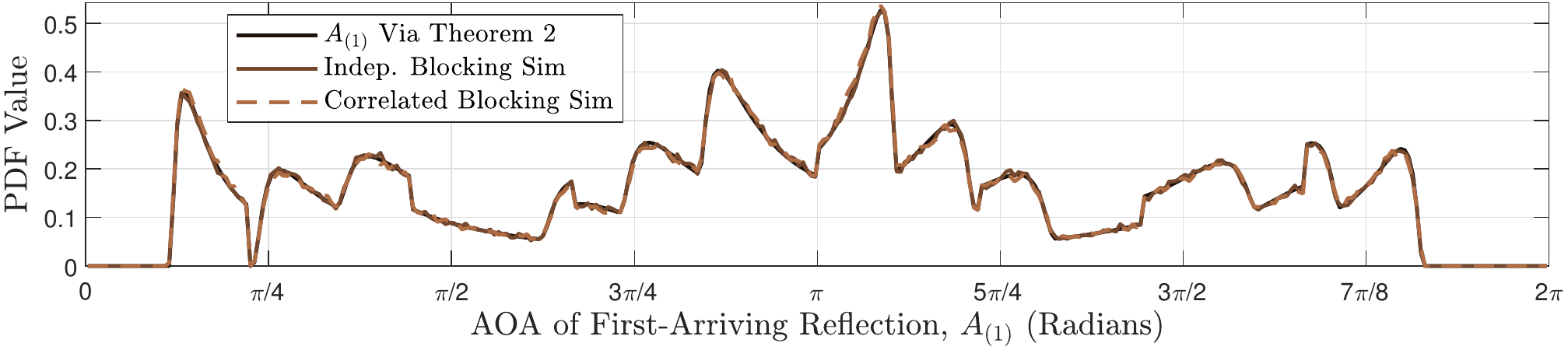}
\vspace{-20pt}
\caption{Reflector orientations: $\textbf{Supp}(\Theta) = \{20^\circ, 40^\circ, 60^\circ\}$ \\[1.5ex]} \label{AOA:b}
\end{subfigure}%
\\  % maximizeseparation between the subfigures
\begin{subfigure}{0.83\textwidth}
\centering
\includegraphics[width=\linewidth]{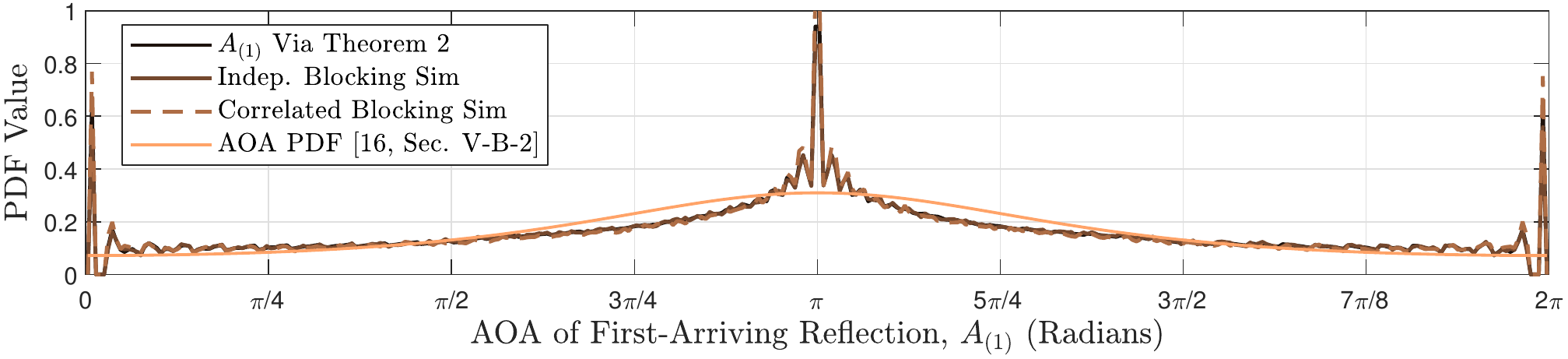}
\vspace{-20pt}
\caption{Reflector orientations: $\textbf{Supp}(\Theta) = \{1^\circ, 5^\circ, 9^\circ, \dots, 81^\circ, 85^\circ, 89^\circ \}$ \\[-1.5ex]} \label{AOA:c}
\end{subfigure}

\caption{\textsc{AOA Marginal PDFs of the 1$^\text{st}$-Arriving Reflection.}  PDFs were generated for a test link setup of $d = 350m$, a reflector density of $\lambda = 30$ reflectors per $km^2$, reflector widths sampled from $f_W = \text{unif}(w_{min} =10m, w_{max} = 40m, n_w = 4)$, and reflector orientations given under each plot.  Simulated PDFs were generated over 500,000 Boolean model realizations.
\\[-8.7ex]} \label{Fig_AOA}
\end{figure}
% --------------------------------	

\vspace{-23pt}
\subsection{Numerical Results} \label{Sec_Numerical_Results2}
\vspace{-3pt}

	This section compares our analytically derived AOA distribution of the first-arriving reflection (Theorem \ref{DistA_1}) against two AOA PDFs generated via simulation, labeled `\emph{Indep. Blocking Sim}' and `\emph{Correlated Blocking Sim}'.  The two simulated PDFs were generated in the exact same manner as in Sec. \ref{Sec_Numerical_Results1}, with the only difference being that the AOA of the first-arriving reflection was recorded rather than its path length.  First, note the remarkable overlap between Theorem \ref{DistA_1} and `Correlated Blocking Sim' which reveals how well the independent blocking assumption holds.  Next, the unique form of the distributions in Figs. \ref{AOA:a} and \ref{AOA:b} reveal that when reflectors take on one, or a few, orientations, such as buildings in a homogeneous city block for example, unique AOA profiles result, with certain AOAs being very prominent and others simply non-existent.  This highlights how Theorem \ref{DistA_1} can be used to optimize beam sweeping, for example, by greatly reducing the angular search space depending on the environment.  Lastly, we compare Theorem \ref{DistA_1} with an elliptical, omni-directional scattering model from the traditional channel modeling literature.  The `\emph{AOA PDF}' from \cite[Sec. V-B-2]{Ertel} in Fig. \ref{AOA:c} was generated under the same base station-mobile separation distance, and assumes there is one omni-directional point scatterer uniformly distributed over an $s_{max}$-ellipse (see Def. \ref{NLOSBE}), where $s_{max}$ was chosen s.t. $P[S_{(1)} \,|\, \{V_\infty \geq 1\} \leq s_{max}] \approx 0.75$, \emph{i.e.}, the analogous VRP associated with the first-arriving reflection in our model in Fig. \ref{AOA:c} would fall within $\mathcal{E}_{s_{max}}$ $75\%$ of the time.  It is fascinating to see that as we increase the number of reflector orientations, our AOA distribution in Theorem \ref{DistA_1} begins to approach that of the omni-directional scattering model, with the notable exception being the pronounced peaks coming from both in front of and behind the mobile.  To conclude, the distribution of $A_{(1)}$ presented here, derived under the Boolean model, is the first to capture the impact of environmental obstacles (\emph{i.e.} buildings) at mm-wave frequencies, which is insight that can not be gleaned from the more elementary omni-directional scattering models where blocking, the Specular Reflection Law, and reflectors with non-zero area are not considered.

% *******  CONCLUSION  *******
\vspace{-12pt}
\section{Conclusion}
\vspace{-4pt}

	Under a Boolean model of reflectors, which can facilitate or block reflections, and assuming NLOS propagation is due to first-order reflections, this paper presented the first analytical derivation of the TOA and AOA distributions of the first-arriving multipath component (MPC) experienced on a single link in outdoor, mm-wave (\emph{e.g.}, 5G) networks. 
	%of path length (\emph{i.e.}, time delay) distributions of the first $n$ arriving multipath components in outdoor, 5G, mm-wave networks.  
In so doing, the TOA of the first-arriving MPC was used to derive the distribution of the bias experienced on NLOS range measurements for localization.  It was shown that this analytically derived NLOS bias distribution: 1) matches closely with the decades-old exponential and gamma models assumed in the localization literature, thus offering the first support of these NLOS bias models based on the more accurate first-arriving MPC approach; and 2) gives intuition into how bias behaves when the environment of reflectors/blockages changes.  
Finally, numerical analysis of the AOA distribution revealed: 1) how reflector (\emph{e.g.}, building) orientations impact this distribution; and 2) how this distribution approaches the form of the AOA distribution of the first-arriving MPC from an elliptical, omni-directional scattering model as the number of reflector orientations increases.

% *******  APPENDICES  *******
\appendices

% --- Convergence of Integrals ---
\vspace{-23pt}
\section{\vspace{-4pt}Proof Extension of Lemma \ref{Lemma_Vs}: Convergence of Integrals in Intensity Measures}  \label{LemmaProofExt}
\vspace{-5pt}

	Here, we consider the worst case, \emph{i.e.}, $d = s_1$ and $s_2 = \infty$, and show that the integrals in the following intensity measures converge:
\vspace{-5pt}
\begin{align}\label{Int_1}
\Lambda_{\text{I}, v, w_i, \theta_j}(\Omega_\text{I}) = \!\frac{\lambda w_i}{n_w n_\theta} \! \int_{[\mathbf{R}_{\theta_j} \mathbf{m}]_1}^{\infty}  \!\!\!\!\!\!\!\!\!\!\!\!\!\!\!\!  \rho\Big( \mathbf{R}^{-1}_{\theta_j} \mathbf{g}^*_\text{I}(x_{\theta_j}) \Big) \text{d}x_{\theta_j}, ~\,  \Lambda_{\text{II}, v, w_i, \theta_j}(\Omega_\text{II}) = \frac{\lambda w_i}{n_w n_\theta} \! \int_{[\mathbf{R}_{\theta_j} \mathbf{b}]_2}^{\infty}  \!\!\!\!\!\!\!\!\!\!\!\!\!\!\! \rho\Big( \mathbf{R}^{-1}_{\theta_j} \mathbf{g}^*_\text{II}(y_{\theta_j}) \Big) \, \text{d}y_{\theta_j}. \\[-6.5ex] \nonumber
\end{align}

	Consider first the integral in $\Lambda_{\text{I}, v, w_i, \theta_j}\!(\Omega_\text{I})$.  We begin with two helpful derivations.  The first is:
	
\vspace{-27pt}
\begin{align} \label{1st_deriv}
\mathbf{R}_{\theta_j}^{-1} \mathbf{g}^*_\text{I} (x_{\theta_j}) = \big(4 x_{\theta_j}\big)^{-1}
\Big[(4 x_{\theta_j}^2 \!\!+\! d^2\!\sin^2\!\!\theta_j) \cos \theta_j, \, (4 x_{\theta_j}^2 \!\!-\! d^2\!\cos^2\!\!\theta_j) \sin \theta_j\Big]^T,\\[-5.5ex] \nonumber
\end{align}
%\begin{bmatrix} 
%(4 x_{\theta_j}^2 \!\!+\! d^2\!\sin^2\!\!\theta_j) \cos \theta_j \\
%(4 x_{\theta_j}^2 \!\!-\! d^2\!\cos^2\!\!\theta_j) \sin \theta_j 
%\end{bmatrix},\\[-5.5ex] \nonumber
%\end{align}
and the second is: $\rho(\mathbf{r}) \stackrel{(a)}{=}  e^{-\frac{\lambda}{n_w n_\theta} \sum_{i=1}^{n_w} \sum_{j=1}^{n_\theta} \Big[ \mu_2 \big(\mathcal{L}_{[\mathbf{b}, \mathbf{r}]} \oplus \mathcal{R}_{w_i, \theta_j} \big) + \mu_2 \big(\mathcal{L}_{[\mathbf{r}, \mathbf{m}]} \oplus \mathcal{R}_{w_i, \theta_j} \big)    \Big] }$
\vspace{-3pt}
\begin{align} \label{2nd_deriv}
 \stackrel{(b)}{<} e^{-\frac{\lambda}{n_w n_\theta} \sum_{i=1}^{n_w} \sum_{j=1}^{n_\theta} \big[ w_i \lVert \mathbf{b} - \mathbf{r} \rVert + w_i \lVert \mathbf{r} - \mathbf{m} \rVert \big] } \stackrel{(c)}{<} e^{-\frac{2 \lambda \mathbb{E}[W]}{n_\theta} \sum_{j=1}^{n_\theta} \lVert \mathbf{r} - \mathbf{m} \rVert }, \\[-6.5ex] \nonumber 
\end{align}
% = e^{-\frac{\lambda \mathbb{E}[W]}{n_\theta} \sum_{j=1}^{n_\theta} \big[ \lVert \mathbf{b} - \mathbf{r} \rVert + \lVert \mathbf{r} - \mathbf{m} \rVert \big] } 
where (a) follows from Lemma \ref{Lemma_PRPV} and Definition \ref{Def_BM}, (b) from Lemma \ref{Lemma_PRPV} where we have that $\forall \mathbf{p}, \mathbf{q} \in \mathbb{R}^2, \mu_2 \big(\mathcal{L}_{[\mathbf{p}, \mathbf{q}]} \oplus \mathcal{R}_{w, \theta} \big) \geq w \lVert \mathbf{p} - \mathbf{q} \rVert +w^2 >  w \lVert \mathbf{p} - \mathbf{q} \rVert$, and (c) from $\mathbb{E}[W] = \Big(\sum_{i=1}^{n_w} w_i \Big)/n_w$ and from the fact that $\mathbf{r} \in Q_\text{I} \cup \{\mathbf{m}\} \implies \lVert \mathbf{r} - \mathbf{m} \rVert < \lVert \mathbf{b} - \mathbf{r} \rVert$.  
	
	Next, substituting (\ref{1st_deriv}) into (\ref{2nd_deriv}) yields: $\rho \big( \mathbf{R}_{\theta_j}^{-1} \mathbf{g}^*_\text{I} (x_{\theta_j}) \big) <$
\vspace{-4pt}
\begin{align} \label{sqrt_1}
\exp \left[ -\frac{2 \lambda \mathbb{E}[W]}{n_\theta} \sum_{j=1}^{n_\theta} \textstyle \sqrt{ \! \left(  \frac{ \big(4 x_{\theta_j}^2 \! + d^2\sin^2\theta_j \big) \cos\theta_j } { 4 x_{\theta_j} } - \frac{d}{2}  \right)^2 \!\!\!+ \frac{ \big(4 x_{\theta_j}^2 \!- d^2\cos^2\theta_j \big)^2 \!\!\sin^2\theta_j } { 16 x_{\theta_j}^2 } } ~\right]. \\[-6.0ex] \nonumber 
\end{align}
The large square root term in the summand can be conveniently reduced to the following
\vspace{-3pt}
\begin{align}
%\frac{1}{4 x_{\theta_j}} &\sqrt{ \Bigg( x_{\theta_j} \!\!- \frac{d}{2} \cos \theta_j \Bigg)^2 \Bigg(\!16 x_{\theta_j}^2 \!\!+ 4 d^2 \! \sin^2 \! \theta_j \!\Bigg)  } =  
\Bigg( x_{\theta_j} \!\!- \frac{d}{2} \cos \theta_j \Bigg)  \sqrt{ \!1 \!+ \frac{d^2 \sin^2 \! \theta_j} {4 x_{\theta_j}^2}   } \geq  x_{\theta_j}\!\! - \frac{d}{2} \cos \theta_j  \label{sqrt_2} \\[-5.5ex] \nonumber
\end{align} 
where the inequality follows from the fact that $\sqrt{ 1 + (d^2 \sin^2 \! \theta_j) / (4 x_{\theta_j}^2) } \geq 1$, since $x_{\theta_j} \geq (d/2) \cos \theta_j = [\mathbf{R}_{\theta_j} \mathbf{m}]_1$ (lower limit of integral in $\Lambda_{\text{I}, v, w_i, \theta_j}(\Omega_\text{I})$ in (\ref{Int_1})).

	Finally, substituting (\ref{sqrt_2}) in for the square root term in (\ref{sqrt_1}) and integrating both sides yields
\vspace{-4pt}
\begin{align} \label{Int_Bnd_1}
\int_{[\mathbf{R}_{\theta_j} \mathbf{m}]_1}^{\infty}  \!\!\!\!\!\!\!\!\!\!\!\!\!\!\!\!  \rho\Big( \mathbf{R}^{-1}_{\theta_j} \mathbf{g}^*_\text{I}(x_{\theta_j}) \Big)  \text{d}x_{\theta_j} < \! \int_{[\mathbf{R}_{\theta_j} \mathbf{m}]_1}^{\infty} \!\!\!\!\!\!\!\!\!\!\!\! e^{ -\frac{2 \lambda \mathbb{E}[W]}{n_\theta} \sum_{j=1}^{n_\theta} \big[ x_{\theta_j} - \frac{d}{2} \cos \theta_j \big] } \text{d}x_{\theta_j} = \frac{1}{2 \lambda \mathbb{E}[W]}, \\[-6.5ex] \nonumber
%\frac{-1}{2 \lambda \mathbb{E}[W]} e^{ -\frac{2 \lambda \mathbb{E}[W]}{n_\theta} \sum_{j=1}^{n_\theta} \big[ x_{\theta_j} - \frac{d}{2} \cos \theta_j \big] } \Bigg|_{[\mathbf{R}_{\theta_j} \mathbf{m}]_1}^\infty
\end{align}
and thus, the integral in $\Lambda_{\text{I}, v, w_i, \theta_j}(\Omega_\text{I})$ in (\ref{Int_1}) is bounded.
	
	Following the same strategy, the integral in $\Lambda_{\text{II}, v, w_i, \theta_j}(\Omega_\text{II})$ in (\ref{Int_1}) can be shown to have the same bound as in (\ref{Int_Bnd_1}).  Thus, when $s_2=\infty$, $\Lambda_{\text{I}, v, w_i, \theta_j}(\Omega_\text{I})$ and $\Lambda_{\text{II}, v, w_i, \theta_j}(\Omega_\text{II})$ are bounded.
%\vspace{-8pt}
%\begin{align} \label{Int_Bnd_2}
%\int_{[\mathbf{R}_{\theta_j} \mathbf{b}]_2}^{\infty}  \!\!\!\!\!  \rho\Big( \mathbf{R}^{-1}_{\theta_j} \mathbf{g}^*_\text{II}(y_{\theta_j}) \Big) \, \text{d}y_{\theta_j} <  \frac{1}{2 \lambda \mathbb{E}[W]}.
%\end{align}

% --- Proof of S_(1), With blocking ---
\vspace{-14pt}
\section{\vspace{-4pt}Proof of Theorem \ref{S1wB}}  \label{PfS1wB}
\vspace{-37pt}
\begin{align*}
F_{S_{(1)}} &(s_{(1)} \,|\, V_\infty \geq 1) = P[S_{(1)} \leq s_{(1)} \,|\, V_\infty \geq 1] = P[V_{s_{(1)}} \geq 1 \,|\, V_\infty \geq 1] = \frac{P[V_{s_{(1)}} \geq 1, V_\infty \geq 1]} {P[V_\infty \geq 1]} \\
&= \frac{P[V_\infty \geq 1 \,|\, V_{s_{(1)}} \geq 1] \, P[V_{s_{(1)}} \geq 1]} {P[V_\infty \geq 1]} \stackrel{(a)}{=} \frac{ P[V_{s_{(1)}} \geq 1]} {P[V_\infty \geq 1]} \stackrel{(b)}{=}\frac{1}{1 - P[V_\infty = 0]} \bigg( 1 - P[V_{s_{(1)}} = 0] \bigg), \\[-6.4ex]
\end{align*}
where (a) follows from $P[V_\infty \geq 1 ~|~ V_{s_{(1)}} \geq 1] = 1$, and (b) yields the CDF in the theorem since $V_{s_{(1)}}$ and $V_\infty$ are Poisson distributed according to Lemma \ref{Lemma_Vs}.

	The PDF is obtained by noting the following: $f_{S_{(1)}} (s_{(1)} | V_\infty \! \geq 1) = $
\vspace{-5pt}
\begin{align*}
&\frac{\partial}{\partial s_{(1)}} F_{S_{(1)}} (s_{(1)} \,|\, V_\infty \geq 1) \,=\,\frac{1}{1 - e^{-\hat{\lambda}(\infty)}} e^{-\hat{\lambda}\big(s_{(1)}\big)} \frac{\partial}{\partial s_{(1)}} \Big[\hat{\lambda}\big(s_{(1)}\big)\Big] \stackrel{(a)}= \frac{2 \lambda \mathbb{E}[W]} {n_\theta \Big(1 - e^{-\hat{\lambda}(\infty)} \Big)} e^{-\hat{\lambda}\big(s_{(1)}\big)}  \times \\ 
&\sum_{j=1}^{n_\theta} \! \Bigg[ \! \rho\bigg(\!\mathbf{R}_{\theta_j}^{-1} \mathbf{g}_\text{I}^* \Big(\! \big[ \mathbf{R}_{\theta_j} \mathbf{h}_\text{I}(s_{(1)}\!) \big]_1 \Big) \!\bigg) \frac{\partial}{\partial s_{(1)}} \!\big[ \mathbf{R}_{\theta_j} \mathbf{h}_\text{I}(s_{(1)}\!) \big]_1 \!+\! \rho\bigg(\!\mathbf{R}_{\theta_j}^{-1} \mathbf{g}_\text{II}^* \Big( \!\big[ \mathbf{R}_{\theta_j} \mathbf{h}_\text{II}(s_{(1)}\!) \big]_2 \Big) \!\bigg)  \frac{\partial}{\partial s_{(1)}}\! \big[ \mathbf{R}_{\theta_j} \mathbf{h}_\text{II}(s_{(1)}\!) \big]_2 \! \Bigg], \\[-7ex]
\end{align*}
where (a) follows from Leibnitz's Rule \cite[Theorem 2.4.1] {Stats_Book}.  Then, the PDF expression stated in the theorem is achieved from (a) via the following simplifications:
\vspace{-9pt}
\begin{align*}
\rho\bigg(\mathbf{R}_{\theta_j}^{-1} \mathbf{g}_\text{I}^* \Big( \big[ \mathbf{R}_{\theta_j} \mathbf{h}_\text{I}(s_{(1)}) \big]_1 \Big) \!\bigg) = \rho\big(\mathbf{h}_{\text{I}, \theta_j}(s_{(1)}) \big), ~~~~~~~
\rho\bigg(\mathbf{R}_{\theta_j}^{-1} \mathbf{g}_\text{II}^* \Big( \big[ \mathbf{R}_{\theta_j} \mathbf{h}_\text{II}(s_{(1)}) \big]_2 \Big) \!\bigg) = \rho\big(\mathbf{h}_{\text{II}, \theta_j}(s_{(1)}) \big), \\
\frac{\partial}{\partial s_{(1)}} \big[ \mathbf{R}_{\theta_j} \mathbf{h}_\text{I}(s_{(1)}) \big]_1 = \frac{s_{(1)}} {2\sqrt{s_{(1)}^2 - d^2 \sin^2\theta_j }}, ~~\text{and} ~~~
\frac{\partial}{\partial s_{(1)}} \big[ \mathbf{R}_{\theta_j} \mathbf{h}_\text{II}(s_{(1)}) \big]_2 = \frac{s_{(1)}} {2\sqrt{s_{(1)}^2 - d^2 \cos^2\theta_j }},\\[-7ex]
\end{align*}
where $\theta_j$ was added to $\mathbf{h}_\text{I}$, $\mathbf{h}_\text{II}$ to emphasize the dependence.  The support follows from Def. \ref{DefRVS1}.% and from the fact that these distributions were derived via $V_s$ from Lemma \ref{Lemma_Vs}.

% --- Proof of A_(1), With blocking ---
\vspace{-9pt}
\section{\vspace{-4pt}Proof of Theorem \ref{DistA_1}}  \label{PfDistA_1}
\vspace{-37pt}
\begin{align*}
&f_{\! A_{(1)}} (\alpha_{(1)} \,|\, V_\infty \! \geq \! 1) =\lim_{s_{max} \to \infty} \int_d^{s_{max}} \!\!\!\!\!\!\!\! f_{A_{(1)}} (\alpha_{(1)} \, | \, S_{(1)}, V_\infty \geq 1 ) \, f_{S_{(1)}} (s_{(1)} \, | \, V_\infty \geq 1) \,  \text{d}s_{(1)} \\
&\stackrel{(a)}{=} \sum\limits_{j = 1}^{n_\theta} \sum\limits_{q \in \mathcal{Q}}  \lim_{s_{max} \to \infty} \int_d^{s_{max}}  \frac{\omega_{q, \theta_j} (s_{(1)}) f_{S_{(1)}}( s_{(1)} \, |\, V_\infty \! \geq \! 1 )}{\sum_{j^\prime = 1}^{n_\theta} \sum_{q^\prime \in \mathcal{Q}} \omega_{q^\prime, \theta_{j^\prime}} (s_{(1)})} \delta \Big( \alpha_{(1)} - \psi_{q, \theta_j} (s_{(1)}) \Big) \, \text{d} s_{(1)} \\
&\stackrel{(b)}{=} \sum\limits_{j = 1}^{n_\theta} \sum\limits_{q \in \mathcal{Q}}  \lim_{s_{max} \to \infty} \int_{\psi_{q, \theta_j}(d)}^{\psi_{q, \theta_j}(s_{max})}  \frac{\Big(\psi_{q, \theta_j}^{-1}\Big)^\prime \!\! (\alpha) \, \omega_{q, \theta_j} \Big( \psi_{q, \theta_j}^{-1} (\alpha) \! \Big) f_{S_{(1)}}\Big( \psi_{q, \theta_j}^{-1} (\alpha) \, \Big|\, V_\infty \! \geq \! 1 \Big)}{\sum_{j^\prime = 1}^{n_\theta} \sum_{q^\prime \in \mathcal{Q}} \omega_{q^\prime, \theta_{j^\prime}} \Big( \psi_{q, \theta_j}^{-1} (\alpha_{(1)}) \Big)} \delta \big( \alpha_{(1)} - \alpha \big) \, \text{d} \alpha,
\end{align*}
where (a) follows from Lemma \ref{CondAOA} and by finite additivity of limits and integrals, and (b) from the substitution: $\alpha = \psi_{q, \theta_j}(s_{(1)}) \!\implies\! \psi_{q, \theta_j}^{-1}(\alpha) = s_{(1)} \!\implies\! \Big( \psi_{q, \theta_j}^{-1} \Big)^\prime \! (\alpha) \,\text{d} \alpha = \text{d} s_{(1)}$.  The theorem follows by expanding the sum over $q \in \mathcal{Q}$, computing the integral limits via Table \ref{sAOA_functions}, and computing the integrals by noting $\delta(\cdot)$ and the $\alpha_{(1)}$ values for which the integrals are non-zero.

% *******   REFERENCES   *******


\vspace{-14pt}
\begin{thebibliography}{99}
\vspace{-3pt}
\footnotesize
%\onehalfspacing
%\setstretch{1.00}

% *** TO DO ***
% 2) Perhaps also reference the francescettii paper with all the reflections (referenced in Heath I believe)


\bibitem{Globecom} C. E. O'Lone, H. S. Dhillon, and R. M. Buehrer, ``A mathematical justification for exponentially distributed NLOS bias,'' in \emph{Proc. of the IEEE Global Commun. Conf.}, Waikoloa, HI, USA, Dec. 2019, pp. 1-6.




% *** INTRODUCTION ***

\bibitem{Stoyan} S. N. Chiu, D. Stoyan, W. S. Kendall, and J. Mecke, \emph{Stochastic Geometry and its Applications}, 3$^\text{rd}$ ed., West Sussex, UK: Wiley, 2013.


% Boolean Model 
\bibitem{Heath} T. Bai, R. Vaze, and R. W. Heath, ``Analysis of blockage effects on urban cellular networks," in \emph{IEEE Trans. on Wireless Commun.}, vol. 13, no. 9, pp. 5070-5083, Sept. 2014.


% THIS IS A REALLY GOOD OVERVIEW PAPER!
\bibitem{AndrewsHeath} J. G. Andrews, T. Bai, M. N. Kulkarni, A. Alkhateeb, A. K. Gupta, and R. W. Heath, ``Modeling and analyzing millimeter wave cellular systems," in \emph{IEEE Trans. on Commun.}, vol. 65, no. 1, pp. 403-430, Jan.$\,\,$2017.
% Jan. 2017

% Nor
\bibitem{Nor} N. A. Muhammad, P. Wang, Y. Li, and B. Vucetic, ``Analytical model for outdoor millimeter wave channels using geometry-based stochastic approach," in \emph{IEEE Trans. on Veh. Technol.}, vol. 66, no. 2, pp. 912-926, Feb. 2017.


% G. Das
\bibitem{GDasConf} R. T. Rakesh, G. Das, and D. Sen, ``An analytical model for millimeter wave outdoor directional non-line-of-sight channels," in \emph{Proc. of the IEEE Int. Conf. on Commun. (ICC)}, Paris, France, May 2017, pp. 1-6.

%\bibitem{GDasJournal} R. T. Rakesh, D. Sen, and G. Das, ``A novel geometry-based stochastic double directional analytical model for millimeter wave outdoor NLOS channels,'' in \emph{arXiv:1804.02831}, 2018.


\bibitem{Miaomiao} M. Dong and T. Kim, ``Interference analysis for millimeter-wave networks with geometry-dependent first-order reflections," in \emph{IEEE Trans. on Veh. Technol.}, vol. 67, no. 12, pp. 12404-12409, Dec. 2018.


\bibitem{Comms_Letter} C. E. O'Lone, H. S. Dhillon, and R. M. Buehrer, ``Single-anchor localizability in 5G millimeter wave networks," in \emph{IEEE Wireless Commun. Lett.}, vol. 9, no. 1, pp. 65-69, Jan. 2020.


\bibitem{Aroon} A. Narayanan, S. T. Veetil, and R. K. Ganti, ``Coverage analysis in millimeter wave cellular networks with reflections," in \emph{Proc. of the IEEE Global Commun. Conf.}, Singapore, Dec. 2017, pp. 1-6.


% Positive uniform (Ch. 6)
%\bibitem{DrB} R. Zekavat and R. M. Buehrer, \emph{Handbook of Position Location: Theory, Practice and Advances}. Hoboken, NJ, USA: Wiley, 2012.
\bibitem{DrB} \emph{Handbook of Position Location: Theory, Practice, and Advances}, 2$^\text{nd}$ ed., Wiley, Hoboken, NJ, 2019, pp. 224.


% Exponential model and is improved with a priori knowledge of the bias statistics
\bibitem{Cong} L. Cong and W. Zhuang, ``Nonline-of-sight error mitigation in mobile location," in \emph{IEEE Trans. on Wireless Commun.}, vol. 4, no. 2, pp. 560-573, Mar. 2005.

\bibitem{Reza} R. M. Vaghefi and R. M. Buehrer, ``Cooperative sensor localization with NLOS mitigation using semidefinite programming," in \emph{Proc. of the Workshop on Positioning, Navigation, and Commun.}, Dresden, Germany, Mar. 2012, pp. 13-18.

% Qi NLOS Journal
\bibitem{Qi_NLOS_Journal} Y. Qi, H. Kobayashi, and H. Suda, ``Analysis of wireless geolocation in a non-line-of-sight environment,'' in \emph{IEEE Trans. on Wireless Commun.}, vol. 5, no. 3, pp. 672-681, Mar. 2006.





\bibitem{Mailaender} L. Mailaender, ``On the geolocation bounds for round-trip time-of-arrival and all non-line-of-sight channels,'' in \emph{EURASIP J. on Advances in Signal Process.}, vol. 2008, no. 1, pp. 1-10, Oct. 2007.

%Gaussian mixture where bias is a shifted gaussian or rayleigh
\bibitem{GMix1} F. Yin, C. Fritsche, F. Gustafsson, and A. M. Zoubir, ``TOA-based robust wireless geolocation and Cram\'{e}r-Rao lower bound analysis in harsh LOS/NLOS environments," in \emph{IEEE Trans. on Signal Process.}, vol. 61, no. 9, pp. 2243-55, May 2013. 


\bibitem{Unif} S. Nawaz and N. Trigoni, ``Robust localization in cluttered environments with NLOS propagation," in \emph{Proc of the IEEE Int.l Conf. on Mobile Ad-hoc and Sensor Systems}, San Fran., CA, USA, Nov. 2010, pp. 166-175.


% Gamma (2002)
\bibitem{Qi_2} Y. Qi and H. Kobayashi, ``On geolocation accuracy with prior information in non-line-of-sight environment," in \emph{Proc. of the IEEE 56th Veh. Technol. Conf.}, Vancouver, BC, Canada, Sept. 2002, pp. 285-288.





\bibitem{Ertel}  R. B. Ertel and J. H. Reed, ``Angle and time of arrival statistics for circular and elliptical scattering models,'' in \emph{IEEE J. Select. Areas in Commun.}, vol. 17, no. 11, pp. 1829-1840, Nov. 1999.

\bibitem{Wu} J.-F. Kiang and C.-W. Wu, ``NLOS effects on position location techniques," in \emph{IEEE Int. Conf. on Networking, Sensing and Control}, Taipei, Taiwan, Mar. 2004, pp. 305-308


% Chen Model
\bibitem{Chen} P. Chen, ``A non-line-of-sight error mitigation algorithm in location estimation,'' in \emph{Proc. of the IEEE Wireless Commun. and Networking Conf.}, New Orleans, LA, Sept. 1999, pp. 316-320.

\bibitem{Swaroop} S. Venkatesh, ``The design and modeling of ultra-wideband position-location networks,'' Ph.D Dissertation, Dept. of Elec. and Comp. Eng., Virginia Tech, Blacksburg, VA, 2007.

% Multipath Components used in TOA
\bibitem{Qi_Multipath} Y. Qi, H. Suda, and H. Kobayashi, ``On time-of-arrival positioning in a multipath environment,'' in \emph{Proc. of the IEEE 60th Veh. Technol. Conf.}. Los Angeles, CA, USA, Sept. 2004, pp. 3540-3544.






% Initial Access Paper
\bibitem{IAPaper} Y. Li, J. G. Andrews, F. Baccelli, T. D. Novlan, and C. J. Zhang, ``Design and analysis of initial access in millimeter wave cellular networks,'' in \emph{IEEE Trans. on Wireless Commun.}, vol. 16, no. 10, pp. 6409-6425, Oct. 2017.


% Omni-directional circular and elliptical AOA scattering models
\bibitem{Swamy} R. Janaswamy, ``Angle and time of arrival statistics for the Gaussian scatter density model," in \emph{IEEE Trans. on Wireless Commun.}, vol. 1, no. 3, pp. 488-497, July 2002.


% THz, discusses smooth surfaces for reflections
\bibitem{SmoothSurfaces} S. Ju \emph{et al.}, ``Scattering mechanisms and modeling for terahertz wireless communications,'' in \emph{Proc. of the IEEE Int. Conf. on Commun. (ICC)}, Shanghai, China, May  2019, pp. 1-7.












% *** SYSTEM MODEL ***

%\bibitem{Henk} R. Mendrzik, H. Wymeersch, G. Bauch, and Z. Abu-Shaban, ``Harnessing NLOS components for position and orientation estimation in 5G millimeter wave MIMO," in \emph{IEEE Trans. on Wireless Commun.}, vol. 18, no. 1, pp. 93-107, Jan. 2019.

% Other paper that considers independent-independent blocking
\bibitem{IndepIndep} A. A. AbdelNabi, V. Mancuso, and M. A. Marsan, ``On the outage probability of millimeter wave links with quasi-deterministic propagation,'' in \emph{Proc. of the 3$^{\text{rd}}$ ACM Workshop on Millimeter-wave Networks and Sensing Systems}, Los Cabos, Mexico, Oct. 2019, pp. 1-6.



% Correlated Blocking
\bibitem{Aditya} S. Aditya, H. S. Dhillon, A. F. Molisch, and H. M. Behairy, ``A tractable analysis of the blind spot probability in localization networks under correlated blocking," in \emph{IEEE Trans. on Wireless Commun.}, vol. 17, no. 12, pp. 8150-64, Dec. 2018.

% ---- Decided to get rid of this to save space and just cited my conference paper in these instances instead ---
%\bibitem{Stoyan} S. N. Chiu, D. Stoyan, W. S. Kendall, and J. Mecke, \emph{Stochastic Geometry and its Applications}, 3$^\text{rd}$ ed., West Sussex, UK: Wiley, 2013.

%\bibitem{LittleBook} I. Molchanov, \emph{Statistics of the Boolean Model for Practitioners and Mathematicians}. West Sussex,  UK: John Wiley \& Sons, Ltd., 1997.



% Diffraction Negligible
%\bibitem{First-Order_Reflection} Hao Xu, V. Kukshya, and T. S. Rappaport, ``Spatial and temporal characteristics of 60-GHz indoor channels," in \emph{IEEE J. on Sel. Areas in Commun.}, vol. 20, no. 3, pp. 620-630, April 2002.

%\bibitem{Javier_Journal} J. Schloemann, H. S. Dhillon, and R. M. Buehrer, ``Towards a tractable analysis of localization fundamentals in cellular networks," in \emph{IEEE Trans. on Wireless Commun.}, vol. 15, no. 3, pp. 1768-1782, Mar. 2016.






% *** APPROXIMATIONS ***

\bibitem{CoverAndThomas} T. M. Cover and J. A. Thomas, \emph{Elements of Information Theory}, $2^{\text{nd}}$ ed. Hoboken, NJ, USA:Wiley, 2006.





% *** NUMERICAL RESULTS ***

% BS Deployment Densities in 5G Networks
%\bibitem{David} M. Ding, P. Wang, D. L\'{o}pez-P\'{e}rez, G. Mao, and Z. Lin, ``Performance impact of LoS and NLoS transmissions in dense cellular networks," in \emph{IEEE Trans. on Wireless Commun.}, vol. 15, no. 3, pp. 2365-2380, Mar. 2016.


% NLOS is killer issue
\bibitem{killer_issue} M. P. Wylie and J. Holtzman, ``The non-line of sight problem in mobile location estimation," in \emph{ Proc. of ICUPC - 5th Int. Conf. on Universal Personal Commun.}, Cambridge, MA, USA, Oct. 1996, pp. 827-831.





% *** DISCUSSION ***


% NLOS Bias Characterization Doesn't Exist
\bibitem{GSM} M. I. Silventoinen and T. Rantalainen, ``Mobile station emergency locating in GSM,'' in \emph{Proc. of the IEEE Int. Conf. on Personal Wireless Commun.}, New Delhi, India, Feb. 1996, pp. 232-238.

% Channel Sounding
\bibitem{George_Channel_Sounder} G. R. MacCartney and T. S. Rappaport, ``A flexible millimeter-wave channel sounder with absolute timing," in \emph{IEEE J. on Sel. Areas in Commun.}, vol. 35, no. 6, pp. 1402-1418, June 2017.

\bibitem{Samimi} M. K. Samimi and T. S. Rappaport, ``3-D millimeter-wave statistical channel model for 5G wireless system design,'' in \emph{IEEE Trans. on Microwave Theory and Techn.}, vol. 64, no. 7, pp. 2207-2225, July 2016.

% NYUSIM
%\bibitem{NYUSIM1} S. Sun, G. R. MacCartney Jr., and T. S. Rappaport, ``A novel millimeter-wave channel simulator and applications for 5G wireless communications,'' in \emph{Proc. of the IEEE Global Commun. Conf.}, Paris, France, May 2017, pp. 1-7.

\bibitem{NYUSIM2} S. Ju, O. Kanhere, Y. Xing, and T. S. Rappaport, ``A millimeter-wave channel simulator NYUSIM with spatial consistency and human blockage,'' in \emph{Proc. of the IEEE Global Commun. Conf.}, Waikoloa, HI, USA, Dec. 2019, pp. 1-6.

% Useful papers I never mentioned:
% 1) Modeling of The Distance Error for Indoor Geolocation 


% Reasoning to argue for exponentially distributed NLOS bias (Similar to Swaroop)
%\bibitem{Mitsubishi} S. Gezici and Z. Sahinoglu, ``UWB geolocation for IEEE 802.15.4a personal area networks,'' Mitsubishi Elect. Res. Lab., Cambridge, MA, USA, Rep. TR-2004-110, Aug. 2004.







% *** APPENDIX ***

\bibitem{Stats_Book} G. Casella and R. L. Berger, \emph{Statistical Inference}, $2^{\text{nd}}$ ed. Belmont, CA, USA:Brooks/Cole, 2002.



% *** 1) Delete this if I need space ***
%\bibitem{GDasJournal} R. T. Rakesh, D. Sen, and G. Das, ``A novel geometry-based stochastic double directional analytical model for millimeter wave outdoor NLOS channels,'' in \emph{arXiv:1804.02831}, 2018.

% ---NOTES ---


% Aroon (First arriving path under different assumptions, not relevant for localizability probability.)


% Specular Reflection Law
%\bibitem{BeamBased} C. Tatino, I. Malanchini, D. Aziz, and D. Yuan, ``Beam based stochastic model of the coverage probability in 5G millimeter wave systems," in \emph{15th International Symposium on Modeling and Optimization in Mobile, Ad Hoc, and Wireless Networks (WiOpt)}, Paris, France, May 2017, pp. 1-6.





% --- Boolean Model Books/Papers

%\bibitem{StochasticGeometry} M. Haenggi, \emph{Stochastic Geometry for Wireless Networks}. New York: Cambridge University Press, 2013.

%\bibitem{Stoyan_Paper} D. Stoyan, ``On some qualitative properties of the Boolean model of stochastic geometry,'' in \emph{Journal of Applied Mathematics and Mechanics}, vol. 59, no. 9, pp. 447-454, 1979.

%\bibitem{Cowan} R. Cowan, ``Objects randomly distributed in space: an accessible theory,'' in \emph{Advances in Applied Probability}, vol. 21, no. 3, pp. 543-569, Sept. 1989.

%\bibitem{Calka} P. Calka \emph{et al.}, ``Stochastic geometry: Boolean model and random geometric graphs,'' in \emph{ESAIM: Proceedings and Surveys}, vol. 51, pp. 175-192, Oct. 2015.

%\bibitem{PPTheory} D. J. Daley and D. Vere-Jones, \emph{An Introduction to the Theory of Point Processes: Volume 1: Elementary Theory and Methods, Second Edition}. New York: Springer, 2003.





% Haenggi coverage analysis...The LOS ball model is used here



% --- Assumption 5 (Diffraction) ---
%\bibitem{RayTracing} Z. Zhang, J. Ryu, S. Subramanian, and A. Sampath, ``Coverage and channel characteristics of millimeter wave band using ray tracing," in \emph{IEEE International Conference on Communications (ICC)}, London, UK, June 2015, pp. 1380-1385.






\end{thebibliography}
\end{document}